\documentclass[a4paper,11pt]{article}
\usepackage{jcappub}
\usepackage{amsmath}
\usepackage{amssymb}
\usepackage[T1]{fontenc}
\usepackage{siunitx}
\usepackage{listings}
\usepackage[normalem]{ulem}
\usepackage{bm}
\usepackage{cases}
\usepackage{upgreek}
\usepackage{multirow}
\usepackage{enumitem}
\usepackage{lineno}
\usepackage{xcolor}
\usepackage{cancel}
\usepackage{makecell}
\usepackage{float}
\allowdisplaybreaks

\newcommand{\Pgw}{\mathcal{P}_{\text{gw}}}

\newcommand{\tv}{\tilde{v}}
\newcommand{\tlambda}{\tilde{\lambda}}
\newcommand{\detav}{\Delta\eta_{\text{v}}}
\newcommand{\abs}[1]{\left\vert #1 \right\vert}

\title{Acoustic gravitational waves beyond leading-order in bubble over Hubble radius}

\author[a]{Lorenzo Giombi,}
\author[a]{Jani Dahl,}
\author[a,b]{Mark Hindmarsh}

\affiliation[a]{Department of Physics and Helsinki Institute of Physics,\\PL 64, FI-00014 University of Helsinki, Finland}
\affiliation[b]{Department of Physics and Astronomy, University of Sussex,\\Brighton BN1 9QH, United Kingdom}

\emailAdd{lorenzo.giombi@helsinki.fi}
\emailAdd{jani.dahl@helsinki.fi}
\emailAdd{mark.hindmarsh@helsinki.fi}

\abstract{We calculate the gravitational wave power spectrum from sound waves in a cosmological first-order phase transition in the unexplored regime of large bubbles, by which we mean that the mean bubble spacing $R_*$ is a non-negligible fraction of the Hubble length $\mathcal{H}_*^{-1}$, i.e.  $R_*\mathcal{H}_* \lesssim \mathcal{O}(1)$. Since the amplitude of the gravitational wave signal increases with $R_*\mathcal{H}_*$, this is also the loud signal regime. In this regime the effects of gravity, hitherto neglected,  become relevant. We carry out the calculation in cosmological perturbation theory expanding in the parameter $R_*\mathcal{H}_*$, or bubble over Hubble radius. The leading-order term is the standard result for acoustic production of gravitational waves. At next-to-leading-order we find three novel contributions: two contributions arise from general relativistic corrections to the dynamics of both  sound waves and gravitational waves. A third contribution  comes from secondary gravitational waves induced by curvature perturbations. These contributions suppress the gravitational wave peak amplitude. The suppression factor relative to the leading-order contribution scales as $(R_*\mathcal{H}_*)^2$, and also depends on other transition parameters, such as the sound speed $c_s$, the duration of the acoustic source, and the peak wavenumber of the velocity field $k_p$. We investigate the range $0.3 \lesssim R_*\mathcal{H}_* \lesssim 0.7$ in a simplified model of the velocity field, finding that the suppression factor lies between $2\%$ and $15\%$ when $R_*\mathcal{H}_* \simeq 0.5$, but is independent of the root mean squared fluid velocity. We provide analytical approximations to the next-to-leading-order corrections, and a recipe to join them smoothly across different frequency regimes. Our work improves the precision of current estimates of the gravitational wave power spectrum in the relatively unexplored regime of phase transitions with large bubbles. 
}

\begin{document}
\maketitle
\flushbottom

\section{Introduction}\label{sec:introduction}
Cosmological gravitational waves are unique probes of the Universe before recombination. The weak gravitational coupling allows gravitational waves to stream freely throughout the Universe, forming a stochastic background that inherits the features of the early Universe and of the physical mechanisms that produced them~\cite{Maggiore:1999vm}. In June 2023, the NANOGrav~\cite{NANOGrav:2023gor, NANOGrav:2023hvm} collaboration, as well as the European Pulsar Timing Array (EPTA)~\cite{EPTA:2023fyk, EPTA:2023xxk}, the Parkes Pulsar Timing Array (PPTA)~\cite{Reardon:2023gzh}, and the Chinese Pulsar Timing Array (CPTA)~\cite{Xu:2023wog},  independently reported evidence for the presence of a stochastic background of gravitational waves at $\si{\nano\Hz}$ frequencies. 
A higher frequency stochastic gravitational wave background will be within reach of the next generation of gravitational wave interferometers such as LISA (Laser Interferometer Space Antenna)~\cite{LISA:2017pwj, LISACosmologyWorkingGroup:2022jok}, possibly revealing a large population of compact binary systems around the milli-Hertz frequency band, both galactic~\cite{Evans:1987qa, Bender:1997hs, Cornish:2017vip} and extragalactic~\cite{Regimbau:2011rp}.

In addition, LISA will offer an unprecedented opportunity to investigate the early Universe at the electroweak scale. One possible source within LISA’s frequency window is a first-order phase transition resulting from the breaking of the electroweak symmetry~\cite{Caprini:2015zlo, Caprini:2019egz}. This process naturally arises in many theories beyond the Standard Model at high energy. One of the biggest challenges for LISA will be to disentangle all the individual sources that contribute to the stochastic gravitational wave background, and in particular to distinguish the signal of a first-order phase transition from the strong astrophysical foreground. 
Precise theoretical models of phase transitions are therefore crucial for discriminating the features of the emitted gravitational wave signal from other sources. Finally, through robust models, it may be possible to connect gravitational wave observations with the underlying particle theory. 

Cosmological first-order phase transitions~\cite{Kirzhnits:1972iw, Kirzhnits:1972ut, Kirzhnits:1976ts, Kibble:1980mv} proceed with the nucleation, expansion, and percolation of bubbles of the stable (low-temperature) phase~\cite{Witten:1980ez, Guth:1981uk, Steinhardt:1980wx}.
The potential energy difference between the two phases is converted, throughout the expansion, into kinetic and thermal energy of the fluid around bubbles, which thereby becomes a source of anisotropic stress~\cite{Steinhardt:1981ct}, and thus gravitational radiation~\cite{Witten:1984rs, Hogan:1986qda}.

Numerical simulations of weak phase transitions indicate that the expanding bubbles generate mainly compressional modes in the surrounding fluid~\cite{Hindmarsh:2013xza, Hindmarsh:2015qta, Hindmarsh:2017gnf}.
Therefore, right after the merging of the new-phase bubbles, the bulk fluid motion consists mainly of compressional sound waves in the plasma (\textit{acoustic phase})~\cite{Hindmarsh:2013xza,Hindmarsh:2015qta,Hindmarsh:2016lnk,Hindmarsh:2017gnf,Hindmarsh:2019phv, Jinno:2020eqg, Jinno:2022mie, Cai:2023guc}, unless the transition has significant latent heat and proceeds via deflagrations~\cite{Cutting:2019zws}. 
Afterwards, due to the large Reynolds number of the cosmic fluid in the early Universe~\cite{Arnold:2000dr}, the fluid evolves non-linearly and can develop and maintain turbulence, both acoustic~\cite{Pen:2015qta, RoperPol:2019wvy, Dahl:2021wyk, Dahl:2024eup, Caprini:2024gyk} and vortical ~\cite{Kosowsky:2001xp, Gogoberidze:2007an, Caprini:2009fx, Caprini:2009yp, RoperPol:2019wvy, Auclair:2022jod, RoperPol:2022iel, Caprini:2024gyk}. 
In acoustic turbulence, the non-linearities are in the form of shocks, which are the principal sites of dissipation of kinetic energy. 
Shocks in the fluid form on a characteristic timescale estimated as $\eta_{\rm{sh}} \sim R_*/v_{\text{rms}}$,  turning the acoustic phase into a turbulent phase~\cite{Pen:2015qta, Dahl:2021wyk}. Here $R_*$ is the bubble mean spacing, corresponding to the typical correlation length of the fluid flow, and $v_{\text{rms}}$ is the root-mean-squared velocity of the fluid motion. 
Acoustic and vortical turbulence plays a significant role in the evolution of the fluid flow for strong phase transitions, where the initial value of $v_{\text{rms}}$ is typically $\mathcal{O}(1)$ and the onset of turbulence comes soon after the transition. 


Other sources of gravitational waves that are not discussed in this work are the collisions between bubbles' interfaces~\cite{Kosowsky:1991ua, Kosowsky:1992rz, Kosowsky:1992vn, Huber:2008hg, Jinno:2017fby, Cutting:2018tjt}. Gravitational waves from bubble collisions are typically subdominant compared to acoustic production in transitions proceeding at subluminal speeds. On the contrary, they can constitute a significant fraction of the total energy of gravitational waves when most of the energy released by the transition is used to expand the bubble wall, as often found in strongly supercooled phase transitions~\cite{Ellis:2019oqb, Lewicki:2019gmv, Lewicki:2020jiv, Ellis:2020nnr, Lewicki:2022pdb}. 

In this work we focus on the production of gravitational wave from the acoustic motion in a weak first-order phase transition.
The power spectrum of gravitational waves from sound waves has been investigated by numerical simulations of the fluid flow in some simple scenarios, neglecting both the expansion of the background Universe and the back-reaction of the fluid on the metric~\cite{Hindmarsh:2013xza, Hindmarsh:2015qta, Hindmarsh:2017gnf, Tian:2024ysd}. Neglecting spacetime curvature effects in the bubble evolution is typically a good approximation when the phase transition proceeds fast compared to the expansion rate of the Universe, so that the comoving mean bubble spacing $R_*$ is much smaller than the comoving Hubble radius $R_* \ll \mathcal{H}_*^{-1}$, where $\mathcal{H}_* \equiv \mathcal{H}(\eta_*)$, and $\eta_*$ is the conformal time when sound waves start to propagate. However, even in such cases, the acoustic phase that follows the bubble collisions persists until the energy in the compressional modes is fully dissipated into shocks, and can last for many Hubble times. The effects of the expansion of a radiation-dominated Universe on the power spectrum of acoustically generated gravitational waves were first estimated in Ref.~\cite{Hindmarsh:2015qta}, where the authors described the cosmic fluid dynamics with a conformal rescaling of the energy-momentum tensor simulated in flat spacetime. 

Other spacetime curvature effects in the acoustic source beyond the Universe expansion depend on the relative length scales of the fluid flow $R_*$ and of the expanding Universe $\mathcal{H}^{-1}$, and have so far been ignored by numerical simulations of phase transitions.
However, if the bubble nucleation process is slow enough, it is possible to realize first-order phase transitions with large bubbles $R_* \mathcal{H}_*\lesssim \mathcal{O}(1)$, where 
general relativistic curvature effects cannot be neglected~\cite{Giombi:2023jqq}. Large bubbles also correspond to the regime of strong gravitational signal, since the power spectrum of acoustic gravitational waves grows approximately linearly with the mean bubble spacing $R_*\mathcal{H}_*$~\cite{Caprini:2015zlo, Caprini:2019egz,Hindmarsh:2020hop}. Large bubbles are therefore the configurations that maximize the efficiency of acoustic production of gravitational waves and the resulting signals at LISA and other future interferometers. 

In a thermal transition, the bubble formation proceeds by thermal nucleation with a nucleation rate $\Gamma$ that increases exponentially as the temperature of the Universe decreases~\cite{Affleck:1980ac}. The nucleation rate and the speed of expansion of the bubbles of the stable phase determine the transition rate $\beta = \frac{d}{d\eta}\ln \Gamma(\eta)\vert_{ \eta=\eta_f}$, where the conformal time $\eta_f$ is implicitly defined as the time at which a fraction $1/e$ of the Universe remains in the metastable phase~\cite{Hindmarsh:2020hop}. Finally, we define the mean bubble spacing $R_*$ as inverse cubic root of the bubble number density at asymptotic infinite time, and is related to $\beta$ and $v_w$ as
\begin{equation}\label{RH}
    R_*\mathcal{H}_* = (8\pi)^{1/3} v_w \left( \frac{\beta}{\mathcal{H}_*}\right)^{-1}.
\end{equation} 
Large bubbles are expected to form in slow phase transitions, by which we mean $1 \lesssim \beta/\mathcal{H}_* \lesssim 10$. The lower bound $1 \lesssim \beta/\mathcal{H}_*$ is set by the requirement that the stable phase  percolates~\cite{Guth:1982pn}. Tighter bounds come from
recent studies of primordial black holes seeded during a first-order phase transition. In fact, if the supercooled phase of the transitions persists for too long, large curvature fluctuations would yield the overproduction of long-lived primordial black holes and dark matter~\cite{Lewicki:2023ioy, Lewicki:2024sfw, Lewicki:2024ghw}. Currently, there is not a strong consensus on the precise values of these bounds, as different communities disagree on the correct cosmological gauge to adopt to describe the collapse of super-horizon curvature fluctuations (see for example Refs.~\cite{Lewicki:2023ioy, Lewicki:2024sfw, Lewicki:2024ghw} and Ref.~\cite{Franciolini:2025ztf}).
The most conservative bound is currently 
$\beta/\mathcal{H}_* \geq 3.8$~\cite{Lewicki:2023ioy}, which, by equation~\eqref{RH}, implies $R_*\mathcal{H}_* \lesssim 0.77 \,v_w$. This indicates that large bubbles are likely to be found when the bubbles' phase boundaries propagate at relativistic speeds $v_w \simeq 1$, as for example in strongly supercooled detonations. 
On the opposite, the speed of propagation of deflagration fronts is bounded from above by the Chapman-Jouguet speed~\cite{Kurki-Suonio:1995yaf}. Nonetheless, deflagration fronts can also yield large $R_*\mathcal{H}_*$, as the front heats up the fluid ahead of the bubble wall, reducing the nucleation rate and thus the number density of bubbles~\cite{Ajmi:2022nmq}.  
Hubble-sized bubbles can also be expected in scenarios where bubbles nucleate on a pre-existing network of domain walls or strings from an earlier phase transition~\cite{Blasi:2023rqi}.

Semi-analytic models, such as the sound shell model~\cite{Hindmarsh:2016lnk, Hindmarsh:2019phv}, have been developed to support and interpret numerical simulations. In the original papers, the sound shell model considered the fluid dynamics and the gravitational wave production  neglecting all spacetime curvature effects, such as the expansion of the Universe. It further limited the analysis to comoving gravitational wave wavenumber $k$ around and greater than the typical energy injection scale, $k\gtrsim R_*^{-1}$, obtaining results compatible with numerical simulations of weak phase transitions~\cite{Hindmarsh:2017gnf, Correia:2025qif}. In recent years, different works have extended the sound shell model to the low-frequency regime $k\lesssim R_*^{-1}$~\cite{Sharma:2023mao, RoperPol:2023dzg}, and improved it by accounting for the leading-order general relativistic effect in $R_* \mathcal{H}_* \lesssim \mathcal{O}(1)$, that is the expansion of the Universe with either a pure radiation equation of state (speed of sound $c_s=1/\sqrt{3}$)~\cite{Guo:2020grp, Sharma:2023mao, RoperPol:2023dzg} or a soft equation of state ($c_s<1/\sqrt{3}$)~\cite{Guo:2020grp, Giombi:2024kju, Wang:2021dwl}. 

In the present work we aim to extend the current model of gravitational wave production from sound waves by computing the first general relativistic contributions beyond the leading-order in $R_* \mathcal{H}_* \lesssim \mathcal{O}(1)$, that is beyond the Universe expansion. To pursue a semi-analytical calculation we restrict our analysis to some specific scenarios. In particular we focus on weak phase transitions, and assume that fluid perturbations during the acoustic phase can be approximated as linear and Gaussian. The fluid flow in weak phase transitions typically moves at non-relativistic speed~\cite{Hindmarsh:2013xza, Hindmarsh:2015qta}. To neglect higher order contributions in $R_* \mathcal{H}_* \lesssim \mathcal{O}(1)$, we will make the assumption that $v_{\rm{rms}} \ll R_*\mathcal{H}_*$. This implies that the acoustic flow lasts for a long time before  shocks form, typically longer than a Hubble time. In fact, non-linearities become important on the time scale of shock formation $\eta_{\rm{sh}}$~\cite{Cutting:2019zws, Gogoberidze:2007an, Caprini:2009yp, RoperPol:2019wvy}, and in our scenario $\mathcal{H}_*\eta_{\rm{sh}} \sim (R_*\mathcal{H}_*)/v_{\rm{rms}} \gg 1$. However, in an expanding Universe, the gravitational wave power spectrum is generally almost entirely sourced within one Hubble time $\mathcal{H}_*^{-1}$~\cite{Sharma:2023mao}. We therefore expect that the energy dissipation of sound waves through shocks and non-linear motion has little effect on the shape of the gravitational wave power spectrum in the regimes under consideration. We approximate the energy decay of the acoustic flow as a sharp cut-off of the source at a timescale determined by $\eta_{\rm{sh}}$, which determines the end of the acoustic phase, neglecting the shape evolution of the spectrum due to shocks (see Refs.~\cite{Dahl:2021wyk, Dahl:2024eup} for details on the energy dissipation via shocks). General relativistic effects beyond the leading-order in $R_* \mathcal{H}_* \lesssim \mathcal{O}(1)$ can be significant also in the non-linear regime, but we leave this analysis to future development. 

As the energy density of acoustically generated gravitational waves increases quadratically with the kinetic energy of the fluid~\cite{Hindmarsh:2015qta, Hindmarsh:2017gnf}, acoustic flows with small average velocity $v_{\rm{rms}}$ might lead to weak gravitational wave signals. We stress however that the condition $v_{\rm{rms}} \ll R_*\mathcal{H}_*$ is not too restrictive in scenarios with large bubbles. As an example, we obtain a signal within LISA's sensitivity curve when $R_*\mathcal{H}_* \simeq 0.524$ and $v_{\rm{rms}} \simeq 0.085$. 

Within our framework, we review the calculations of the sound shell model and solve the dynamic equations for gravitational waves and sound waves in cosmological perturbation theory with the expansion parameter $R_* \mathcal{H}_* \lesssim \mathcal{O}(1)$.  
At the next-to-leading-order (NLO), we find general relativistic corrections to the dynamics of both gravitational waves and sound waves. Moreover, while the shear stress components of the energy momentum tensor at the leading-order are sourced only by the fluid velocity perturbations, at the next-to-leading-order in $R_* \mathcal{H}_*$ they get an additional contribution from curvature perturbations. These are indeed a source of secondary (scalar-induced) gravitational waves, which have not been fully understood in the context of first-order phase transitions yet. The amplitude of the novel contributions at the NLO grows quadratically with $R_* \mathcal{H}_*$ with respect to the leading-order term, so that they become relevant only when the typical size of the new-phase bubble is a significant fraction of the causal Universe. 

Secondary gravitational wave production has been considered in the context of primordial black holes formation at strongly supercooled phase transitions~\cite{Lewicki:2024sfw, Lewicki:2024ghw, Cai:2024nln, Franciolini:2025ztf}, where the gravitational waves are sourced by curvature perturbations induced by the bubbles of the stable phase during the phase transition.  In our approach, the power spectrum of scalar-induced gravitational waves is a term beyond the NLO in $R_*\mathcal{H}_*$, and is therefore neglected.

The article is divided as follows: in Section~\ref{sec:GW_PS} we review the calculation of the gravitational wave power spectrum from sound waves in the framework of the sound shell model~\cite{Hindmarsh:2016lnk, Hindmarsh:2019phv}. We additionally include curvature perturbations to the shear stress (secondary gravitational waves) and the NLO corrections in $R_*\mathcal{H}_*$ to the equations of motion of both gravitational waves and sound waves. In Section~\ref{sec:analytic} we provide the analytic approximation for each contribution to the power spectrum in three different frequency regimes: low-frequency regime, where $k\eta_*<k\eta_{\rm{end}}\ll 1 $ and $kR_* \ll 1$; intermediate-frequency regime, with $1\ll k\eta_*\ll k_p\eta_*$ and $kR_* \ll 2\pi$; and high-frequency regime, that is $(kR_* \sim 2\pi) \cup (kR_*> 2\pi)$. Finally, we present and comment our results in Section~\ref{sec:results}. 

Throughout this paper we use units $c = \hbar = 1$. We further adopt the mostly positive signature for the metric $(-, +,+,+)$. 
Greek letters $(\mu, \nu, \dots)$ will be used for four-dimensional  tensorial indices, while Latin letters $(i, j, \dots)$ for three-dimensional indices. We use conformal time $\eta$ as the time coordinate,  $a(\eta)$ for the scale factor in a Friedmann-Lema\^{i}tre-Robertson-Walker Universe, and  use a prime to denote derivative with respect to $\eta$, e.g. $a^\prime = da/d\eta$.


\section{Generation of gravitational wave from sound waves}\label{sec:GW_PS}
We consider scalar and tensor perturbations around a Friedmann-Lema\^{i}tre-Robertson-Walker (FLRW) background Universe in Poisson gauge and conformal coordinates~\cite{Durrer:2004fx, Brandenberger:2003vk}, where the line element is
\begin{equation}\label{metric}
    ds^2 = a^2(\eta)\left[ -(1+2\Phi) d\eta^2 + (1-2\Psi)\gamma_{ij}dx^i dx^j\right],
\end{equation}
with 
\begin{equation}
    \gamma_{ij} = \delta_{ij} + h_{ij}.
\end{equation}
In the line element, $\Phi$ and $\Psi$ are the scalar Bardeen potentials, and $h_{ij}$ are the gauge-invariant tensor perturbations. We describe the energy content of the Universe during the transition as a perfect fluid with energy momentum 
\begin{equation}\label{eq:T_munu}
    T^{\mu\nu} = w u^\mu u^\nu + pg^{\mu\nu},
\end{equation}
where $u^\mu =a^{-1}(1-\Phi, v^i)$ is the fluid four-velocity and $w=e+p$ the fluid enthalpy, being $e$ the energy density and $p$ the pressure of the fluid. In a perfect fluid, anisotropic perturbations vanish at linear order, and the Bardeen potentials coincide $\Phi = \Psi$~\cite{Durrer:2004fx}. Moreover, in linearised gravity, Bardeen potentials and energy fluctuations in the fluid are related through the Bardeen equation~\cite{Baumann:2007zm}
\begin{equation}\label{eq:Bardeen_eq}
    \nabla^2\Psi - 3\mathcal{H}(\Psi^\prime +\mathcal{H}\Psi) = 6\mathcal{H}^2 \frac{\delta e}{\bar{e}}. 
\end{equation}

Gravitational waves are sourced instead by quadrupole moments, that is by shear stress $\mathcal{S}_{ij}(\eta, \bm{x})$. The Einstein equations for tensor perturbations yield~\cite{Baumann:2007zm}
\begin{equation}\label{eq:GW}
    \left(\partial_\eta^2 + 2\mathcal{H}\partial_\eta - \nabla^2 \right) h_{ij} (\eta, \bm{x}) = 16\pi G a^2 \bar{w} \mathcal{S}^{(\text{TT})}_{ij} (\eta, \bm{x}),
\end{equation}
where $\bar{w}$ is the background enthalpy of the cosmic fluid.  The physical degrees of freedom of gravitational waves are recovered by projecting the source onto the transverse and traceless subspace as $\mathcal{S}^{(\text{TT})}_{ij} = \Lambda_{ij, k\ell} \mathcal{S}_{k\ell}$, and we defined
\begin{equation}
    \Lambda_{ij, k\ell} = P_{ik}P_{j\ell} - \frac{1}{2}P_{ij}P_{k\ell},
\end{equation}
with $P_{ij}= \delta_{ij} - \hat{n}_i\hat{{n}}_j$ and $\hat{{n}}_i$ the direction of propagation of the gravitational wave~\cite{Hindmarsh:2019phv, Maggiore:2007ulw}.

The source of shear stress is given by (see Ref.~\cite{Baumann:2007zm} for a derivation)
\begin{equation}\label{eq:S_ij_x}
    \mathcal{S}_{ij} (\eta, \bm{x}) =  v_i v_j + v_i v_j \frac{\delta w}{\bar{w}} + \frac{1}{4\pi G a^2 \bar{w}} \partial_i \Psi \partial_j \Psi, 
\end{equation} 
where $\delta w (\bm{x}, \eta) \equiv w(\bm{x}, \eta) - \bar{w}(\eta)$. Here,  the first two terms correspond to the shear motion of the fluid (\textit{primary gravitational waves})~\cite{Hindmarsh:2019phv}, while the last term derives from the gradients of the Bardeen potentials (\textit{secondary gravitational waves})~\cite{Domenech:2021ztg}. We anticipate that in the scope of this work, the second term $v_iv_j \, \delta w/\bar{w}$ is deemed as a subdominant contribution to the energy density of gravitational waves in the limit $v_{\rm{rms}} \ll R_*\mathcal{H}_*$, and is therefore neglected. We will justify this statement at the beginning of Section~\ref{sec:greensub}. In the previous literature~\cite{Hindmarsh:2016lnk, Guo:2020grp, Sharma:2023mao, RoperPol:2023dzg, Giombi:2024kju}, also the secondary source, that is the last term in equation~\eqref{eq:S_ij_x}, has been neglected in the approximation of small bubbles, that is $R_*\mathcal{H}_* \ll 1$. In the present work, we aim to go beyond this approximation, and include the leading corrections to the gravitational wave energy density originating from curvature perturbations.

We remark that the condition $v_{\rm{rms}} \ll R_*\mathcal{H}_*$ is not too restrictive, especially in a scenario of weak transition with large bubbles where $R_*\mathcal{H}_* \lesssim \mathcal{O}(1)$. In Section~\ref{sec:comparison} we will mention an example of gravitational wave signal from a weak phase transition with $v_{\rm{rms}} \ll R_*\mathcal{H}_*$ which could potentially be detected by LISA. 
 
For later convenience, in order to align our notation with that of Ref.~\cite{Giombi:2024kju}, we define the rescaled field $\ell_{ij} \equiv a(\eta)h_{ij}$ and rewrite the wave equation~\eqref{eq:GW} as
\begin{equation}\label{eq:ell}
    \left(\partial_\eta^2 -\nabla^2 -\frac{a^{\prime\prime}}{a} \right) \ell_{ij} (\eta, \bm{x}) = 16\pi Ga^3 \bar{w}\mathcal{S}_{ij}^{(\rm{TT})} (\eta, \bm{x}).
\end{equation}
In this form, the term $a^{\prime\prime}/a$ vanishes in a radiation dominated Universe, since by the Friedmann equation $a^{\prime\prime}/a = 4\pi G (\bar{e} - 3\bar{p})/3$.


\subsection{Energy density of gravitational waves}
We define the energy density of gravitational waves as the time-time component of the energy momentum tensor of gravitational waves $e_{\rm{gw}} \equiv - T^{\phantom{g} 0}_{\rm{gw}\; 0}$, which is computed by averaging the covariant time derivative of tensor perturbations over several wavelengths
\begin{equation}
    T^{\phantom{g} \mu}_{\rm{gw}\; \nu} = \frac{1}{32 \pi G}\langle \mathcal{D}^\mu h^{*}_{ij} \mathcal{D}_\nu h^{ij}\rangle,
\end{equation}
with $\mathcal{D}$ the Levi-Civita connection. The Christoffel symbols inside the covariant derivatives bring terms beyond the analytical precision of this work. We therefore compute the energy density of gravitational waves as~\footnote{Expanding the covariant derivative yields (see ref.~\cite{ValbusaDallArmi:2024hwm} for details) 
\begin{equation}
    e_{\text{gw}} = \frac{1}{32 \pi G a^2}\left\langle  h^{\prime *}_{ij} h^{ij\,\prime}\left(1 + 2\Psi  + \mathcal{O}(\Psi^2)\right)\right\rangle .
\end{equation}
Recalling equation~\eqref{eq:GW}, we estimate $\left\langle  h^{\prime *}_{ij} h^{ij\,\prime}\right\rangle\sim 32\pi G \, \bar{e} (R_*\mathcal{H}_*)^4 \left[v^4 + v^2\delta^2 (R_*\mathcal{H}_*)^2 + \mathcal{O}\left(\delta^4 (R_*\mathcal{H}_*)^4\right)\right] $, with $\delta = \delta e/\bar{e}$. Similarly, recalling the Bardeen equation~\eqref{eq:Bardeen_eq}, we estimate the first correction to the Isaacson formula as $\left\langle  h^{\prime *}_{ij} h^{ij\,\prime} \Psi\right\rangle \sim 32\pi G \,\bar{e} (R_*\mathcal{H}_*)^4 \left[v^4 \delta (R_*\mathcal{H}_*)^2 + \mathcal{O}\left(v^2\delta^3 (R_*\mathcal{H}_*)^4\right)\right] $. Assuming that density and velocity perturbations in sound waves are of the same order, we neglect this term in the limit $\sqrt{v_iv^i} \ll (R_*\mathcal{H}_*)^2$. 
} 
\begin{equation}\label{eq:Isaacson}
    e_{\text{gw}} = \frac{1}{32 \pi G a^2}\langle h^{\prime\,*}_{ij} h^{ij\,\prime}\rangle \, ,
\end{equation}
which recovers the Isaacson formula in the case of a static flat background~\cite{Isaacson:1968hbi, Isaacson:1968zza}. 
Since the sound waves are generated by the collisions between the fronts of randomly distributed bubbles, we assume that fluid perturbations are statistically homogeneous and isotropic. In momentum space, the two point correlation function is then entirely described by the spectral density $P_{h^\prime}(\eta, \vert\bm{k}\vert)$ as
\begin{equation}\label{eq:h_correlator}
    \langle \tilde{h}_{ij}^{\prime \, *} (\bm{k}_1, \eta) \tilde{h}^{ij\, \prime}(\bm{k}_2, \eta)\rangle = (2\pi)^3  \delta(\bm{k}_1 - \bm{k}_2) P_{h^{\prime}}(\vert\bm{k}_1\vert, \eta),
\end{equation}
where $\tilde{h}_{ij} (\bm{k}, \eta)$ is the Fourier transform of the gravitational wave field
\begin{equation}
    \Tilde{h}_{ij}(\eta, \bm{k}) = \int d^3\bm{x} \, h_{ij}(\bm{x},\eta) e^{-i\bm{k} \cdot \bm{x}}.
\end{equation}
The gravitational wave power spectrum normalized to the background energy density $\bar{e}$ is then
\begin{equation}
    \Pgw \equiv \frac{1}{\bar{e}}\frac{\operatorname{d} e_{\text{gw}}}{\operatorname{d}\ln k} =  \frac{1}{32 \pi G a^2 \bar{e}}\frac{k^3}{2\pi^2}P_{h^\prime}.
\end{equation}

\subsection{Gravitational wave power spectrum with a soft equation of state}\label{sec:greensub}

During the acoustic phase, sound waves dissipate energy via shocks. While this is a dynamical process that can extend over a long time~\cite{Caprini:2024gyk}, we approximate the process of energy dissipation as a sharp cut-off of the acoustic source at the time of shock formation, estimated as $\eta_{\rm{sh}}\sim R_*/v_{\rm{rsm}}$. 

We assume that the cosmic fluid during and after the transition can be described by a barotropic equation of state $p(e)=\omega e$, with $\omega$ a constant parameter ($\omega =1/3$ for an ultra-relativistic fluid). 
The symmetry-breaking process that drives the transition induces a softening of the equation of state inside the bubbles in the broken phase~\cite{Ares:2020lbt, Tenkanen:2022tly} that can persist for a long time, even after the dissipation of energy in the sound waves. In fact, the equation of state does not depend on the fluid flow, but only on the temperature of the cosmic fluid and on the underlying particle theory that triggered the transition.

We consider here the scenario discussed in our previous work in Ref.~\cite{Giombi:2024kju}: at time $\eta_*$, after the collisions between the bubbles' interfaces, the Universe is in the broken phase and described by a \textit{soft} equation of state $\omega \leq 1/3$, and sound waves start propagating across the thermal plasma. Non-linear dissipative processes quench the sound wave energy over a time scale $\eta_{\rm{sh}}\sim R_*/v_{\rm{rms}}$. The evolution of the equation of state is instead independent of the fluid motion, and remains ``soft'' until the Universe smoothly evolves towards the exact radiation era with $\omega = 1/3$.

To pursue an analytic calculation of the gravitational wave power spectrum we assume that sound waves remain nearly stationary during the acoustic phase until $\eta_\text{end}\sim \eta_{\rm{sh}}$, when they abruptly disappear, modelling their dissipation by shock formation. Moreover, we approximate the relaxation of the equation of state, which in a real scenario varies smoothly from the ``soft'' state toward the radiation state,  as an instantaneous change at an arbitrary radiation time $\eta_\text{r}$. Therefore, we approximate that the equation of state remains constant and ``soft'' until $\eta_\text{r}$, when it suddenly moves to the exact radiation era. While the end time of the source $\eta_{\text{end}}$ is set by the timescale of acoustic dissipation $\eta_{\text{sh}} \sim R_*/v_{\text{rms}}$, the return to radiation time $\eta_{\text{r}}$ is  a separate physical scale. The only constraint on $\eta_{\text{r}}$ is that radiation must dominate prior to Big Bang nucleosynthesis ($T\sim 1 \;\text{MeV}$). For simplicity, in this work we assume $\eta_{\text{r}} > \eta_{\text{end}}$, so that we can neglect the time evolution of the equation of state during the activity of the source. The case $\eta_{\text{r}} < \eta_{\text{end}}$ can also be treated within our framework, but it requires a smooth matching between the sound-wave and gravitational-wave solutions at the transition in the equation of state.

The scenario of a constant equation of state is compatible with a thermal phase transition in an expanding Universe if $\omega$ evolves sufficiently slowly in time. It is possible to improve this approximation by allowing for a temperature-dependent equation of state, which is particularly important in the case of a long-lived flow $\mathcal{H}_*\eta_{\rm{sh}} \gg 1$. This, however, would introduce a complicated time dependence in fluid shear stress, hindering the analytical estimation of the gravitational wave power spectrum. We therefore leave this analysis for future developments of this work. We further recall that in this work we account for the  dissipation of energy in sound waves via shocks only as a time cut-off of the acoustic source. This means that we also neglect the spectral shape of the source due to acoustic turbulence in the fluid, which has been observed for example in Refs.~\cite{Dahl:2021wyk, Dahl:2024eup, Caprini:2024gyk}. 

The evolution of the equation of state parameter is then summarized as
\begin{equation}
  \omega(\eta) =
    \begin{cases}
      \omega & \eta_* < \eta < \eta_{\text{r}},\\
      1/3, &  \eta \geq \eta_{\text{r}}.
    \end{cases}       
    \label{eq:eospar}
\end{equation}
The limit $\eta_{\text{r}} \rightarrow \eta_*$ corresponds to assuming a pure radiation equation of state, that is $\omega = 1/3$, throughout the transition. This scenario represents a special case of our analysis. 
The equation of state~\eqref{eq:eospar} allows us to find a simple solution to the system of Friedmann and continuity equations 

\begin{equation}
    3\mathcal{H}^2 = 8\pi G \bar{e}a^2, \qquad \bar{e}^\prime + 3\mathcal{H}(\bar{e}+\bar{p}) = 0,
\end{equation}
describing the time evolution of the homogeneous and isotropic background Universe. In particular we find   
\begin{equation}
    a(\eta) = a_* \left(\frac{\eta}{\eta_*} \right)^{1+\nu} , \qquad \bar{w}(\eta) = \bar{w}_* \left(\frac{a}{a_*} \right)^{-2\frac{2+\nu}{1+\nu}} ,  \qquad \nu =  \frac{1-3\omega}{1+3\omega}
\end{equation}
with $a_* \equiv a(\eta_*)$,  $\bar{w}_* \equiv \bar{w}(\eta_*)$. The parameter $\nu \in [0, 1)$ measures the deviation from the case of pure radiation, where $\omega = 1/3$ and the Universe expands with $a^{\prime\prime} = 0$.  

Under the approximation of stationary source and piecewise constant equation of state, the power spectrum of acoustically generated gravitational waves at late time $\eta >\eta_{\text{r}}$ was computed in Ref.~\cite{Giombi:2024kju} as
\begin{samepage}
\begin{align}\label{P_hp}
    \Pgw &=  3(1+\omega)^2\mathcal{H}_*^2 \left(\frac{a_*}{a_{\text{r}}}\right)^{\frac{2\nu}{1+\nu}}\frac{k^3}{2\pi^2} \iint_{\eta_*}^{\eta_{\text{end}}} d\eta_1  d\eta_2 \left(\frac{\eta_*^2}{\eta_1\eta_2}\right)^{1-\nu} \times \qquad\qquad\nonumber\\
    & \qquad\qquad\qquad\qquad\qquad\qquad\qquad \times\; G_{k}^\prime(\eta, \eta_1)G_{k}^\prime(\eta, \eta_2) U_{\mathcal{S}}(k, \eta_1, \eta_2),
\end{align}
\end{samepage}
with $U_{\mathcal{S}}$ the unequal time correlator  of the fluid shear stress 
\begin{equation}\label{eq:Us}
    \langle \tilde{\mathcal{S}}^{*\, (TT)}_{ij}(\eta_1, \bm{k}_1) \tilde{\mathcal{S}}^{(TT) \, ij}(\eta_2, \bm{k}_2)\rangle = U_{\mathcal{S}}(\bm{k}_1, \eta_1, \eta_2) (2\pi)^3 \delta(\bm{k}_1 - \bm{k}_2),
\end{equation}
and $G_k(\eta, \eta_1)$ the Green's function for the wave equation~\eqref{eq:ell}. The Green's function method is described in more detail in Appendix~\ref{app:Green_function}, where we find 
\begin{equation}\label{eq:Green}
    G_k(\eta, \eta_1) = -k \eta \eta_1 \left[ j_{\nu}(k\eta)y_{\nu}(k\eta_1) - j_{\nu}(k\eta_1)y_{\nu} (k\eta) \right]\Theta(\eta - \eta_1), \qquad \nu =  \frac{1-3\omega}{1+3\omega},
\end{equation}
where $j_\nu$ and $y_\nu$ are the spherical Bessel functions of the first and second kind, and $\Theta(\eta -\eta_1)$ the Heaviside step function~\footnote{This expression of the Green's function for tensor perturbations coincides with the expression given in Ref.~\cite{Giombi:2024kju}, equation 2.11. The Green's function in equation~\eqref{eq:Green} is here written in a more compact way by using the fact that the Wronskian of spherical Bessel functions can be simplified as $\mathcal{W}\big(j_\nu(z_1), y_\nu (z_1)\big) \equiv \partial_{z_1} j_\nu(z_1)y_\nu (z_1) - j_\nu (z_1) \partial_{z_1} y_\nu (z_1) = z_1^{-2}$ with $z_1 = k\eta_1$~\cite[\href{http://dlmf.nist.gov/10.50.E1}{(10.50.1)}]{NIST:DLMF}. }. In radiation, that is when $\nu =0$, the spherical Bessel functions revert to linear combinations of the trigonometric functions, and
\begin{equation}\label{eq:G_rad}
    G_k^{\nu= 0}(\eta, \eta_1) = \frac{1}{k}  \sin\left[k(\eta-\eta_1)\right]\Theta(\eta - \eta_1).
\end{equation}

\subsection{The shear stress unequal time correlator beyond the leading-order in the bubble-over-Hubble radius expansion}

Numerical simulations of weak phase transitions show that the expanding bubbles generate mainly non-relativistic compressional modes, indicating that the linear motion dominates the fluid flow in the initial stages of the gravitational wave production~\cite{Hindmarsh:2015qta}. 
We restrict our analysis to the compressional components of the fluid velocity, assuming that $v_{\rm{rms}} \ll R_*\mathcal{H}_*$. Shocks and non-linearities develop at later times, on the timescale $\eta_{\rm{sh}}\sim R_*/v_{\rm{rms}}$. We approximate the effect of non-linearities as an instantaneous dissipation of the energy of sound waves into turbulence at a time $\eta_{\rm{end}} \propto \eta_{\rm{sh}}$. In the large bubble regime $R_*\mathcal{H}_*\lesssim \mathcal{O}(1)$, this also means that the acoustic flow can last for many Hubble times.

We then assume that, during the acoustic phase, fluid perturbations can be treated as linear to a good approximation. 
Due to the stochastic nature of sound waves, we further assume that the perturbations can be described as Gaussian random fields with zero mean. Finally, we assume that in the linear regime velocity perturbations and energy fluctuations in sound waves are of the same order, so  that $\vert \delta w\vert /\bar{w} \sim v$.  

Since fluid perturbations are correlated up to the characteristic length scale $R_*$, the relevant wavenumbers of the causal flow are $p\gtrsim1/R_*$, with most of the energy being released around the energy injection scale $p\sim 1/R_*$~\cite{Hindmarsh:2013xza, Hindmarsh:2015qta, Hindmarsh:2017gnf}. We aim to perform a perturbative calculation of the gravitational wave power spectrum by expanding in the small bubble-over-Hubble radius parameter $R_* \mathcal{H}_* \ll \mathcal{O}(1)$. General relativistic effects are enhanced in the large-bubble regime, where the perturbative parameter grows to order $R_* \mathcal{H}_* \lesssim \mathcal{O}(1)$. 

In the regimes under consideration, we can simplify the source of shear stress in equation~\eqref{eq:S_ij_x} by neglecting terms $v_iv_j\, \delta w/\bar{w}$, that is by considering 
\begin{equation}\label{eq:S_new}
    \mathcal{S}_{ij} (\eta, \bm{x}) \simeq  v_i v_j  + \frac{1}{4\pi G a^2 \bar{w}} \partial_i \Psi \partial_j \Psi.
\end{equation} 
Let us justify this point by a simple dimensional analysis. From the Bardeen equation~\eqref{eq:Bardeen_eq}, we estimate $\Psi \sim (R_*\mathcal{H}_*)^2 \delta w/\bar{w}$, and the second term on the right hand side of equation~\eqref{eq:S_new} is estimated as
\begin{equation}\label{eq:term_2}
    \frac{1}{4\pi G a^2 \bar{w}} \partial_i \Psi \partial_j \Psi \sim (R_*\mathcal{H}_*)^2 \left(\frac{\delta w}{\bar{w}}\right)^2. 
\end{equation}
Recalling equation~\eqref{P_hp}, the energy density of gravitational waves is proportional to the two-point correlator of shear stress.  By expressing the two-point correlator with the source~\eqref{eq:S_new}, and expanding it in power of the bubble-over-Hubble parameter $R_*\mathcal{H}_* \lesssim\mathcal{O}(1)$, we find:
\begin{itemize}
    \item[-] at the leading order $\langle \mathcal{S}_{ij} \mathcal{S}^{ij} \rangle^{\rm{lo}} \sim \langle v_iv_j v^i v^j \rangle \sim v^4$.
    \item [-] At the next-to-leading order $\langle \mathcal{S}_{ij} \mathcal{S}^{ij} \rangle^{\rm{nlo}} \sim \mathcal{H}_*^{-2}\langle v^i v^j \partial_i\Psi \partial_j\Psi \rangle \sim (R_*\mathcal{H}_*)^2 v^2 \delta^2$, where $\delta \equiv \delta w/\bar{w}$, and in the second step we used the estimate in equation~\eqref{eq:term_2}.

\item[-] At the next-to-next-to-leading order  $\langle \mathcal{S}_{ij} \mathcal{S}^{ij} \rangle^{\rm{nnlo}} \sim \mathcal{H}_*^{-4} \langle \partial_i \Psi \partial_j \Psi \partial^i \Psi \partial^j \Psi \rangle \sim (R_*\mathcal{H}_*)^4 \delta^4$. 
\end{itemize}
If in addition we consider terms $v_iv_j\, \delta w /\bar{w}$ as in equation~\eqref{eq:S_ij_x}, we find new contributions. However, since we assumed that the perturbations are Gaussian fields with zero mean, by applying the Wick's theorem one can see that odd-point correlators vanish, and in particular $\langle v_iv_j v^i v^j \,\delta w/\bar{w} \rangle =0$. Therefore, the only non vanishing additional contribution from the terms $v_iv_j\, \delta w /\bar{w}$ is 
\begin{itemize}
    \item [-]  $\langle \mathcal{S}_{ij} \mathcal{S}^{ij} \rangle^{vv\delta} \sim \langle v_iv_jv^i v^j (\delta w/\bar{w})^2 \rangle \sim v^4 \delta^2$.
\end{itemize}
Assuming that $\vert \delta w\vert /\bar{w} \sim v$, this contribution is subdominant with respect to $\langle \mathcal{S}_{ij} \mathcal{S}^{ij} \rangle^{\rm{nlo}}$ in the limit $v_{\rm{rms}}\ll R_*\mathcal{H}_*$. We thus neglect this contribution and consider the source of shear stress as in equation~\eqref{eq:S_new}.

\bigskip

In momentum space, using the background Friedmann equation $3\mathcal{H}^2 = 8\pi G\bar{e}a^2$, the source is written as
\begin{equation}\label{eq:S}
    \tilde{\mathcal{S}}_{ij}(\eta, \bm{k}) = 
      \int \frac{d^3 \bm{p}}{(2\pi)^3} \hat{p}_i \hat{q}_j \left[  \tv(\eta, \bm{p}) \tv(\eta, -\bm{q}) + \frac{2}{3(1+\omega)} \frac{p}{\mathcal{H}}  \tilde{\Psi}(\eta, \bm{p}) \frac{q}{\mathcal{H}} \tilde{\Psi}(\eta, -\bm{q})\right],
\end{equation}
with $\bm{q} = \bm{p} - \bm{k}$ and $\tv(\eta,\bm{p}) = i \hat{p}_i \tv^i(\eta,\bm{p})$, being $\hat{p}_i$ the unit normal direction of the sound wave wavenumber, $\hat{p}_i = p_i/p$. To be concise, in the following we will suppress trivial functional time dependence and use the notation $\tv^i_{\bm{p}} \equiv \tv^i(\eta,\bm{p})$.

The linearized Einstein equations relate curvature perturbations to density and velocity fluctuations as~\cite{Durrer:2004fx} 
\begin{equation}\label{eq:Einstein_eqs}
    \tilde{\Psi}_{\bm{p}} = -\dfrac{3}{2} (1+\omega)\left(\dfrac{\mathcal{H}}{p}\right)^2 \left( \tlambda_{\bm{p}} + 3 \dfrac{\mathcal{H}}{p} \tv_{\bm{p}}\right),
\end{equation}
with $\lambda(\bm{x}, \eta) = [e(\bm{x}, \eta) - \bar{e}(\eta)]/\bar{w}(\eta)$,
and allow us to write the source tensor~\eqref{eq:S} in terms of fluid perturbations only. 
Considering $p \sim R_*^{-1}$ and $\mathcal{H}/p \sim R_*\mathcal{H}_*$, at the next-to-leading-order in the bubble-to-Hubble radius expansion $R_*\mathcal{H}_* \lesssim \mathcal{O}(1)$ we find
\begin{equation}\label{eq:S_ij}
    \tilde{\mathcal{S}}_{ij}(\eta, \bm{k}) = 
    \int \frac{d^3 \bm{p}}{(2\pi)^3} \hat{p}_i \hat{q}_j \left[ \tv_{\bm{p}} \tv_{-\bm{q}} + \frac{3}{2}(1+\omega) \dfrac{\mathcal{H}}{p}\dfrac{\mathcal{H}}{q}  \tlambda_{\bm{p}} \tlambda_{-\bm{q}} \right] + \mathcal{O}\left(\frac{\mathcal{H}}{p}\right)^3.
\end{equation}

The shear stress unequal time correlator~\eqref{eq:Us} is then a combination of 4-point correlator functions. 
Since we assumed the perturbations to be Gaussian random fields, we can use Wick's theorem to reduce the shear stress unequal time correlator to the combination of two-point correlators of fluid perturbations~\cite{Kosowsky:2001xp, Gogoberidze:2007an, Caprini:2009fx}. 
For compressional modes we define
\begin{subequations}\label{eqs:GG}
    \begin{eqnarray}
        \langle \tv_{1,\bm{p}} \, \tv^*_{2,\bm{q}} \rangle &=& C_{\tv \tv}(p, \eta_1, \eta_2)(2\pi)^3 \delta^3(\bm{p} - \bm{q}),\\
        \langle \tv_{1,\bm{p}} \,  \tlambda^*_{2,\bm{q}} \rangle &=& C_{\tv\tlambda}(p, \eta_1, \eta_2)(2\pi)^3 \delta^3(\bm{p} - \bm{q}).
    \end{eqnarray}
\end{subequations}
Additional terms $C_{\tlambda\tlambda}$ originate from the correlators of density fluctuations $\langle \tlambda_{1,\bm{p}} \,  \tlambda^*_{2,\bm{q}} \rangle$. These are however neglected as they bring contributions beyond the next-to-leading-order in the bubble-over-Hubble radius expansion $R_*\mathcal{H}_* \lesssim \mathcal{O}(1)$. 
In this calculation we also neglect the contributions from the cross-correlator $\langle v_{\bm{p}_1}^i v_{-\bm{p}_2}^{j}\rangle$, since they are argued to be subdominant compared to~\eqref{eqs:GG}~\cite{Hindmarsh:2016lnk, Hindmarsh:2019phv}.
The spectral densities $C_{\tv \tv}(p, \eta_1, \eta_2)$ and $C_{\tv\tlambda}(p, \eta_1, \eta_2)$ depend on the initial conditions at the time $\eta_*$ when sound waves set in the fluid, and they will be discussed in detail in Section~\eqref{sec:correlators}. The leading-order contribution to the shear stress unequal time correlator~\eqref{eq:Us} was obtained already in our previous paper (Ref.~\cite{Giombi:2024kju}, Section 2.1), where we found
\begin{equation}
    \begin{split}
    \langle \tilde{\mathcal{S}}^*_{ij}(\bm{k_1}, \eta_1) \tilde{\mathcal{S}}^{ij}(\bm{k_2}, \eta_2)\rangle^{\text{lo}} =&  \Lambda_{ij, k\ell}(\bm{k})  \Lambda_{ij,mn}(\bm{k}^\prime) \times\nonumber\\
    &\times \int \frac{d^3 \bm{p_1}}{(2\pi)^3} \int \frac{d^3 \bm{p_2}}{(2\pi)^3} \hat{p}_1^k \hat{q}_1^\ell \hat{p}_2^m \hat{q}_2^n  \langle \tilde{v}^*_{1,\bm{p_1}} \tilde{v}^*_{1,-\bm{q_1}} \tilde{v}_{2,\bm{p_2}} \tilde{v}_{2,-\bm{q_2}}\rangle ,
    \end{split}
\end{equation}
where $\bm{q_1} = \bm{p_1} - \bm{k_1}$ and $\bm{q_2} = \bm{p_2} - \bm{k_2}$, leading to 
\begin{equation}\label{eq:U}
    U^{\text{lo}}_{\mathcal{S}}(k, \eta_1, \eta_2) = \frac{1}{4\pi^2 k} \int_0^\infty dp \int_{\vert p -k\vert}^{p+k}dq (1-\mu_p^2)^2 \frac{p^3}{q}   C_{\tilde{v}\tilde{v}}(p, \eta_1, \eta_2)C_{\tilde{v}\tilde{v}}(q, \eta_1, \eta_2), 
\end{equation}
with $\mu_p \equiv \hat{\bm{p}}\cdot \hat{\bm{k}} = (p^2 +k^2 -q^2)/(2pk)$, and $\text{d}\mu_p = - q/(pk) \text{d}q$.

For the sake of brevity, here we just outline the new calculation of the next-to-leading-order (NLO) contributions. Let us then consider
\begin{equation}
\begin{split}
    \langle \tilde{\mathcal{S}}^*_{ij}(\bm{k_1},& \eta_1) \tilde{\mathcal{S}}^{ij}(\bm{k_2}, \eta_2)\rangle^{\text{nlo}} =   \Lambda_{ij, k\ell}(\bm{k_1})  \Lambda_{ij,mn}(\bm{k_2}) \int \frac{d^3 \bm{p_1}}{(2\pi)^3} \int \frac{d^3 \bm{p_2}}{(2\pi)^3} \hat{p}_1^k \hat{q}_1^\ell \hat{p}_2^m \hat{q}_2^n \times \qquad\\
    & \times  \frac{3}{2}(1+\omega)\left[ \frac{\mathcal{H}_1^2}{p_1 q_1} \langle \tilde{v}^*_{2,\bm{p_2}} \tilde{v}^*_{2,-\bm{q_2}} \tilde{\lambda}_{1,\bm{p_1}} \tilde{\lambda}_{1,-\bm{q_1}} \rangle + \frac{\mathcal{H}_2^2}{p_2 q_2} \langle \tilde{v}^*_{1,\bm{p_1}} \tilde{v}^*_{1,-\bm{q_1}} \tilde{\lambda}_{2,\bm{p_2}} \tilde{\lambda}_{2,-\bm{q_2}} \rangle \right],
    \end{split}
\end{equation}
with $\mathcal{H}_{i} \equiv \mathcal{H}(\eta_i), \; i=1,2$. Using Wick's theorem and the definitions~\eqref{eqs:GG}, we obtain 
\begin{samepage}
\begin{eqnarray}
    \langle \tilde{\mathcal{S}}^*_{ij}(\bm{k_1}, \eta_1) \tilde{\mathcal{S}}^{ij}(\bm{k_2}, \eta_2)\rangle^{\text{nlo}} =   \Lambda_{ij, k\ell}(\bm{k_1})  \Lambda_{ij,mn}(\bm{k_2})  \iint d^3 \bm{p_1}d^3 \bm{p_2} \; \hat{p}_1^k \hat{q}_1^\ell \hat{p}_2^m \hat{q}_2^n \,\delta^3(\bm{k_1} - \bm{k_2}) \times \nonumber\\
    \times  \frac{3}{2}(1+\omega)\bigg\{ \frac{\mathcal{H}_1^2}{p_1 q_1} C_{\tv\tlambda}(p_1, \eta_2, \eta_1)C_{\tv\tlambda}(q_1, \eta_2, \eta_1) \Big[ \delta^3(\bm{p_1} - \bm{p_2}) + \delta^3(\bm{p_2} + \bm{q_1}) \Big] \qquad\nonumber\\
     + \frac{\mathcal{H}_2^2}{p_2 q_2} C_{\tv\tlambda}(p_1, \eta_1, \eta_2)C_{\tv\tlambda}(q_1, \eta_1, \eta_2) \Big[ \delta^3(\bm{p_1} - \bm{p_2}) + \delta^3(\bm{p_2} + \bm{q_1}) \Big] \bigg\}.\qquad
\end{eqnarray}
\end{samepage}
We can now simplify this expression using the properties of the TT-projector $\Lambda_{ij, k\ell}$. Indeed, the terms in the square brackets proportional to $\delta^3(\bm{p_1} -\bm{p_2})$ lead to
\begin{equation}
    \Lambda_{k\ell, mn}(\bm{k_1}) \hat{p}_1^k \hat{q}_1^\ell \hat{p}_1^m \hat{q}_1^n  = \Lambda_{k\ell, mn}(\bm{k_1}) \hat{p}_1^k \hat{q}_1^\ell \hat{q}_1^m \hat{p}_1^n  = \frac{1}{2}(1-\mu_{p_1}^2)^2 \frac{p_1^2}{q_1^2},
\end{equation}
where in the first step we used the definition $\bm{q} = \bm{p} - \bm{k}$ and the fact that $\Lambda_{ij, k\ell}$ is transverse on all its indices, and in the second step we defined $\mu_p \equiv \hat{\bm{p}}\cdot \hat{\bm{k}}$. The calculation for the term in the square brackets proportional to $\delta^3(\bm{p_2} + \bm{q_1})$ proceeds analogously, resulting in exactly the same factor. Therefore
\begin{equation}
\begin{split}
    \langle \tilde{\mathcal{S}}^*_{ij}& (\bm{k_1}, \eta_1) \tilde{\mathcal{S}}^{ij}(\bm{k_2}, \eta_2)\rangle^{\text{nlo}} =   \int d^3 \bm{p}\; (1-\mu_p^2)^2 \frac{p^2}{q^2} \times \\
    &\times  \frac{3}{2}(1+\omega)\bigg[ \frac{\mathcal{H}_1^2}{p q} C_{\tv\tlambda}(p, \eta_2, \eta_1)C_{\tv\tlambda}(q, \eta_2, \eta_1) + \frac{\mathcal{H}_2^2}{p q} C_{\tv\tlambda}(p, \eta_1, \eta_2)C_{\tv\tlambda}(q, \eta_1, \eta_2) \bigg].
    \end{split}
\end{equation}
Notice that the correlators $C_{\tv\tlambda}(p, \eta_1, \eta_2)$ are not required to be symmetric in $\eta_1$ and $\eta_2$.
Finally, adding the leading-order contribution~\eqref{eq:U}, we write the shear stress unequal time correlator~\eqref{eq:Us} at next-to-leading-order in the bubble-over-Hubble radius expansion as 
\begin{equation}\label{eq:unequal time correlator}
\begin{split}
    U_{\mathcal{S}}(k,& \eta_1, \eta_2) =  \frac{1}{4\pi^2 k} \int_0^\infty dp \int_{\vert p -k\vert}^{p+k}dq\, (1-\mu_p^2)^2 \frac{p^3}{q}  \bigg\{ C_{\tv \tv}(p, \eta_1, \eta_2)C_{\tv \tv}(q, \eta_1, \eta_2) \\
     & +\frac{3}{2}(1+\omega)\bigg[ \frac{\mathcal{H}_1^2}{p q} C_{\tv\tlambda}(p, \eta_2, \eta_1)C_{\tv\tlambda}(q, \eta_2, \eta_1) + \frac{\mathcal{H}_2^2}{p q} C_{\tv\tlambda}(p, \eta_1, \eta_2)C_{\tv\tlambda}(q, \eta_1, \eta_2) \bigg],
\end{split}
\end{equation}
where we performed the change of integration variable $\mu_p \rightarrow q$ as in equation\eqref{eq:U}.


\subsection{Shear stress correlators from sound waves}\label{sec:correlators}
Conservation of energy and momentum $\nabla_\mu T^{\mu\nu} = 0$ provides the equations of motion for the fluid variables. These are derived in Appendix~\ref{app:sw}, which the reader can use as a guide for the following calculations. At linear order in cosmological perturbations and at next-to-leading-order in the bubble-over-Hubble radius expansion $R_*\mathcal{H}_* \lesssim \mathcal{O}(1)$ we find
\begin{subequations}\label{eq:eom_0}
\begin{eqnarray}
    \tlambda^\prime_{\bm{p}} + \left[1 - \dfrac{9}{2}(1+\omega) \left(\frac{\mathcal{H}}{p}\right)^2 \right] p \tv_{\bm{p}} = 0, \\
    \tv^\prime_{\bm{p}} + \frac{2\nu}{\eta} \tv_{\bm{p}} - c_s^2 \left[ 1 - \frac{3}{2} \frac{1+\omega}{c_s^2}\left(\frac{\mathcal{H}}{p}\right)^2 \right] p\tlambda_{\bm{p}}  = 0.
\end{eqnarray}
\end{subequations}
At the next-to-leading-order in the bubble-over-Hubble radius expansion we can decouple this system and obtain the equations of motion for fluid density and velocity perturbations separately
\begin{subequations}\label{eq:eom}
\begin{eqnarray}
    \tv^{\prime\prime}_{\bm{p}} + \frac{2\nu}{\eta} \tv^\prime_{\bm{p}} + c_s^2 \left[ 1 - \frac{2}{c_s^2} (1+3\omega)\left(\frac{\mathcal{H}}{p}\right)^2 \right] p^2 \tv_{\bm{p}} = 0, \label{sub:one}\\
    \tlambda^{\prime\prime}_{\bm{p}} + \frac{2\nu}{\eta} \tlambda^\prime_{\bm{p}} + c_s^2 \left[ 1 - \frac{3}{2c_s^2}(1+\omega) (1+3\omega)\left(\frac{\mathcal{H}}{p}\right)^2 \right] p^2 \tlambda_{\bm{p}} = 0.\label{eq:v_diff}
\end{eqnarray}
\end{subequations}
These are the equations for damped harmonic oscillators with a time-dependent effective frequency. Previous works~\cite{Hindmarsh:2016lnk, Hindmarsh:2019phv} only considered the free propagation of sound waves on a flat space-time with pure radiation. Equations~\eqref{eq:eom} recover this case in the limit $\nu\rightarrow 0$ and $\mathcal{H}/p \rightarrow 0$. The effect of damping is carried by first derivative terms, and was analyzed in detail in our previous work~\cite{Giombi:2024kju}. Notice that these vanish in pure radiation when $\nu = 0$. The next-to-leading-order general relativistic corrections, carried by the time-dependent terms inside the square brackets, bring a time modulation of the effective frequency of fluid perturbations that modifies over time the amplitude and the phase of sound waves. Analytic solutions to equations~\eqref{eq:eom} and~\eqref{eq:eom_0} can be expressed, at any order in the bubble-over-Hubble radius expansion, as a superposition of plane waves. At the next-to-leading-order we find
\begin{subequations}\label{eq:sound_waves}
\begin{eqnarray}
    \tv_{\bm{p}}(\eta) &=& \left(\frac{\eta}{\eta_*}\right)^{-\nu} \hat{p}_i\Bigg\{ v^i_{\bm{p}} e^{-ic_s p\eta} \left[ 1 - \frac{i}{2c_s p\eta}\left(4+\nu(3+\nu)\right)\right] + \nonumber\\
    && \quad\qquad\qquad\qquad\qquad\qquad + v^{*i}_{-\bm{p}} e^{ic_s p\eta} \left[ 1+ \frac{i}{2c_s p\eta}\left(4+\nu(3+\nu)\right) \right] \Bigg\},\\
    \tlambda_{\bm{p}}(p\eta) &=& -\frac{i}{c_s}  \left(\frac{\eta}{\eta_*}\right)^{-\nu} \hat{p}_i\Bigg\{v^i_{\bm{p}} e^{-ic_s p\eta} \left[ 1 -\frac{i}{2c_s p\eta}\left(4+\nu(1+\nu)\right) \right] - \nonumber\\
    && \quad\qquad\qquad\qquad\qquad\qquad - v^{*i}_{-\bm{p}} e^{i c_s p\eta} \left[ 1 + \frac{i}{2c_s p\eta}\left(4+\nu(1+\nu)\right) \right] \Bigg\}.
\end{eqnarray}
\end{subequations}
The details on the derivation of these solutions is left in Appendix~\ref{app:sw}. We emphasize the distinction between the plane wave amplitudes $v^i_{\bm{p}}, \lambda_{\bm{p}}$ and the Fourier transform of the fluid variables  $\tv^i_{\bm{p}}, \tlambda_{\bm{p}}$. 
The unequal time correlators of Fourier modes $C_{\tv \tv}(p, \eta_1, \eta_2)$ and $C_{\tv\tlambda}(p, \eta_1, \eta_2)$ are thereby computed and related to the spectral density $P_v$ of plane wave amplitudes 
\begin{eqnarray}
    \langle   v_{\bm{p}_1}^i v_{\bm{p}_2}^{*\, j}\rangle &=& \hat{p}^i_1\hat{p}^j_1 P_v(p_1) (2\pi)^3 \delta^3(\bm{p}_1 - \bm{p}_2). \label{eq:P} 
\end{eqnarray}
This way, the unequal time correlators of Fourier modes~\eqref{eqs:GG} can be written, at next-to-leading-order in the bubble-over-Hubble radius expansion, as
\begin{subequations}\label{subeq:C}
    \begin{eqnarray}
        C_{\tv \tv}(p, \eta_1, \eta_2) &=& 2 \left(\frac{\eta_1 \eta_2}{\eta_*^2}\right)^{-\nu} P_v(p)  \bigg\{ \cos(c_s p\eta_-) - \nonumber\\
        && \qquad\qquad - \frac{1}{2c_s} \left(4+\nu(3+\nu)\right)\left(\frac{1}{p\eta_1} -\frac{1}{p\eta_2}\right) \sin(c_s p\eta_-) \bigg\},\label{eq:Ca}\\
        C_{\tv\tlambda}(p, \eta_1, \eta_2) &=& \frac{2}{c_s}  \left(\frac{\eta_1 \eta_2}{\eta_*^2}\right)^{-\nu} P_v(p)  \bigg\{ \sin(c_s p\eta_-) + \nonumber\\
        && \qquad\qquad + \frac{1}{2c_s}\left(\frac{4+\nu(3+\nu)}{p\eta_1} -\frac{4+\nu(1+\nu)}{p\eta_2}\right)\cos(c_s p\eta_-)\bigg\},
    \end{eqnarray}
\end{subequations}
with $\eta_- = \eta_1 - \eta_2$.

\subsection{The gravitational wave power spectrum}\label{sec:the_gravitational_wave_power_spectrum}
We find convenient at this point to introduce the dimensionless wavenumbers $z=kR_*$, $x = pR_*$, $y = q R_*$ and the dimensionless time variables $\uptau \equiv \eta/R_*$,   $\uptau_* \equiv \eta_*/R_*$, $\uptau_{\text{end}} \equiv \eta_{\text{end}}/R_*$. We further define the dimensionless spectral density of plane wave amplitudes
\begin{equation}
    P_v(p) \equiv v_{\text{rms}}^2 R_*^3 \tilde{P}_v\left(p R_*\right),
\end{equation}
with root mean square (RMS) fluid velocity $v_{\text{rms}}$. Remembering the results on the shear stress unequal time correlator~\eqref{eq:unequal time correlator} with spectral densities~\eqref{subeq:C}, the power spectrum of gravitational waves~\eqref{P_hp} can be written as
\begin{equation}\label{eq:Pgw}
    \Pgw =  3(1+\nu)\left(\Gamma v_{\text{rms}}^2\right)^2 (\mathcal{H}_* R_*) \left(\frac{a_*}{a_{\text{r}}}\right)^{\frac{2\nu}{1+\nu}}  \frac{(kR_*)^3}{2\pi^2} \tilde{P}_{\text{gw}} (kR_*),
\end{equation}
with adiabatic index $\Gamma \equiv \bar{w}/\bar{e} = 1+\omega$.
The shape of the spectrum is controlled by the dimensionless spectral density function
\begin{equation}\label{eq:pgw}
    \tilde{P}_{\text{gw}} (kR_*) = \frac{\uptau_*}{\pi^2 z^3}  \int_0^\infty dx \int_{\vert x - z\vert}^{x+z}dy \, \rho(z, x, y) \tilde{P}_v(x) \tilde{P}_v(y) \Delta \left(z, x, y, \uptau_*, \uptau_{\text{end}}, \uptau\right),
\end{equation}
with
\begin{equation}\label{eq:rho_1}
    \rho(z, x, y) = \frac{\left[ y^2 - (x-z)^2\right]^2\left[ (x+z)^2 -  y^2\right]^2}{16xyz^2} ,
\end{equation}
a geometric function that encodes the projection of the sound wave wavenumbers onto the TT-subspace. We notice that $\rho$ vanishes when $y^2= (x\pm z)^2$, that is at the extrema of the integration. The kernel $\Delta \left(z, x, y, \uptau_*, \uptau_{\text{end}}, \uptau\right)$ is a vertex that encodes the resonance condition for sound and gravitational waves, and captures all the general relativistic corrections in the short sound wavelength expansion. Up to the next-to-leading-order we find 
\begin{equation}\label{eq:Delta}
    \begin{split}
        \Delta &\left(z, x, y, \uptau_*, \uptau_{\text{end}}, \uptau\right) =  \iint_{\uptau_*}^{\uptau_{\text{end}}} \frac{d\uptau_1 d\uptau_2}{\uptau_*^2} \left(\frac{\uptau_*^2}{\uptau_1\uptau_2} \right)^{1+\nu} G^\prime_z(\uptau, \uptau_1)G^\prime_z(\uptau, \uptau_2) \times \\
        & \times \bigg\{ \cos(c_s x \uptau_-) \cos(c_s y\uptau_-)\\
        &\qquad - \frac{4+\nu(3+\nu)}{2c_s} \left(\frac{1}{\uptau_1} - \frac{1}{\uptau_2}\right) \Big[ \frac{\sin(c_s x \uptau_-)\cos(c_s y \uptau_-) }{x} + \frac{\sin(c_s y \uptau_-)\cos(c_s x \uptau_-) }{y}\Big] \\
        & \qquad + \frac{(1+\nu)(2+\nu)}{c_s^2} \bigg(\frac{1}{\uptau_2^2} + \frac{1}{\uptau_1^2}\bigg)\frac{\sin(c_s x \uptau_-)\sin(c_s y \uptau_-)}{xy} \bigg\}.
    \end{split}
\end{equation}
As we aim to quantify the gravitational wave power spectrum after the end of the acoustic phase, we can fix the time $\eta$ such that $\eta > \eta_{\rm{end}}$ and $k\eta \gg 1$ for all scales of interest for the acoustic production. 
The factor $G^\prime_z(\uptau, \uptau_1)G^\prime_z(\uptau, \uptau_2)$ is thus estimated in Appendix~\ref{app:Green_function} by averaging the gravitational wave power spectrum over many oscillations at wavenumber $k$, resulting in (see equation~\eqref{GpGp}) 
\begin{equation}\label{GreenGreen}
        G^\prime_z(\uptau, \uptau_1)G^\prime_z(\uptau, \uptau_2) \overset{z\gg 1}{\longrightarrow}  \frac{z\uptau_1 z\uptau_2}{2} \Big[ y_\nu(z\uptau_1) y_\nu(z\uptau_2) + j_\nu(z\uptau_1) j_\nu(z\uptau_2) \Big].
\end{equation}
The spectral density of gravitational waves~\eqref{eq:pgw} with kernel function~\eqref{eq:Delta} and equation~\eqref{GreenGreen} can be thus computed with a numerical integration in four dimensions.

\subsection{Kernel function of the gravitational wave power spectrum}
For the purpose of this work, we aim to reduce the dimensionality of the integral by performing analytically the time integrations in the kernel~\eqref{eq:Delta}, allowing us to speed up the numerical evaluation of the gravitational wave power spectrum. To this end, we approximate the factor~\eqref{GreenGreen} by considering two different regimes: sub-horizon gravitational wave modes with $k\eta_{\rm{end}}>k\eta_* \gg 1$ and super-horizon gravitational wave modes with $k\eta_* < k\eta_{\rm{end}} \ll 1 \ll k\eta$. In both regimes we expand the Green's functions at the next-to-leading-order in $k\eta_* \gg 1$ or $k\eta_{\rm{end}} \ll 1$ respectively. The results of these expansions can be seen in Appendix~\ref{app:Green_function}. In general we can write
\begin{equation}\label{GreenGreen_approx}
        G^\prime_k(\uptau, \uptau_1)G^\prime_k(\uptau, \uptau_2)\big\vert_{k\eta\gg 1} = G^\prime_k(\uptau, \uptau_1)G^\prime_k(\uptau, \uptau_2)\big\vert_{k\eta\gg 1}^{\rm{lo}} + G^\prime_z(\uptau, \uptau_1)G^\prime_z(\uptau, \uptau_2)\big\vert_{k\eta\gg 1}^{\rm{nlo}} + \dots,
\end{equation}
where the leading-order and the next-to-leading  order terms are determined by comparing equation~\eqref{GreenGreen_approx} with equation~\eqref{eq:gpgp_result} in the sub-horizon regime or equation~\eqref{eq:gg_long} in the super-horizon regime. We will make this expansion more explicit in Sections~\ref{sec:super} and~\ref{sec:sub}.

The expression of the kernel~\eqref{eq:Delta} and the expansion of the Green's functions as in equation~\eqref{GreenGreen_approx} finally allow us to identify four different contributions:
\begin{equation}\label{delta_4}
\begin{split}
    \Delta\left(z, x, y, \uptau_*, \uptau_{\text{end}}\right) = \Delta_{\text{sw}}^{\text{lo}} \left(z, x, y, \uptau_*, \uptau_{\text{end}}\right)  + \Delta^{\text{nlo}}_{\text{gw}} \left(z, x, y, \uptau_*, \uptau_{\text{end}}\right) + \\
    + \Delta^{\text{nlo}}_{\text{sw}} \left(z, x, y, \uptau_*, \uptau_{\text{end}}\right)  + \Delta^{\text{nlo}}_{\Phi} \left(z, x, y, \uptau_*, \uptau_{\text{end}}\right),
\end{split}
\end{equation}
where 
\begin{subequations}\label{eqs:kernel}
\begin{align}
    \Delta_{\rm{sw}}^{\rm{lo}} &\left(z, x, y, \uptau_*, \uptau_{\text{end}}\right) =  \iint_{\uptau_*}^{\uptau_{\text{end}}} \frac{d\uptau_1 d\uptau_2}{\uptau_*^2} \left(\frac{\uptau_*^2}{\uptau_1\uptau_2} \right)^{1+\nu}  G^\prime_k(\uptau, \uptau_1)G^\prime_k(\uptau, \uptau_2)\big\vert_{k\eta\gg 1}^{\rm{lo}} \times \nonumber\\
    & \qquad\qquad\qquad\qquad\qquad\qquad\qquad\qquad\qquad\qquad\qquad \times \cos(c_s x \uptau_-) \cos(c_s y\uptau_-),\label{eq:delta_low_general}\\
    \Delta_{\rm{gw}}^{\rm{nlo}} &\left(z, x, y, \uptau_*, \uptau_{\text{end}}\right) = \iint_{\uptau_*}^{\uptau_{\text{end}}} \frac{d\uptau_1 d\uptau_2}{\uptau_*^2} \left(\frac{\uptau_*^2}{\uptau_1\uptau_2} \right)^{1+\nu}  G^\prime_k(\uptau, \uptau_1)G^\prime_k(\uptau, \uptau_2)\big\vert_{k\eta\gg 1}^{\rm{nlo}} \times \nonumber\\
    & \qquad\qquad\qquad\qquad\qquad\qquad\qquad\qquad\qquad\qquad\qquad \times \cos(c_s x \uptau_-) \cos(c_s y\uptau_-),\\
    \Delta_{\rm{sw}}^{\rm{nlo}} &\left(z, x, y, \uptau_*, \uptau_{\text{end}}\right) = - \frac{4+\nu(3+\nu)}{2c_s} \iint_{\uptau_*}^{\uptau_{\text{end}}} \frac{d\uptau_1 d\uptau_2}{\uptau_*^2} \left(\frac{\uptau_*^2}{\uptau_1\uptau_2} \right)^{1+\nu}  G^\prime_k(\uptau, \uptau_1)G^\prime_k(\uptau, \uptau_2)\big\vert_{k\eta\gg 1}^{\rm{lo}} \times \nonumber\\
    & \qquad\qquad\qquad \times\left(\frac{1}{\uptau_1} - \frac{1}{\uptau_2}\right) \Big[ \frac{\sin(c_s x \uptau_-)\cos(c_s y \uptau_-) }{x} + \frac{\sin(c_s y \uptau_-)\cos(c_s x \uptau_-) }{y}\Big], \\
     \Delta_{\Phi}^{\rm{nlo}} &\left(z, x, y, \uptau_*, \uptau_{\text{end}}\right) = \frac{(1+\nu)(2+\nu)}{c_s^2} \iint_{\uptau_*}^{\uptau_{\text{end}}} \frac{d\uptau_1 d\uptau_2}{\uptau_*^2} \left(\frac{\uptau_*^2}{\uptau_1\uptau_2} \right)^{1+\nu}   G^\prime_k(\uptau, \uptau_1)G^\prime_k(\uptau, \uptau_2)\big\vert_{k\eta\gg 1}^{\rm{lo}} \times \nonumber\\
    & \qquad\qquad\qquad\qquad\qquad\qquad\qquad\qquad\quad \times\bigg(\frac{1}{\uptau_2^2} + \frac{1}{\uptau_1^2}\bigg)\frac{\sin(c_s x \uptau_-)\sin(c_s y \uptau_-)}{xy} \bigg\}.
\end{align}
\end{subequations}
In these expressions, $\Delta_{\text{sw}}^{\text{lo}}$ is the leading-order contribution from sound waves propagating in a flat expanding Universe, $\Delta_{\text{gw}}^{\text{nlo}}$ the next-to-leading-order (NLO) correction coming from the gravitational wave Green's functions,  $\Delta_{\text{sw}}^{\text{nlo}}$ the NLO correction due to a modification of the sound waves effective frequency~\eqref{eq:sound_waves}, and $\Delta^{\text{nlo}}_\Phi$ the contribution from curvature perturbations. In the following we will discuss these terms in different regimes. 


\subsubsection{Gravitational wave mode expansion on super-horizon scales}\label{sec:super}
Gravitational wave modes on super-horizon scales $k\eta_* < k\eta_{\rm{end}}\ll1 \ll k\eta$ were outside the causal horizon during the acoustic phase, and did not have time to complete one period of oscillation within the duration of the source, i.e. $k(\eta_{\text{end}}-\eta_*) \ll 1$. In this regime, the Green's function~\eqref{eq:Green} can be estimated by expanding the argument of the Bessel functions for small argument, as shown in Appendix~\ref{app:Green_function}, resulting in (see equation~\eqref{eq:gg_long})
\begin{subequations}\label{eq:gg_low_1}
    \begin{align}
        G^\prime_k(\uptau, \uptau_1)G^\prime_k(\uptau, \uptau_2)\big\vert_{k\eta\gg 1}^{\rm{lo}} & \overset{k\eta_{\rm{end}}\ll 1}{=} \frac{\Gamma^2 \left(\frac{1}{2} +\nu \right)}{2\pi} \left(\frac{k\eta_1}{2} \right)^{-\nu} \left(\frac{k\eta_2}{2} \right)^{-\nu} \\
        G^\prime_k(\uptau, \uptau_1)G^\prime_k(\uptau, \uptau_2)\big\vert_{k\eta\gg 1}^{\rm{nlo}} & \overset{k\eta_{\rm{end}}\ll 1}{=} - \sin(\pi\nu) \frac{\Gamma \left(\frac{1}{2} +\nu \right) \Gamma \left(-\frac{1}{2} -\nu \right)}{2\pi} \times\qquad\qquad\qquad\qquad\qquad\qquad \nonumber\\
        & \times \left[ \left(\frac{k\eta_1}{2} \right)^{1+\nu} \left(\frac{k\eta_2}{2} \right)^{-\nu} + \left(\frac{k\eta_1}{2} \right)^{-\nu}\left(\frac{k\eta_2}{2} \right)^{1+\nu} \right],    \quad \nu \not\in \frac{\mathbb{N}^0}{2}.
    \end{align}
\end{subequations}
This expression provides a good approximation for all cases with $\nu\not\in\mathbb{N}^0/2$. The case $\nu = 1/2$ can be studied using the complete expression~\eqref{eq:gg_long}. For the sake of simplicity, in this work we only consider equation~\eqref{eq:gg_low_1}, and we understand the case $\nu = 1/2$ with a limit procedure. We refer to Appendix~\ref{app:kernel} for a more detailed explanation.   
We also recall the limit of validity of equation~\eqref{eq:gg_low_1}, which holds for wavenumbers $k$ on super-horizon scales throughout the acoustic production $k\eta_{\rm{end}} \ll 1$ but that are well inside the horizon at the time $\eta$, that is $k\eta \gg 1$, as required by equation~\eqref{GreenGreen}. 

The properties of the trigonometric functions inside the kernel~\eqref{eq:Delta} allow us to separate the integration variables $\uptau_1$ and $\uptau_2$ and perform the integration analytically. The details of the integration are given in Appendix~\ref{app_low}, resulting in
\begin{subequations}\label{eq:kernel_super}
    \begin{align}
    \Delta^{\text{lo}}_{\text{sw}} &\overset{k\eta_{\rm{end}}\ll 1}{=} \left(\frac{k\eta_*}{2}\right)^{-2\nu}\frac{\Gamma^2 \left( \frac{1}{2} +\nu\right)}{4\pi} \sum_{m = \pm}  \left\vert \omega_m\uptau_*\right\vert^{4\nu} \times \nonumber\\
    & \qquad\qquad\qquad\qquad\qquad\qquad \times \left[\left(\operatorname{ci}_{-2\nu}(\omega_m\uptau)\Big\vert^{\uptau_{\text{end}}}_{\uptau_*} \right)^2  + \left( \operatorname{si}_{-2\nu}(\omega_m\uptau)\Big\vert^{\uptau_{\text{end}}}_{\uptau_*} \right)^2 \right],\label{eq:delta_lo}\\ 
    \Delta^{\text{nlo}}_{\text{gw}} &\overset{k\eta_{\rm{end}}\ll 1}{=} -\left(\frac{k\eta_*}{2} \right)\sin(\pi\nu)  \frac{\Gamma\left(-\frac{1}{2}-\nu\right)\Gamma \left(\frac{1}{2}+\nu\right)}{2\pi} \sum_{m= \pm} \vert \omega_m \uptau_* \vert^{-1+2\nu} \times \nonumber\\
    & \qquad\qquad \times\left[\operatorname{ci}_{-2\nu}(\omega_m\uptau)\Big\vert^{\uptau_{\text{end}}}_{\uptau_*} \sin(\omega_m\uptau)\Big\vert^{\uptau_{\text{end}}}_{\uptau_*}  - \operatorname{si}_{-2\nu}(\omega_m\uptau)\Big\vert^{\uptau_{\text{end}}}_{\uptau_*} \cos(\omega_m\uptau) \Big\vert^{\uptau_{\text{end}}}_{\uptau_*}\right], \\
    \Delta^{\text{nlo}}_{\text{sw}} &\overset{k\eta_{\rm{end}}\ll 1}{=} -\left(\frac{k\eta_*}{2}\right)^{-2\nu}\frac{\Gamma^2 \left( \frac{1}{2} +\nu\right)}{2\pi} \frac{4+\nu(3+\nu)}{2c_s^2 \uptau_*^2 xy } \sum_{m =\pm} m    \left\vert \omega_m \uptau_*\right\vert^{2+4\nu} \times \nonumber\\
    & \times\Big[\operatorname{si}_{-1-2\nu}(\omega_m\uptau)\Big\vert^{\uptau_{\text{end}}}_{\uptau_*} \operatorname{ci}_{-2\nu}(\omega_m\uptau)\Big\vert^{\uptau_{\text{end}}}_{\uptau_*} - \operatorname{ci}_{-1-2\nu}(\omega_m\uptau)\Big\vert^{\uptau_{\text{end}}}_{\uptau_*} \operatorname{si}_{-2\nu}(\omega_m\uptau)\Big\vert^{\uptau_{\text{end}}}_{\uptau_*} \Big] ,\\
    \Delta^{\text{nlo}}_\Phi &\overset{k\eta_{\rm{end}}\ll 1}{=} -\left(\frac{k\eta_*}{2}\right)^{-2\nu}\frac{\Gamma^2 \left( \frac{1}{2} +\nu\right)}{2\pi}  \frac{(1+\nu)(2+\nu)}{c_s^2 xy\uptau_*^2} \sum_{m = \pm} m \left\vert \omega_m \uptau_* \right\vert^{2+4\nu} \times  \nonumber\\
    & \quad \times \Big[\operatorname{ci}_{-2-2\nu}(\omega_m\uptau)\Big\vert^{\uptau_{\text{end}}}_{\uptau_*} \operatorname{ci}_{-2\nu}(\omega_m\uptau)\Big\vert^{\uptau_{\text{end}}}_{\uptau_*} + \operatorname{si}_{-2-2\nu}(\omega_m\uptau)\Big\vert^{\uptau_{\text{end}}}_{\uptau_*} \operatorname{si}_{-2\nu}(\omega_m\uptau)\Big\vert^{\uptau_{\text{end}}}_{\uptau_*} \Big],
\end{align}
\end{subequations}
with $\omega_\pm = c_s (x\pm y)$ and $\nu = (1-3\omega)/(1+3\omega)$. To simplify the above expressions, we introduced a new notation such that $f(\uptau)\vert^{\uptau_{\text{end}}}_{\uptau_*} \equiv f(\uptau_{\text{end}}) - f(\uptau_*)$ for any arbitrary function $f(\uptau)$. The time integration is thereby completely performed analytically in the low-frequency regime by writing the kernel in terms of the generalized sine~\cite[\href{https://dlmf.nist.gov/8.21.E4}{(8.21.4)}]{NIST:DLMF} and cosine~\cite[\href{https://dlmf.nist.gov/8.21.E5}{(8.21.5)}]{NIST:DLMF} integral functions, defined respectively as
 \begin{equation}\label{eqs:trigo}
     \operatorname{si}(\nu, x) =  \int_x^\infty \frac{\sin(t)}{t^{1-\nu}}dt ,\qquad\quad \operatorname{ci}(\nu, x) =  \int_x^\infty \frac{\cos(t)}{t^{1-\nu}}dt.
\end{equation}
We adopt the sign convention of Ref.~\cite{NIST:DLMF}, which implies $\operatorname{ci}(0,x) = - \operatorname{Ci}(x)$ and $\operatorname{si}(0,x) = - \operatorname{si}(x)$. For convenience, in this article we prefer to use a more compact notation, such that $\operatorname{ci}_\nu(x) \equiv \operatorname{ci}(\nu,x)$ and $\operatorname{si}_\nu(x) \equiv \operatorname{si}(\nu,x)$. 

We previously studied the effects of a softer equation of state on the gravitational wave power spectrum from sound waves in Ref.~\cite{Giombi:2024kju}. The leading-order term~\eqref{eq:delta_low_general} with $G^\prime_k(\uptau, \uptau_1)G^\prime_k(\uptau, \uptau_2)\big\vert_{k\eta\gg 1}^{\rm{lo}}$ given by equation~\eqref{eq:gg_low_1} should then reproduce the results of Ref.~\cite{Giombi:2024kju}, in particular equation (B.4) therein, in the same super-horizon regime. The two results are however not compatible in general due to the fact that in Ref.~\cite{Giombi:2024kju} the sub-horizon expansion $k\eta_* \gg 1$ is used across all gravitational wave frequencies, ignoring the modification of the gravitational wave Green's function in the super-horizon regime $k\eta_*< k\eta_{\rm{end}} \ll 1$ and $k\eta \gg1$. The discrepancy between equation~\eqref{eq:delta_lo} and our previous result increases with the softening of the equation of state, and vanishes in radiation. In fact, the two results are identical in pure radiation, where $\nu = 0$ and $G^\prime_k(\uptau, \uptau_1)G^\prime_k(\uptau, \uptau_2)\big\vert_{k\eta\gg 1}^{\rm{lo}} \overset{\nu \rightarrow0}{\longrightarrow} 1/2$, and the kernel term~\eqref{eq:delta_low_general} in this limit recovers equation (B.4) of Ref.~\cite{Giombi:2024kju}. This is also the result found in Refs.~\cite{Sharma:2023mao, RoperPol:2023dzg}.
We finally point out that $\Delta_{\text{gw}}^{\text{nlo}}$ vanishes in pure radiation, where $\nu= 0$.

\subsubsection{Gravitational wave mode expansion on sub-horizon scales}\label{sec:sub}
For gravitational wave modes on sub-horizon scales $k\eta_{\rm{end}} > k\eta_* \gg 1$ we estimate the gravitational wave Green's function~\eqref{eq:Green} by taking the limit of large arguments. At the NLO in the short gravitational wave wavelength expansion we find (see equation~\eqref{eq:gpgp_result})
\begin{subequations}\label{eq:GG_sub}
    \begin{align}
        G^\prime_k(\uptau, \uptau_1)G^\prime_k(\uptau, \uptau_2)\big\vert_{k\eta\gg 1}^{\rm{lo}} & \overset{k\eta_*\gg 1}{\simeq} \frac{1}{2} \cos\left(k\eta_-\right) \, ,\\
        G^\prime_z(\uptau, \uptau_1)G^\prime_z(\uptau, \uptau_2)\big\vert_{k\eta\gg 1}^{\rm{nlo}} & \overset{k\eta_*\gg 1}{\simeq} - \frac{\nu(1+\nu)}{4}\sin\left(k\eta_-\right) \left(\frac{1}{k\eta_1} -\frac{1}{k\eta_2}\right)  \, ,
    \end{align}
\end{subequations}
with $\eta_- = \eta_1 - \eta_2$. The details on the analytical integration of the kernel~\eqref{eq:Delta} are left in Appendix~\ref{app_large}, where we find
\begin{subequations}\label{eq:kernel_sub}
    \begin{align}
        \Delta^{\text{lo}}_{\text{sw}} &\overset{k\eta_*\gg 1}{=} \frac{1}{8} \sum_{m, n = \pm 1} \left\vert\omega_{mn} \uptau_*\right\vert^{2\nu} \left[\left(\operatorname{ci}_{-\nu}(\omega_{m n} \uptau)\Big\vert^{\uptau_{\text{end}}}_{\uptau_*} \right)^2  + \left( \operatorname{si}_{-\nu}(\omega_{m n} \uptau)\Big\vert^{\uptau_{\text{end}}}_{\uptau_*} \right)^2 \right], \label{eq:delta_lo_sw}\\ 
        \Delta^{\text{nlo}}_{\text{gw}} &\overset{k\eta_*\gg 1}{=} -\frac{\nu (1+\nu)}{8}  \sum_{m, n = \pm 1}  \frac{\left\vert\omega_{mn} \uptau_*\right\vert^{1+2\nu}}{z\uptau_*} \operatorname{sign}(\omega_{mn})\times  \nonumber\\
        & \;\times\left[ \operatorname{si}_{-1-\nu}(\omega_{m n} \uptau)\Big\vert^{\uptau_{\text{end}}}_{\uptau_*} \operatorname{ci}_{-\nu}(\omega_{m n} \uptau)\Big\vert^{\uptau_{\text{end}}}_{\uptau_*} - \operatorname{ci}_{-1-\nu}(\omega_{m n} \uptau)\Big\vert^{\uptau_{\text{end}}}_{\uptau_*} \operatorname{si}_{-\nu}(\omega_{m n} \uptau)\Big\vert^{\uptau_{\text{end}}}_{\uptau_*} \right], \label{eq:delta_nlo_gw}\\
        \Delta^{\text{nlo}}_{\text{sw}} &\overset{k\eta_*\gg 1}{=} -\frac{1}{8} \frac{4+\nu(3+\nu)}{c_s} \sum_{m, n = \pm 1} \left(\frac{n}{y\uptau_*} + \frac{m}{x\uptau_*}\right) \left\vert\omega_{mn} \uptau_*\right\vert^{1+2\nu} \operatorname{sign}(\omega_{mn})\times\qquad\qquad \nonumber\\
        & \;\times\left[ \operatorname{si}_{-1-\nu}(\omega_{m n} \uptau)\Big\vert^{\uptau_{\text{end}}}_{\uptau_*} \operatorname{ci}_{-\nu}(\omega_{m n} \uptau)\Big\vert^{\uptau_{\text{end}}}_{\uptau_*} - \operatorname{ci}_{-1-\nu}(\omega_{m n} \uptau)\Big\vert^{\uptau_{\text{end}}}_{\uptau_*} \operatorname{si}_{-\nu}(\omega_{m n} \uptau)\Big\vert^{\uptau_{\text{end}}}_{\uptau_*} \right], \label{eq:delta_nlo_sw}\\
        \Delta^{\text{nlo}}_{\Phi} &\overset{k\eta_*\gg 1}{=} -\frac{2(1+\nu) (2+\nu)}{8 c_s^2}  \sum_{m, n = \pm 1} mn \frac{\left\vert\omega_{mn} \uptau_*\right\vert^{2+2\nu}}{xy\uptau_*^2}  \times  \nonumber\\
        & \;\times\left[ \operatorname{ci}_{-2-\nu}(\omega_{m n} \uptau)\Big\vert^{\uptau_{\text{end}}}_{\uptau_*} \operatorname{ci}_{-\nu}(\omega_{m n} \uptau)\Big\vert^{\uptau_{\text{end}}}_{\uptau_*} + \operatorname{si}_{-2-\nu}(\omega_{m n} \uptau)\Big\vert^{\uptau_{\text{end}}}_{\uptau_*} \operatorname{si}_{-\nu}(\omega_{m n} \uptau)\Big\vert^{\uptau_{\text{end}}}_{\uptau_*} \right], \label{eq:delta_nlo_phi}
    \end{align}
\end{subequations}
with
\begin{equation}
    \omega_{m n} = z +c_s(mx + ny),
\end{equation}
and $\nu = (1-3\omega)/(1+3\omega)$. 
We notice that, in the conformal limit $\nu = 0$, the leading-order contribution $\Delta_{\text{sw}}^{\text{lo}}$ reverts to the well known expression used for an expanding Universe in pure radiation~\cite{Sharma:2023mao, RoperPol:2023dzg}, and the next-to-leading-order contribution $\Delta_{\text{gw}}^{\text{nlo}}$ vanishes because gravitational waves propagate freely in a conformally expanding Universe.

We finally remark again on the different meaning of the label NLO for the three contributions. While for the terms $\Delta_{\Phi}^{\text{nlo}}$ and $\Delta_{\text{sw}}^{\text{nlo}}$ this indicates the contributions at next-to-leading-order in the short sound wavelength expansion $R_*\mathcal{H}_* \lesssim \mathcal{O}(1)$, for the term $\Delta_{\text{gw}}^{\text{nlo}}$ it has a twofold meaning: the next-to-leading-order in the super-horizon modes ($k\eta_* < k\eta_{\rm{end}} \ll 1 \ll k\eta$) or sub-horizon modes ($k\eta_{\rm{end}} > k\eta_* \gg 1$).

\subsubsection{Immediate return to the radiation era} 
If the equation of state right after the phase transition is again pure radiation, that is $\eta_{\text{r}} \rightarrow \eta_*$, we can model the acoustic phase as an era of pure radiation with $\omega = 1/3$ and $\nu = 0$. The gravitational wave Green's function has a closed expression in terms of trigonometric functions and, as obtained in equation~\eqref{eq:GG_rad} of Appendix~\ref{app:Green_function}, we have
\begin{equation}
        G^\prime_k(\eta, \eta_1)G^\prime_k(\eta, \eta_2)\overset{\nu=0}{\simeq}  \frac{1}{2}  \cos\left[k(\eta_1-\eta_2) \right].
\end{equation}
We notice that this is exactly the leading-order term in the expansion of the Green's function~\eqref{eq:GG_sub}. Therefore, we can follow the same steps outlined in Appendix~\ref{app_large}, setting $\nu = 0$, to obtain the kernel contributions
\begin{subequations}\label{kernel_rad}
    \begin{eqnarray}
        \Delta^{\text{lo}}_{\text{sw}} \Big\vert_{\nu =0}&=& \frac{1}{8} \sum_{m, n = \pm 1} \left[\left(\operatorname{Ci}(\omega_{m n} \uptau)\Big\vert^{\uptau_{\text{end}}}_{\uptau_*} \right)^2  + \left( \operatorname{si}(\omega_{m n} \uptau)\Big\vert^{\uptau_{\text{end}}}_{\uptau_*} \right)^2 \right],\\
        \Delta^{\text{nlo}}_{\text{gw}} \Big\vert_{\nu =0} &=& 0\, , \\
        \Delta^{\text{nlo}}_{\text{sw}} \Big\vert_{\nu =0}&=&  \frac{1}{2c_s} \sum_{m, n = \pm 1} \left(\frac{n}{y} + \frac{m}{x}\right) \left\vert\omega_{mn}\right\vert \operatorname{sign}(\omega_{mn})\times\qquad\qquad\qquad\qquad\qquad\qquad\qquad  \nonumber\\
        && \;\times\left[\operatorname{si}_{-1}(\omega_{m n} \uptau)\Big\vert^{\uptau_{\text{end}}}_{\uptau_*} \operatorname{Ci}(\omega_{m n} \uptau)\Big\vert^{\uptau_{\text{end}}}_{\uptau_*} - \operatorname{ci}_{-1}(\omega_{m n} \uptau)\Big\vert^{\uptau_{\text{end}}}_{\uptau_*} \operatorname{si}(\omega_{m n} \uptau)\Big\vert^{\uptau_{\text{end}}}_{\uptau_*} \right],\label{eq:delta_sw_nlo_nu0}\\ 
        \Delta^{\text{nlo}}_{\Phi} \Big\vert_{\nu =0} &=& \frac{1}{2 c_s^2}  \sum_{m, n = \pm 1} mn \frac{\omega^2_{mn}}{xy} \times  \nonumber\\
        && \;\times\left[ \operatorname{ci}_{-2}(\omega_{m n} \uptau)\Big\vert^{\uptau_{\text{end}}}_{\uptau_*} \operatorname{Ci}(\omega_{m n} \uptau)\Big\vert^{\uptau_{\text{end}}}_{\uptau_*} + \operatorname{si}_{-2}(\omega_{m n} \uptau)\Big\vert^{\uptau_{\text{end}}}_{\uptau_*} \operatorname{si}(\omega_{m n} \uptau)\Big\vert^{\uptau_{\text{end}}}_{\uptau_*} \right].
    \end{eqnarray}
\end{subequations}
We remind the reader about the sign convention on the trigonometric integral functions, discussed below equation~\eqref{eqs:trigo}, such that $\operatorname{si}_0(x) = -\operatorname{si}(x)$ and $\operatorname{ci}_0(x) = -\operatorname{Ci}(x)$.  These expressions of the kernel, which holds only in pure radiation, are valid across the whole gravitational wave frequency spectrum. As a consistency check, while it is clear that, by construction, the kernel functions in the sub-horizon scales~\eqref{eq:kernel_sub} recover the expressions~\eqref{kernel_rad} when $\nu = 0$, 
we point out that also the kernel functions in the super-horizon scales~\eqref{eq:kernel_super} agree with equations~\eqref{kernel_rad} when $\nu=0$ and $k\eta_* = z\uptau_* \ll 1 $. We also recall that all these expressions are always derived in the limit $k\eta \gg 1$.

In the case $\nu = 0$, the NLO contribution  from the gravitational wave Green's function $\Delta_{\text{gw}}^{\text{nlo}}$ vanishes because gravitational waves propagate in a conformally expanding Universe, and the Green's function~\eqref{eq:Green} reverts to the constant-amplitude, oscillatory sine function~\eqref{eq:G_rad}. Instead, even in a conformally expanding Universe, the sound wave NLO contribution $\Delta_{\text{gw}}^{\text{nlo}}$ does not vanish. Indeed, even if background expansion effects  vanish when $\nu= 0$, the propagation of sound waves is additionally influenced by the self gravity of the fluid. We show in Appendix~\ref{app:sw} how curvature perturbations generated by the sound waves modify the dynamics. In the end, these effects give rise to the kernel contribution $\Delta_{\text{sw}}^{\text{nlo}}$ in equation~\eqref{eq:delta_sw_nlo_nu0}.

\section{Acoustic gravitational wave power spectrum at next-to-leading-order}\label{sec:analytic}
In the sound shell model, the spectral density $P_v(k)$ of plane wave amplitudes~\eqref{eq:P} 
is computed from the hydrodynamic solution of a 
single expanding bubble~\cite{Hindmarsh:2016lnk, Hindmarsh:2019phv}. The dynamics of a self-gravitating bubble in a general relativistic framework was studied in Refs.~\cite{Giombi:2023jqq, Jinno:2024nwb}, but the implications on the spectral density $P_v(k)$ are yet to be understood. For the purpose of this work, in analogy with the analysis of Ref.~\cite{Giombi:2024kju}, we then choose to use the ansatz 
\begin{equation}\label{eq:Pv}
    P_v(k) = 3\pi \frac{v_{\text{rms}}^2}{k_p^3}\frac{(k/k_p)^2}{1+ (k/k_p)^6},
\end{equation}
where we define the peak gravitational wavenumber as $k_p R_* = 2\pi$. The ansatz~\eqref{eq:Pv} is physically well motivated for an irrotational causal flow~\cite{Durrer:2003ja} in presence of shocks~\cite{Dahl:2021wyk}. The root mean squared velocity is set by the Parseval's theorem as  
\begin{equation}
    v_{\text{rms}}^2 = \int\frac{d^3\bm{p}}{(2\pi)^3}C_{\tilde{v}\tilde{v}}(p, \eta_*, \eta_*),
\end{equation}
with equal time correlator $C_{\tilde{v}\tilde{v}}(p, \eta_*, \eta_*)$ given by equation~\eqref{eq:Ca}. The parametrization of the velocity spectral density with the analytic function~\eqref{eq:Pv} allows us to pursue the evaluation of the gravitational wave power spectrum analytically in different frequency regimes. The leading-order contribution~\eqref{eq:delta_lo_sw} has been discussed in details in Ref.~\cite{Giombi:2024kju}; in the next sections we will focus on the NLO contributions~\eqref{eq:kernel_super} and~\eqref{eq:kernel_sub}. 

For the sake of clarity and coherence, the calculations leading to the asymptotic expressions of the kernel function terms used in Sections~\ref{sec:low},~\ref{subsec:intermediate},~\ref{sec:high} are presented in Appendix~\ref{app:kernel}.

\subsection{Asymptotic expansions of the gravitational wave power spectrum}\label{sec:table}
The approximations of the gravitational wave Green's function on super-horizon~\eqref{eq:gg_low_1} and sub-horizon~\eqref{eq:GG_sub} scales allow us to perform analytically the time-integrations in the kernel functions~\eqref{eqs:kernel}, resulting in equations~\eqref{eq:kernel_super} and equations~\eqref{eq:kernel_sub} respectively. With these approximations, the calculation of the gravitational wave power spectrum~\eqref{eq:Pgw} reduces to an integration in two dimensions. In Section~\ref{sec:results} we show the results of this two-dimensional integration for some values of the phase transition parameters (speed of sound $c_s$, bubble-over-Hubble radius $R_*\mathcal{H}_*$, source duration $\detav =\eta_{\rm{end}} - \eta_*$ and radiation time $\eta_{\rm{r}}$).

Some additional approximations in different asymptotic regimes allow us to further compute analytically (entirely or partially) the integrations over the sound wave wavenumbers in equation~\eqref{eq:pgw}. These approximations are discussed thoroughly in the next sections. Here we would like to give a brief overview in order to introduce the reader to the different notation.

On super-horizon  scales $k\eta_* <k\eta_{\rm{end}} \ll 1 \ll k\eta$, we find analytical approximations to the gravitational wave power spectrum in the regime of ``\textit{low-frequency}'' $kR_*\sim (R_*\mathcal{H}_*)k\eta_* \ll1$, where the sound wave wavenumbers tend to align $y\approx x$.

On sub-horizon scales $k\eta_{\rm{end}}> k\eta_* \gg 1$, analytical approximations can be found for modes that oscillate fast within the duration of the source, that is for sufficiently long lasting sources such that $k(\eta_{\rm{end}} - \eta_*) \gg 1$. Here we further divide the spectrum into two regimes: first, an ``\textit{intermediate-frequency}'' regime, where $1\ll k\eta_* \ll k_p\eta_*$, that is, recalling that $k_p = 2\pi/R_*$, $kR_* \ll 2\pi$. In this regime the sound wave wavenumber $\bm{y}$ can be approximately evaluated as $y\approx x -\mu z$, with $\mu = \hat{\bm{x}} \cdot \hat{\bm{z}}$. Second, a ``\textit{high-frequency}'' regime for gravitational wave wavenumbers around the energy injection scale $kR_*\sim 2\pi$ and beyond $kR_* > 2\pi$.  In the case of a long-lasting source, high-frequency modes oscillate many times during the acoustic phase, so that sound waves at this frequency decorrelate much before the end of the acoustic phase, allowing to extend the support of the time-integration in the kernel to $\eta_{\rm{end}} \rightarrow\infty$. 

The different approximations are discussed in detail in Appendix~\ref{app:kernel}. To help the reader familiarise themselves with the nomenclature of the different regimes, we provide a summary in Table~\ref{table_approx}. 


\begin{table}[H]
\begin{center}
\begin{tabular}{|c|c|c|}
\hline
\multicolumn{2}{|c|}{\textbf{Frequency regime}} & \textbf{Approximation} \\ \hline
\makecell{Super-horizon\\ $k\eta_* < k\eta_{\rm{end}} \ll 1 \ll k\eta$}  & \makecell{Low-frequency\\ $kR_* \ll 1$\\
$k(\eta_{\rm{end}} -\eta_*) \ll 1$} & $y\approx x$ \\ \hline
\multirow{2}{*}{\makecell{Sub-horizon\\ $k\eta_{\rm{end}} \gg k\eta_* \gg 1$ }} & \makecell{Intermediate-frequency \\ $ kR_* \ll 2\pi$ \\ $k(\eta_{\rm{end}} -\eta_*) \gg 1$} & $y\approx x -\mu z$ \\ \cline{2-3}
 & \makecell{High-frequency\\ $(kR_*\sim 2\pi) \cup (kR_* > 2\pi)$ \\ $k(\eta_{\rm{end}} -\eta_*) \gg 1$} & \makecell{$\eta_{\rm{end}} \rightarrow \infty$ \\ $\uptau_- \ll \uptau_+$}  \\ \hline
\end{tabular}
\caption{Summary of the analytic approximations to the gravitational wave power spectrum in different asymptotic regimes. Here $\eta_*$ is the conformal time at which the acoustic motion starts generating gravitational waves, and $\eta_{\rm{end}}$ the time at which  the energy in the acoustic motion is completely dissipated (recall discussion in Section~\ref{sec:greensub}). The length scale of fluid perturbations is the mean bubble spacing $R_*$, and the dimensionless wavenumbers $x$ and $y$ are defined at the beginning of Section~\ref{sec:the_gravitational_wave_power_spectrum}. We also defined $\uptau_\pm \equiv \uptau_1 \pm \uptau_2$. The approximation $\uptau_- \ll \uptau_+$ stems from the fact that gravitational waves are mostly sourced within the first Hubble time, while sound waves rapidly decorrelate afterwards~\cite{Guo:2020grp, RoperPol:2023dzg}. More details are found in Section~\ref{sec:high} and in Appendix~\ref{sec:appendix_high}. Recall that many other assumptions are included in our calculation of the gravitational wave power spectrum~\eqref{eq:Pgw}. We in fact assume that compressional fluid perturbations are linear, stationary, stochastic, and Gaussian, and that $v_{\rm{rms}} \ll R_*\mathcal{H}_*$.}\label{table_approx}
\end{center}
\end{table}

\subsubsection{Low-frequency regime: causal tail of the spectrum}\label{sec:low}
Gravitational wave modes on super-horizon scale $k\eta_* < k\eta_{\rm{end}} \ll 1 \ll k\eta$ do not have time to complete one period of oscillation within the time duration of the acoustic source $k(\eta_{\text{end}}-\eta_*) \ll 1$. The kernels functions for these modes can be approximated as in equation~\eqref{eq:kernel_super}. 
This also implies the ``\textit{low-frequency}'' regime $kR_*\sim (R_*\mathcal{H}_*) k\eta_* \ll 1$, corresponding to gravitational wave modes that are outside the typical correlation length of sound waves. 

In the regime $z\ll 1 $  the wavenumeber $\bm{y} = \bm{x} - \bm{z}$ tends to align to $\bm{x}$, that is $\bm{y} \rightarrow \bm{x}$; in the same way, $\omega_\pm \rightarrow c_s x(1\pm 1)$. Since the kernel functions~\eqref{eq:kernel_super} are now independent of $y$, it is possible to perform the $y$-integration in the spectral density function~\eqref{eq:pgw} analytically. Using $dy \, \rho(z, x, y)  = d\mu \,(1-\mu^2)^2 x^4 z^3 /y^2$, we perform the trivial integration over the azimuth angle $\mu = \hat{\bm{k}}\cdot \hat{\bm{p}}$ and get
\begin{equation}\label{eq:spec_low}
    \tilde{P}_{\text{gw}}^{\text{low}} (kR_*) = \underset{z\rightarrow 0}{\operatorname{lim}}  \;\, \frac{16 \uptau_*}{15 \pi^2} \int_0^\infty dx x^2 \tilde{P}_v^2(x) \Delta (z, x, \uptau_*, \uptau_{\text{end}}).
\end{equation}
Using now the kernel terms in equation~\eqref{kenerl_super}, and defining
\begin{equation}\label{eq:I}
    \mathcal{I}_v \equiv \frac{1}{2\pi^2} \int_0^\infty dx x^2 \tilde{P}_v^2(x), \qquad \mathcal{J}_v \equiv \frac{1}{2\pi^2} \int_0^\infty dx \tilde{P}_v^2(x),
\end{equation}
we find, neglecting the oscillatory and decaying terms of the kernel, the following contributions to the gravitational wave power spectrum
\begin{subequations}\label{eq:P_gw_low}        
    \begin{align}
    &\mathcal{P}^{\text{lo, low}}_{\text{gw, sw}} (kR_*) \overset{kR_*\ll 1}{\simeq}  3(1+\nu)\left(\Gamma v_{\text{rms}}^2\right)^2 (\mathcal{H}_* R_*)^{2\nu} \left(\frac{a_*}{a_{\text{r}}}\right)^{\frac{2\nu}{1+\nu}}  \frac{(kR_*)^{3-2\nu}}{2\pi^2} \left(\frac{1+\nu}{2}\right)^{-2\nu} \times \nonumber\\
    & \qquad\qquad\qquad\qquad\qquad\qquad\qquad\qquad\qquad\times  \frac{\Gamma^2\left(\frac{1}{2}+\nu \right)}{2\pi} (1+\nu) \frac{16}{15} \mathcal{I}_v \Upsilon^2_{2\nu}\left(\frac{\eta_*}{\eta_{\text{end}}} \right), \\
    &\mathcal{P}^{\text{nlo, low}}_{\text{gw, gw}} (kR_*)  \overset{kR_*\ll 1}{\simeq}  - 3(1+\nu)^3\left(\Gamma v_{\text{rms}}^2\right)^2 (\mathcal{H}_* R_*)^{-1} \left(\frac{a_*}{a_{\text{r}}}\right)^{\frac{2\nu}{1+\nu}}  \frac{(kR_*)^4}{2\pi^2} \nonumber\\
    & \qquad\qquad\qquad\qquad \times \sin(\pi\nu)  \frac{\Gamma\left(-\frac{1}{2}-\nu\right)\Gamma \left(\frac{1}{2}+\nu\right)}{2\pi}  \left[\frac{\uptau_{\text{end}}}{\uptau_*} -1\right] \Upsilon_{2\nu}\left(\frac{\eta_*}{\eta_{\text{end}}}\right) \frac{16}{15}\mathcal{I}_v , \label{eq:P_gwgw_low}\\
    &\mathcal{P}^{\text{nlo, low}}_{\text{gw, sw}} (kR_*) \overset{kR_*\ll 1}{\simeq} - 3(1+\nu)^2\left(\Gamma v_{\text{rms}}^2\right)^2 (\mathcal{H}_* R_*)^{2\nu} \left(\frac{a_*}{a_{\text{r}}}\right)^{\frac{2\nu}{1+\nu}}  \frac{(kR_*)^{3-2\nu}}{2\pi^2} \left(\frac{1+\nu}{2}\right)^{-2\nu} \times \nonumber\\
    & \qquad\qquad\qquad\qquad \times  \frac{\Gamma^2\left(\frac{1}{2}+\nu \right)}{2\pi} (4+\nu(3+\nu)) \frac{32}{15\pi^2} \int_0^\infty dx\, x^2 \tilde{P}^2_v(x) (2c_sx\uptau_*)^{4\nu}\times\nonumber\\
    & \quad\times \left[\operatorname{si}_{-1-2\nu} (2c_s x\uptau) \Big\vert^{\uptau_{\text{end}}}_{\uptau_{*}} \operatorname{ci}_{-2\nu} (2c_s x\uptau) \Big\vert^{\uptau_{\text{end}}}_{\uptau_{*}}  -  \operatorname{ci}_{-1-2\nu} (2c_s x\uptau) \Big\vert^{\uptau_{\text{end}}}_{\uptau_{*}} \operatorname{si}_{-2\nu} (2c_s x\uptau) \Big\vert^{\uptau_{\text{end}}}_{\uptau_{*}}\right], \\
    &\mathcal{P}^{\text{nlo, low}}_{\text{gw, }\Phi} (kR_*) \overset{kR_*\ll 1}{\simeq} 3(1+\nu)\left(\Gamma v_{\text{rms}}^2\right)^2 (\mathcal{H}_* R_*)^{2+2\nu} \left(\frac{a_*}{a_{\text{r}}}\right)^{\frac{2\nu}{1+\nu}}  \frac{(kR_*)^{3-2\nu}}{2\pi^2} \left(\frac{1+\nu}{2}\right)^{-2\nu} \times\nonumber\\
    & \qquad\qquad\qquad\qquad\qquad\qquad \times\frac{\Gamma^2\left(\frac{1}{2}+\nu \right)}{\pi}\frac{(2+\nu)}{c_s^2} \Upsilon_{2\nu}\left(\frac{\eta_*}{\eta_{\text{end}}} \right)  \Upsilon_{2+2\nu}\left(\frac{\eta_*}{\eta_{\text{end}}} \right)\frac{16}{15}\mathcal{J}_v, 
    \end{align}
\end{subequations}
where we used $\uptau_* = (1+\nu)/(R_*\mathcal{H}_*)$, and captured the dependence on the duration of the stationary source inside the function
\begin{equation}\label{upsilon}
    \Upsilon_\ell \left(\frac{\eta_*}{\eta_{\rm{end}}}\right) = \frac{1}{\ell(\nu)}\left[1- \left(\frac{\eta_*}{\eta_{\rm{end}}} \right)^{\ell(\nu)}\right].
\end{equation}
With the analytic expression of the fluid velocity spectral density in equation~\eqref{eq:Pv} we can evaluate $\mathcal{I}_v = 1/32\pi^2$ and $\mathcal{J}_v = \mathcal{I}_v/(2\pi)^2$. 
The above estimation of the gravitational wave power spectrum at small wavenumber $k\eta_*< k\eta_{\rm{end}} \ll 1 \ll k\eta$ predicts the scaling 
\begin{equation}\label{eq:power_law_low}
        \mathcal{P}^{\text{lo, low}}_{\text{gw, sw}}(k), \;\mathcal{P}^{\text{nlo, low}}_{\text{gw, sw}} (k), \; \mathcal{P}^{\text{nlo, low}}_{\text{gw, $\Phi$}}(k) \sim  k^{3-2\nu}, \quad\qquad \;\mathcal{P}^{\text{nlo, low}}_{\text{gw, gw}}(k) \sim k^{4}.
\end{equation}
The scaling of the causality tail of the gravitational wave power spectrum is therefore affected by the expansion rate of the Universe, and reverts to the known $\Pgw \propto k^{3}$ profile when $\nu =0$, that is in pure radiation. As we will comment later on, the contribution of $\mathcal{P}^{\text{nlo, low}}_{\text{gw, gw}}$ is subdominant compared to the other terms, so we expect to find an overall scaling $\mathcal{P}^{\text{nlo, low}}_{\text{gw, total}}(k) \sim k^{3-2\nu}$ of the total gravitational wave power spectrum. This power law scaling agrees with the general expectation on the stochastic gravitational wave background for modes that enter the horizon when the Universe is described by a constant equation of state parameter $\omega$~\cite{Cai:2019cdl, Hook:2020phx, Domenech:2021ztg}.

We also remark that the power scaling computed in this section applies for gravitational wave modes such that $k(\eta_{\rm{end}} - \eta_*)\ll 1$ and $k\eta\gg1$. Indeed, the latter condition is necessary for our approximation on the Green's functions of gravitational waves introduced in equation~\eqref{GreenGreen}. On larger scales, such that $k\eta \ll 1$, any causal field whose correlation in physical space decays faster than any power law at large distance $\vert \bm{x}\vert \rightarrow\infty$ is instead expect to lead to the scaling $\mathcal{P}_{\rm{gw}} \sim k^3$, as it was proved in Ref.~\cite{Durrer:2003ja}.

The generalized sine and cosine functions provide oscillatory and decaying contributions that become subdominant when the source lasts for many gravitational wave oscillation periods $k(\eta_{\text{end}} -\eta_*)\gg 1$. The NLO contribution to the sound wave propagation, $\mathcal{P}^{\text{nlo, low}}_{\text{gw, sw}}$, only contains oscillatory decaying terms, and thus rapidly becomes subdominant in this frequency regime. 

The NLO contribution to the gravitational wave Green's function $\mathcal{P}^{\text{nlo, low}}_{\text{gw, gw}}$ vanishes at $\nu = 0$ and is largely suppressed by the hierarchy of scales $(kR_*) \ll (\mathcal{H}_*R_*) \lesssim \mathcal{O}(1)$, valid in this frequency regime. Therefore, at low wavenumber $k$, the steep power law scaling $\mathcal{P}^{\text{nlo, low}}_{\text{gw, gw}}(k) \sim k^4$ makes this contribution rapidly negligible compared to the leading-order term. We also notice that this contribution increases, in absolute value, with the duration of the source. However, we recall that the expansion of the Green's function has been performed in the small parameter $k\eta_*< k\eta_{\text{end}} \ll 1$,  so that the above expression~\eqref{eq:P_gwgw_low} cannot be trusted for indefinitely long-lasting sources, but only as long as this hierarchy is maintained. 
All the other contributions in equations~\eqref{eq:P_gw_low} converge instead to a limiting profile as $\uptau_*/\uptau_{\text{end}}\rightarrow 0$ with either a power law, or, when $\nu =0$, a logarithm, as seen in Refs.~\cite{RoperPol:2023dzg, Giombi:2024kju}.

The largest correction to the leading-order term in the low-frequency regime comes from curvature perturbations $\mathcal{P}^{\text{nlo, low}}_{\text{gw, }\Phi}$, which brings a positive contribution  suppressed by a factor $(R_*\mathcal{H}_*)^2$ with respect to $\mathcal{P}^{\text{lo, low}}_{\text{gw, sw}}$. We also notice that  $\mathcal{P}^{\text{lo, low}}_{\text{gw, sw}} \sim (\mathcal{H}_* R_*)^{2\nu}$. This factor introduces an equation of state dependent suppression on the gravitational wave power spectrum at small wavenumbers $k$ in all cases where $\nu >0$, while it reverts to unity in the conformal radiation case where $\nu =0$.

\subsubsection{Intermediate-frequency regime: shallow growth}\label{subsec:intermediate}
Next we consider gravitational wave modes in the ``\textit{intermediate-frequency}'' range, $1\ll k\eta_*\ll k_p\eta_*$. Assuming that these modes oscillate several times within the acoustic phase  $k\detav = k(\eta_{\text{end}}-\eta_*) \gg 1 $  we find (a detailed derivation can be found in Appendix~\ref{sec:append_int})
\begin{subequations}\label{eqs:pgw_int}
    \begin{eqnarray}
        \mathcal{P}^{\text{lo, int}}_{\text{gw, sw}} (kR_*) &\simeq &  3\left(\Gamma v_{\text{rms}}^2\right)^2 (\mathcal{H}_* R_*)^2 \left(\frac{a_*}{a_{\text{r}}}\right)^{\frac{2\nu}{1+\nu}}  \times \nonumber\\
        && \qquad\qquad\;\;\; \times \frac{z}{\pi^2}  \frac{2}{3c_s^4}  \left[ 3-2c_s^2 - \frac{3}{c_s}(1-c_s^2) \operatorname{arctanh}(c_s) \right]   \mathcal{I}_v,\\
        \mathcal{P}^{\text{nlo, int}}_{\text{gw, gw}} (kR_*) &\simeq & - 3\left(\Gamma v_{\text{rms}}^2\right)^2 (\mathcal{H}_* R_*)^4  \left(\frac{a_*}{a_{\text{r}}}\right)^{\frac{2\nu}{1+\nu}}  \times \nonumber\\
        && \qquad\qquad\quad \times \frac{1}{\pi^2 z} \frac{\nu }{c_s^4 (1+\nu)}    \left[ 3   - \frac{1}{c_s}(3-c_s^2) \operatorname{arctanh}(c_s) \right] \mathcal{I}_v, \\
        \mathcal{P}^{\text{nlo, int}}_{\text{gw, sw}} (kR_*) &\simeq & - 3\left(\Gamma v_{\text{rms}}^2\right)^2 (\mathcal{H}_* R_*)^4 \left(\frac{a_*}{a_{\text{r}}}\right)^{\frac{2\nu}{1+\nu}}  \times \nonumber\\
        && \times \frac{z}{\pi^2 } \frac{4+\nu(3+\nu)}{3c_s^6 (1+\nu)^2}  \left[ 15 - 4c_s^2  - \frac{3}{c_s}(5-3c_s^2) \operatorname{arctanh}(c_s) \right] \mathcal{J}_v, \\
        \mathcal{P}^{\text{nlo, int}}_{\text{gw}, \Phi} (kR_*) &\simeq & 3\left(\Gamma v_{\text{rms}}^2\right)^2 (\mathcal{H}_* R_*)^4  \left(\frac{a_*}{a_{\text{r}}}\right)^{\frac{2\nu}{1+\nu}}  \times \nonumber\\
        &&  \times \frac{z}{\pi^2}4\frac{ (2+\nu)}{3 c_s^6 (1+\nu)}   \left[ 3-2c_s^2 - \frac{3}{c_s}(1-c_s^2) \operatorname{arctanh}(c_s) \right] \mathcal{J}_v,
    \end{eqnarray}
\end{subequations}
with $\mathcal{I}_v$ and $\mathcal{J}_v$ defined in equation~\eqref{eq:I}.
All the general NLO contributions are suppressed by a factor $(\mathcal{H}_*R_*)^{2}$ with respect to the leading-order term, which therefore dominates the gravitational wave power spectrum. The contribution from the modified propagation of gravitational waves decays linearly with the frequency $\tilde{P}^{\text{nlo, int}}_{\text{gw, gw}} (kR_*) \propto k^{-1}$, while all the other terms follow the shallow growth  $\mathcal{P}^{\text{lo, int}}_{\text{gw, sw}} (k)\sim \mathcal{P}^{\text{nlo, int}}_{\text{gw, sw}}(k)\sim \mathcal{P}^{\text{nlo, int}}_{\text{gw}, \Phi} (k)\sim k^1$, found for the leading-order term~\cite{Sharma:2023mao, RoperPol:2023dzg}. We finally emphasize that, contrary to our results in the low-frequency regime, the dependence of each term in equation~\eqref{eqs:pgw_int} on the bubble-over-Hubble radius parameter $R_*\mathcal{H}_*$ does not vary with the equation of state.

\subsubsection{High-frequency regime: spectral peak amplitude}\label{sec:high}

Let us now discuss gravitational wave modes around and beyond the peak scale $k \gtrsim k_p$ that oscillate several times during the acoustic phase $k(\eta_{\rm{end}} - \eta_*) \gg 1$, which we denote as the ``\textit{high-frequency}'' regime. An extended calculation of the gravitational wave power spectrum in this regime is left in Appendix~\ref{sec:appendix_high}. Due to the fast oscillations, the periodic functions in the kernel~\eqref{eqs:kernel} decorrelate very rapidly. However, the emitted gravitational wave is found in phase with the incoming sound waves when $z -c_s(x+y) = 0$, at which the power spectrum develops a strong resonance~\cite{Hindmarsh:2019phv, Sharma:2023mao, RoperPol:2023dzg}. This allows us to approximate the integral of periodic functions with a Dirac delta functions centred at the location of the resonance. Following the analysis outlined in Refs.~\cite{Hindmarsh:2016lnk, Hindmarsh:2019phv}  we find
\begin{subequations}\label{eq:P_gw_high}
    \begin{align}
        \mathcal{P}^{\text{lo, high}}_{\text{gw, sw}} (kR_*) & = \phantom{-}  3(1+\nu)\left(\Gamma v_{\text{rms}}^2\right)^2 (\mathcal{H}_* R_*) \left(\frac{a_*}{a_{\text{r}}}\right)^{\frac{2\nu}{1+\nu}} \Upsilon_{1+2\nu}\left(\frac{\eta_*}{\eta_{\text{end}}} \right)\times\nonumber \\
        & \times \frac{z^2}{2\pi^2} \frac{1}{4\pi c_s } \left(\frac{1-c_s^2}{c_s^2}\right)^2 \int_{x_-}^{x_+} dx \,  \frac{(x-x_+)^2(x-x_-)^2}{xy} \tilde{P}_v(x) \tilde{P}_v(y) , \label{eq:P_sw_high} \\
    \mathcal{P}^{\text{nlo, high}}_{\text{gw, gw}} (kR_*) & =  -3\left(\Gamma v_{\text{rms}}^2\right)^2 (\mathcal{H}_* R_*)^3  \left(\frac{a_*}{a_{\text{r}}}\right)^{\frac{2\nu}{1+\nu}} \Upsilon_{3+2\nu}\left(\frac{\eta_*}{\eta_{\text{end}}}\right) \times  \nonumber\\
    &  \times \frac{z}{2\pi^2}\frac{\nu}{8 \pi c_s^2} \left(\frac{1-c_s^2}{c_s^2}\right)^2  \int_{x_-}^{x_+} dx \frac{(x-x_+)^2(x-x_-)^2}{xy^2} \tilde{P}_v(x) \tilde{P}_v(y) \times \nonumber\\
    & \times \left\{ \frac{7y^4 - 6y^2(x^2+z^2) - (x^2-z^2)^2}{\left[y^2 -(x-z)^2\right] \left[y^2 - (x+z)^2\right]}  +2\left[1-\frac{3(y/2\pi)^6}{1+(y/2\pi)^6} \right]\right\}, \label{eq:p_gw}\\
    \mathcal{P}^{\text{nlo, high}}_{\text{gw, sw}} (kR_*) & =  \phantom{-}3 \left(\Gamma v_{\text{rms}}^2\right)^2 (\mathcal{H}_* R_*)^3  \left(\frac{a_*}{a_{\text{r}}}\right)^{\frac{2\nu}{1+\nu}} \Upsilon_{3+2\nu}\left(\frac{\eta_*}{\eta_{\text{end}}} \right) \times \nonumber \\
    &  \times \frac{z^3}{2\pi^2}  \frac{4+\nu(3+\nu)}{8 \pi c_s^4 (1+\nu)} \left(\frac{1-c_s^2}{c_s^2}\right)^2 \int_{x_-}^{x_+} dx \frac{(x-x_+)^2(x-x_-)^2}{x^2y^3}  \tilde{P}_v(x) \tilde{P}_v(y) \times \nonumber \\
    & \times \left\{ \frac{7y^4 - 6y^2(x^2+z^2) - (x^2-z^2)^2}{\left[y^2 -(x-z)^2\right] \left[y^2 - (x+z)^2\right]} -\frac{x}{x+y} +2\left[1-\frac{3(y/2\pi)^6}{1+(y/2\pi)^6} \right]\right\}, \label{eq:p_sw}\\
        \mathcal{P}^{\text{nlo, high}}_{\text{gw}, \Phi}  (kR_*) & = - 3\left(\Gamma v_{\text{rms}}^2\right)^2 (\mathcal{H}_* R_*)^3  \left(\frac{a_*}{a_{\text{r}}}\right)^{\frac{2\nu}{1+\nu}} \Upsilon_{3+2\nu}\left(\frac{\eta_*}{\eta_{\text{end}}}\right)  \times \nonumber \\
        & \times \frac{z^2}{2\pi^2} \frac{(2+\nu)}{2\pi c_s^3}  \left(\frac{1-c_s^2}{c_s^2}\right)^2  \int_{x_-}^{x_+} dx \,  \frac{(x-x_+)^2(x-x_-)^2}{x^2y^2} \tilde{P}_v(x) \tilde{P}_v(y) ,\label{eq:P_phi_high}
    \end{align}
\end{subequations}
where we defined $x_\pm = z(1\pm c_s)/(2c_s)$, so that $y = x_+ +x_- -x$, and $\Upsilon_\ell$ is the lifetime suppression factor defined in equation~\eqref{upsilon}, that is 
\begin{equation}\label{Upsilon_lf}
    \Upsilon_\ell \left(\frac{\eta_*}{\eta_{\rm{end}}}\right) = \frac{1}{\ell(\nu)} \left[1- \left(\frac{\eta_*}{\eta_{\text{end}}}\right)^{\ell(\nu)}\right],
\end{equation}
with $\ell(\nu) = 1+2\nu$ for the leading-order term, and $\ell(\nu) = 3+2\nu$ for the NLO terms. 
The suppression factor of the leading-order term $\Upsilon_{1+2\nu}$ recovers the lifetime suppression factor defined in Ref.~\cite{Guo:2020grp} in the case studied therein, that is in the case of pure radiation ($\omega = 1/3$ and $\nu = 0$), also recently confirmed in Ref.~\cite{RoperPol:2023dzg}, and in the case of pressureless matter ($\omega = 0$ and $\nu = 1$). Our expression generalizes the notion of the lifetime suppression factor to generic equation of state parameter $\omega$. 

Considering expression~\eqref{eq:Pv} of the velocity spectral density, we can infer the power-law indices of the gravitational wave power spectrum. If the velocity spectral density scales as $P_v\sim k^n$, with $n = 2$ for $k<k_\star$ and $n = -4$ for $k>k_\star$, then 
\begin{equation}\label{eq:power_law}
    \begin{split}
        \mathcal{P}^{\text{lo, high}}_{\text{gw, sw}}(k) \sim k^{2n+5}\sim \left\lbrace\begin{array}{lcr}
        k^{9} & & k<k_\star, \\
        k^{-3} & & k>k_\star,
    \end{array} \right.\quad \;\mathcal{P}^{\text{nlo, high}}_{\text{gw, sw, gw, $\Phi$}}(k) \sim k^{2n+3}\sim \left\lbrace\begin{array}{lcr}
        k^{7} & & k<k_\star, \\
        k^{-5} & & k>k_\star,
    \end{array} \right. 
    \end{split}
\end{equation}
where $k_\star\simeq 2c_s k_p$ denotes the wavenumber corresponding to the peak amplitude of the spectrum~\cite{Giombi:2024kju}.

\section{Numerical analysis of the gravitational wave power spectrum at NLO}\label{sec:results}

With the choice of the spectral density~\eqref{eq:Pv}, we can now carry out the numerical integration of the gravitational wave power spectrum~\eqref{eq:Pgw} using the \texttt{integrate.quad} routine of \texttt{SciPy}~\cite{2020SciPy-NMeth, behnel2011cython}. More details on the strategy of numerical integration can be found in Appendix~\ref{app:num} and in Appendix of Ref.~\cite{Giombi:2024kju}.

Following the discussion on the gravitational wave Green's function of Section~\ref{sec:the_gravitational_wave_power_spectrum}, we divide the support of the gravitational wave power spectrum in two regions: the region $k\eta_* < 1$, where we use the super-horizon ($k\eta_* < k\eta_{\rm{end}} \ll 1 \ll k\eta$) approximation of the kernel functions~\eqref{eq:kernel_super}; and region $k\eta_* \geq 1$, where we use the sub-horizon regime ($k\eta_{\rm{end}} > k\eta_* \gg 1$) approximations~\eqref{eq:kernel_sub}. 
This way, having performed analytically the two time-integrations over $\uptau_1$ and $\uptau_2$, we perform the numerical integration of the gravitational wave power spectrum~\eqref{eq:Pgw} over the two sound wave dimensionless wavenumbers $x$ and $y$.
For the purpose of this paper, we will refer to these integrations in two dimensions as \textit{numerical integrations}. These can be further approximated by our semi-analytical approximations~\eqref{eq:P_gw_low}~\eqref{eqs:pgw_int} and~\eqref{eq:P_gw_high}. Only in the case of pure radiation, where $\nu = 0$, the integration of the kernel function can be performed analytically across the entire spectrum, as long as $k\eta \gg 1$, yielding the results in equations~\eqref{kernel_rad}.

Due to the artificial splitting of the support of the gravitational wave power spectrum, our numerical integrations in two dimensions are inaccurate in the region $k\eta_* \sim 1$, and in general the integrations in the sub-horizon and super-horizon regimes do not match at $k\eta_* = 1$. To understand how the two regimes smoothly join at $k\eta_* \sim 1$ one needs to perform the complete integration of the gravitational wave power spectrum~\eqref{eq:Pgw} in four dimensions, using equation~\eqref{GreenGreen} for the product of time derivatives of gravitational wave Green's functions. We leave the analysis of these fully-numerical integrations to future work. We remark that this discontinuity is not physical, but it arises due to the different limiting procedures with which we computed the gravitational wave Green's function in the sub- and super-horizon regimes. 

In this section we present an analysis of the gravitational wave power spectrum~\eqref{eq:Pgw} scanning through three different parameters: (i) the characteristic length of sound waves compared to the Hubble radius $R_*\mathcal{H}_*$; (ii) the speed of sound $c_s$; and (iii) the  duration of the acoustic phase $\detav \equiv \eta_\text{end} - \eta_*$. We parametrise the duration of the acoustic phase in numbers $N_{\text{sh}}$ of shock formation times $\eta_{\text{sh}}$
\begin{equation}
    \detav = N_{\text{sh}}\eta_{\text{sh}}.
\end{equation}
In analogy with Ref.~\cite{Giombi:2024kju}, we choose to relate $\eta_{\text{sh}}$ to the integral scale $\xi_*$ as
\begin{equation}
    \eta_{\text{sh}} \equiv \frac{\xi_*}{v_{\text{rms}}}, \qquad \xi_* \equiv \frac{1}{v_{\text{rms}}^2} \int\frac{d^3\bm{p}}{(2\pi)^3} p^{-1} P_v(p) = \frac{R_*}{4\pi\sqrt{3}} , 
\end{equation}
where we used $k_p R_* = 2\pi$.
Henceforth we further set the normalization $\eta_* = 1$. 
We assume that the equation of state parameter $\omega$ remains approximately constant throughout the duration of the source, so that $a_{\text{r}} \geq a_{\text{end}}$. Our analysis is therefore limited to two opposite cases:
\begin{enumerate}[label=\roman*)]
    \item The case where the softening of the equation of state lasts for a long period of time, at least longer than the duration of the acoustic source, that is
\begin{equation}
    \frac{a_*}{a_{\text{r}}} \leq \frac{a_*}{a_{\text{end}}} = \left(\frac{\eta_*}{\eta_{\text{end}}}\right)^{1+\nu}.
\end{equation}
For simplicity we choose $a_*/a_{\rm{r}} = 0.1$ in Figure~\ref{fig:convergence}, where we study the gravitational wave power spectrum for different source durations, and $a_{\rm{r}} = a(\eta_{\rm{end}})$ otherwise. To see more deeply the effects of the background energy dilution carried by the scale factor ratio $a_*/{a_\text{r}}$ we refer to  Ref.~\cite{Giombi:2024kju}.\\

\item The case of immediate return to radiation, where $\eta_{\text{r}} \rightarrow \eta_*$. This limit amounts to consider $c_s = 1/\sqrt{3}$ throughout the acoustic phase.
\end{enumerate}


\subsection{Dependence on the time duration of the source and power-law scaling}\label{sec:time}

\begin{figure}
    \centering
    \includegraphics[width=1.0\textwidth]{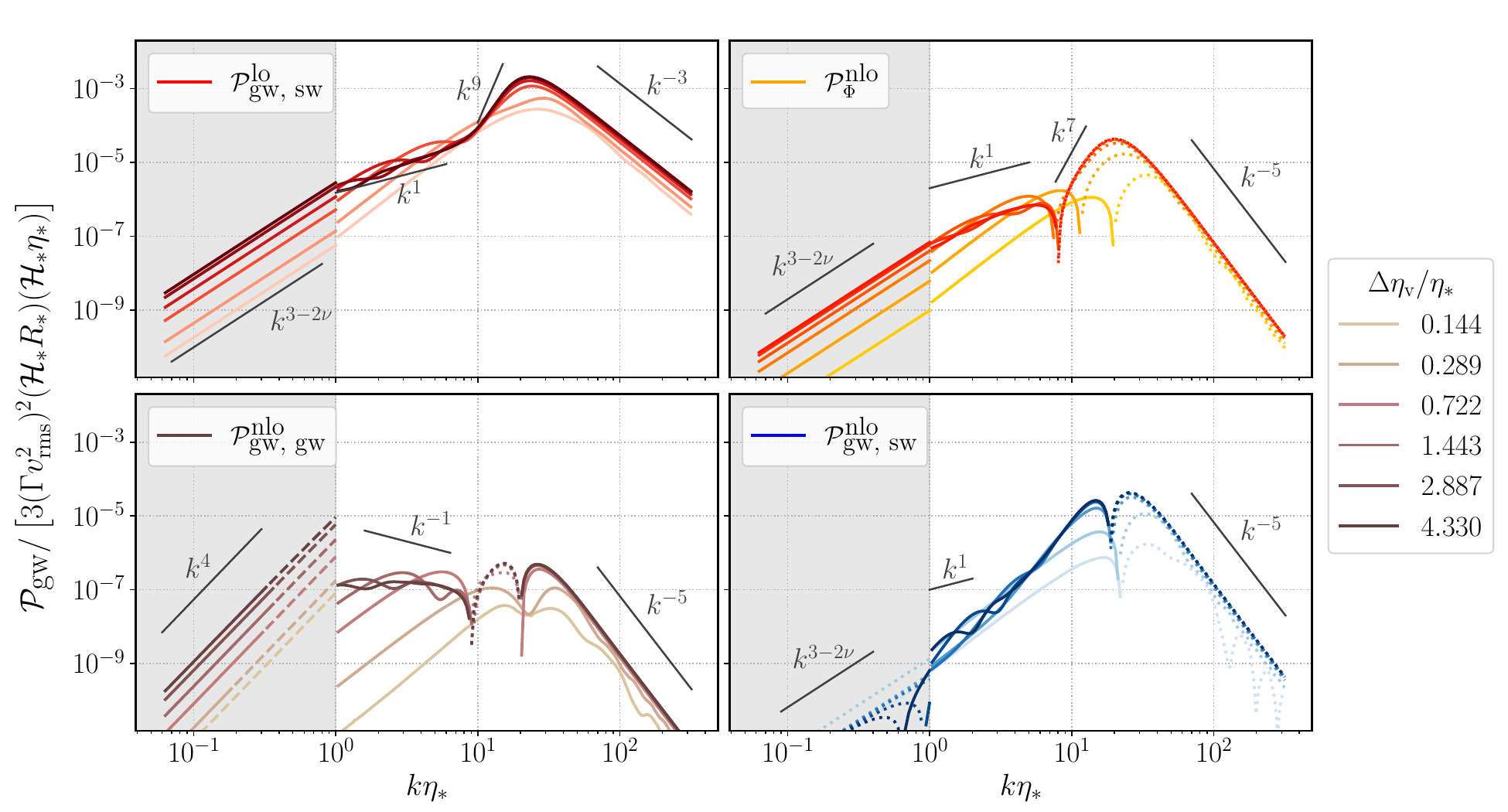}
    \caption{Power spectrum of gravitational waves from sound waves for different duration of the acoustic phase $\detav/\eta_*$. The leading-order contribution from sound waves and the NLO contribution from curvature perturbations are displayed on top-left and top-right panels respectively. Contributions from the  dynamics of gravitational waves and sound waves are shown in the bottom-left and bottom-right panels.  Solid line are used for positive and dotted lines for negative values. Dashed lines are used in the bottom left panel in the low-frequency region when the contribution $\mathcal{P}_{\text{gw,gw}}^{\text{nlo}}$ in the super-horizon regime overcome the value in the sub-horizon regime, that is for every $k$ such that  $\mathcal{P}_{\text{gw,gw}}^{\text{nlo, low}}(k\eta_*) > \mathcal{P}_{\text{gw,gw}}^{\text{nlo, int}}(k\eta_* =1)$. This is a region where we do not trust the low-frequency result, due to the fact that the regime $k\eta_*\sim k\eta_{\text{end}} \sim 1$ breaks the perturbative expansion of the gravitational wave Green's function~\eqref{eq:gg_low_1}. The equation of state parameter in the broken phase is $\omega  = 0.2$, and the time of return to radiation is set to $a_*/a_{\text{r}} = 0.1$. The gravitational wave peak frequency is set to $k_p\eta_* = 20$, which corresponds to a fractional bubble mean spacing compared to the Hubble radius $R_*\mathcal{H}_* = 0.393$.}
    \label{fig:convergence}
\end{figure}

Figure~\ref{fig:convergence} shows the contribution to the power spectrum brought by each individual term of the kernel~\eqref{eq:kernel_super}  and~\eqref{eq:kernel_sub} for different source duration when $\nu\neq 0$. The shaded gray area highlights the region of the spectrum that we computed in the sub-horizon limit, with kernel functions~\eqref{eq:kernel_super}. The contribution $\mathcal{P}^{\text{nlo}}_{\text{gw, gw}}$ converges to a limiting profile after approximately one Hubble time in the intermediate and high-frequency band, where $k\eta_* \gg 1$. In the low-frequency band, instead, the contribution  $\mathcal{P}^{\text{nlo, low}}_{\text{gw, gw}}$ seems to grow with the source duration without ever reaching a convergence. As briefly explained in Section~\ref{sec:low}, this behaviour is expected from the fact that the gravitational wave Green's function in the regime  $k\eta_* <k\eta_{\rm{end}} \ll1$ and $k\eta \gg1$ has been estimated with a polynomial by Taylor expanding around the small parameter $k\eta_* \ll 1$. In this estimation, the term in the Green's function~\eqref{eq:gg_low_1} grows linearly with the source duration and, likewise, so does the power spectrum contribution $\mathcal{P}^{\text{nlo, low}}_{\text{gw, gw}}$. This approximation however breaks down when $k\eta_{\text{end}} \sim 1$. We therefore expect the linear growth with the source duration to be lost for sufficiently long-lasting sources or for wavenumbers $k\eta_* \sim 1$. A deep understanding of the power spectrum in this regime requires the full four-dimensional integration of equation~\eqref{eq:pgw}, which we leave for future developments of this work. All the other contributions, $\mathcal{P}^{\text{lo}}_{\text{gw, sw}}$, $\mathcal{P}^{\text{nlo}}_{\text{gw, sw}}$  and $\mathcal{P}^{\text{nlo}}_{\text{gw,}\Phi}$, seem to tend toward a unique limiting profile within approximately one Hubble time across all gravitational wave wavenumbers. 

Let us now comment on the power law scaling of the spectrum. Our numerical results indicate that the power law exponents that characterize the leading-order contribution to the gravitational wave power spectrum $\mathcal{P}_{\text{gw, sw}}^{\text{lo}}$ are preserved in the large-bubble regime. However, the equation of state describing the thermal plasma during the transition influences the power law scaling of large gravitational wave wavelength modes outside the correlation length of sound waves $kR_* \lesssim k\eta_* \ll 1$ and $k\eta\gg 1$. Indeed, these modes are not affected by the acoustic source, and, while they remain constant outside the causal horizon, they decay by a redshift factor $a(\eta_k)/a(\eta)$ from the moment $\eta_k \sim k^{-1}$ when they cross the horizon. The background evolution of the scale factor $a(\eta)$ is thereby imprinted in the gravitational wave power spectrum, which scales as    $\mathcal{P}_{\text{gw, sw}}^{\text{lo, low}} \propto k^{3-2\nu}$~\cite{Cai:2019cdl, Hook:2020phx, Domenech:2021ztg}. For the same reason, the same scaling profile is found for the next-to-leading-order terms $\mathcal{P}_{\text{gw, sw}}^{\text{nlo, low}}$ and $\mathcal{P}_{\text{gw,} \Phi}^{\text{nlo, low}}$. We remark however that on even larger scales, where $k\eta\ll1$, the causality of the fluid flow is expected to yield a universal scaling $\mathcal{P}_{\text{gw}}\sim k^3$ regardless of the equation of state. This limit has not been address in our analysis, and thus does not appear in our numerical results. In the same frequency regime, with $k\eta_{\rm{end}} \ll 1$ and $k\eta \gg 1$, the NLO contribution from the gravitational wave Green's function scales as $\mathcal{P}_{\text{gw, gw}}^{\text{nlo, low}} \propto k^4$. This steeper scaling behaviour arises from the NLO term in the series expansion of the Green's function~\eqref{eq:gg_low_1} for $k\eta_* \ll 1$ and $k\eta \gg 1$.

In the intermediate-frequency range $1\ll k\eta_*\ll k_p\eta_*$ the leading-order contribution develops the shallow profile $\mathcal{P}_{\text{gw, sw}}^{\text{lo}} \propto k^1$ discussed in Refs.~\cite{Sharma:2023mao, RoperPol:2023dzg, Giombi:2024kju}, where the Universe expansion is taken into account. In the same frequency range, the numerical integrations also indicate the scaling $\mathcal{P}_{\text{gw, gw}}^{\text{nlo}} \propto k^{-1}$ and $\mathcal{P}_{\text{gw,}\,\Phi}^{\text{nlo}} \propto  k^{1}$ of the next-to leading-order terms, which we estimated analytically in Section~\ref{subsec:intermediate}.

In the high-frequency region, at gravitational wave modes around the peak $k\sim k_p$ and beyond $k\gtrsim k_p$, the leading-order sound wave contribution approximately follows the power scaling of the sound shell model, with a steep $\mathcal{P}_{\text{gw, sw}}^{\text{lo}} \propto k^{9}$ growth on the left side of the peak, and a $\mathcal{P}_{\text{gw, gw}}^{\text{nlo}} \propto k^{-3}$ decay on the right side of the peak. For the next-to-leading-order terms instead, we estimated in Section~\ref{sec:high} a $\mathcal{P}_{\text{gw, sw}}^{\text{nlo}}, \mathcal{P}_{\text{gw, gw}}^{\text{nlo}}, \mathcal{P}_{\text{gw}, \, \Phi}^{\text{nlo}} \propto k^{7}$ growth at frequencies just below the peak, and a decay as $\mathcal{P}_{\text{gw, sw}}^{\text{nlo}}, \mathcal{P}_{\text{gw, gw}}^{\text{nlo}}, \mathcal{P}_{\text{gw}, \, \Phi}^{\text{nlo}} \propto k^{-5}$ at frequencies just above the peak. While we observe the correct decay scaling $k^{-5}$ in all the terms, we see the approximate $k^7$ growth only for the contribution of curvature perturbations $\mathcal{P}_{\text{gw}, \, \Phi}^{\text{nlo}}$. In the other two terms, whose expressions were computed in equations~\eqref{eq:p_sw} and~\eqref{eq:p_gw}, the naive power counting that led to the estimate~\eqref{eq:power_law} is spoiled by the modulator factors in the curly brackets. In Appendix~\ref{sec:appendix_high} we show that these factors arise from the particular interference structure of the kernel terms $\Delta_{\text{sw}}^{\text{nlo}}$ and $\Delta_{\text{gw}}^{\text{nlo}}$ in equation~\eqref{eqs:kernel_high_0}. Notably, these are zero when the sound wave leading-order term $\Delta_{\text{sw}}^{\text{lo}}$ and the curvature perturbation term $\Delta_{\Phi}^{\text{nlo}}$ have a maximum. As a consequence, $\mathcal{P}_{\text{gw, sw}}^{\text{nlo}}$ and $\mathcal{P}_{\text{gw, gw}}^{\text{nlo}}$ vanish in correspondence of the peak amplitude of $\mathcal{P}_{\text{gw, sw}}^{\text{lo}}$ and $\mathcal{P}_{\text{gw}, \, \Phi}^{\text{nlo}}$.

\begin{figure}
    \centering
    \includegraphics[width=1.0\textwidth]{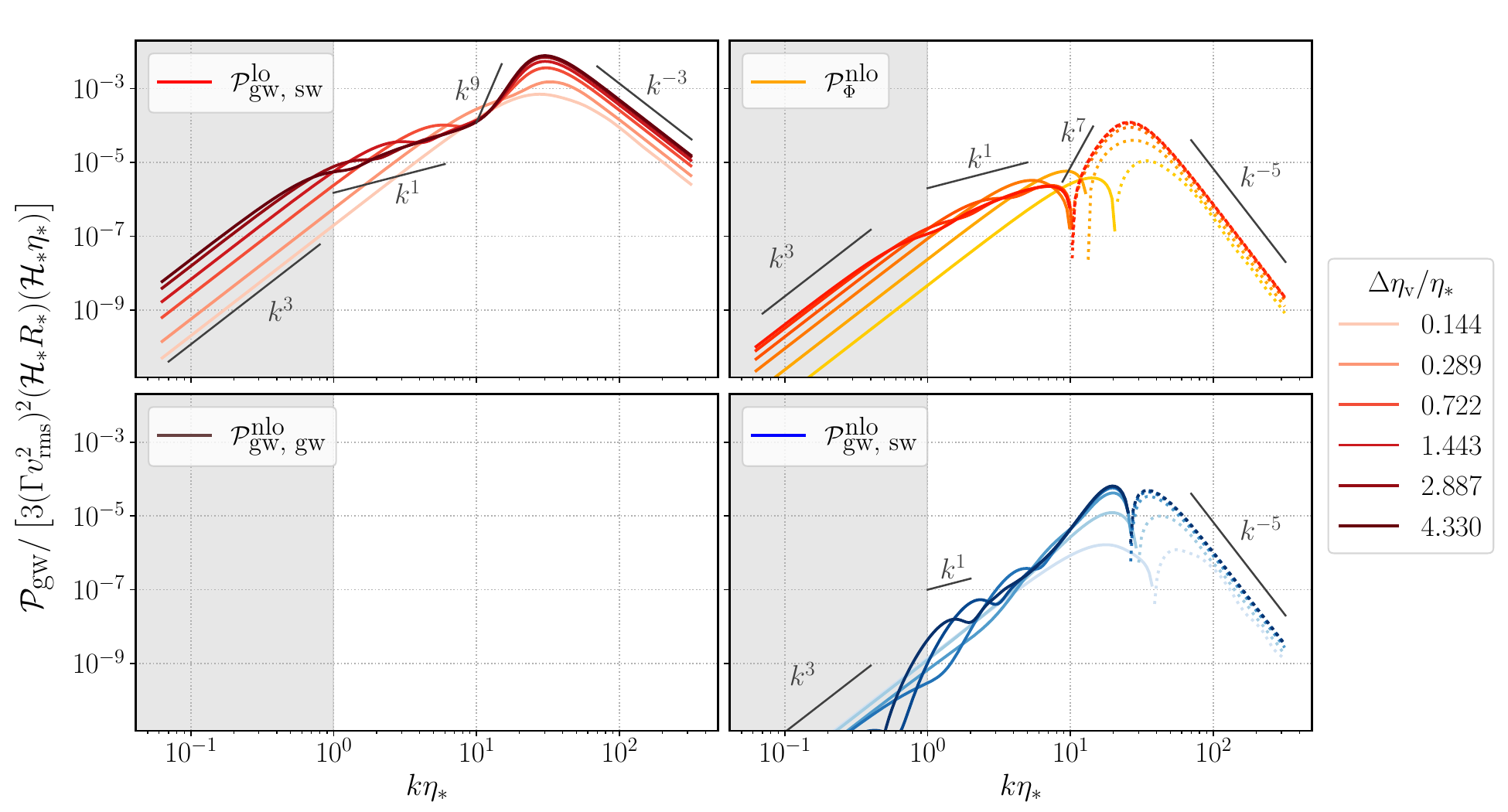}
    \caption{Power spectrum of gravitational waves from sound waves for different duration of the acoustic phase $\detav/\eta_*$ with a pure radiation equation of state $c_s = 1/\sqrt{3}$. The leading-order contribution from sound waves and the NLO contribution from curvature perturbations are displayed in the top-left and top-right panels respectively. The contributions from the  dynamics of gravitational waves and sound waves are shown in the  bottom-left and bottom-right panels. Solid line are used for positive and dotted lines for negative values. The gravitational wave peak frequency is set to $k_p\eta_* = 20$, corresponding to a fractional bubble mean spacing compared to the Hubble radius $R_*\mathcal{H}_* = 0.393$.}
    \label{fig:convergence_rad}
\end{figure}

\paragraph{The case of pure radiation:} In the case of immediate return to radiation, so that one can consider $c_s = 1/\sqrt{3}$ throughout the transition, the kernel functions simplify to the expressions~\eqref{kernel_rad}. The numerical strategy used in Ref.~\cite{Giombi:2024kju} cannot be applied to this case study, since the integer order of the generalized trigonometric integral functions leads to divergences in both the Gamma function and the Krummer confluent hypergeometric function (see equations~\eqref{eqs:gen_trig}). Therefore, for the purpose of the numerical integration, we find convenient to rewrite these functions as in equation~\eqref{eq:kernel_sub_rad} of Appendix~\ref{app:num}. The results of the numeric integration of the gravitational wave power spectrum~\eqref{eq:pgw} with kernel functions~\eqref{eq:kernel_sub_rad} are displayed in Figure~\ref{fig:convergence_rad}.

The effects due to the non-conformal Universe expansion are turned off when $\nu = 0$. We see indeed that the contribution from the NLO term of the gravitational wave Green's function $\mathcal{P}_{\text{gw, gw}}^{\text{nlo}}$ vanishes in this limit. The case of pure radiation allows us to turn off the non-conformal effects of background evolution of the Universe and isolate the effects of the fluid self-gravity.

\subsection{Numerical results and comparison with the analytic approximation}\label{sec:the_grav}
\begin{figure}
    \centering
    \includegraphics[width=1.0\textwidth]{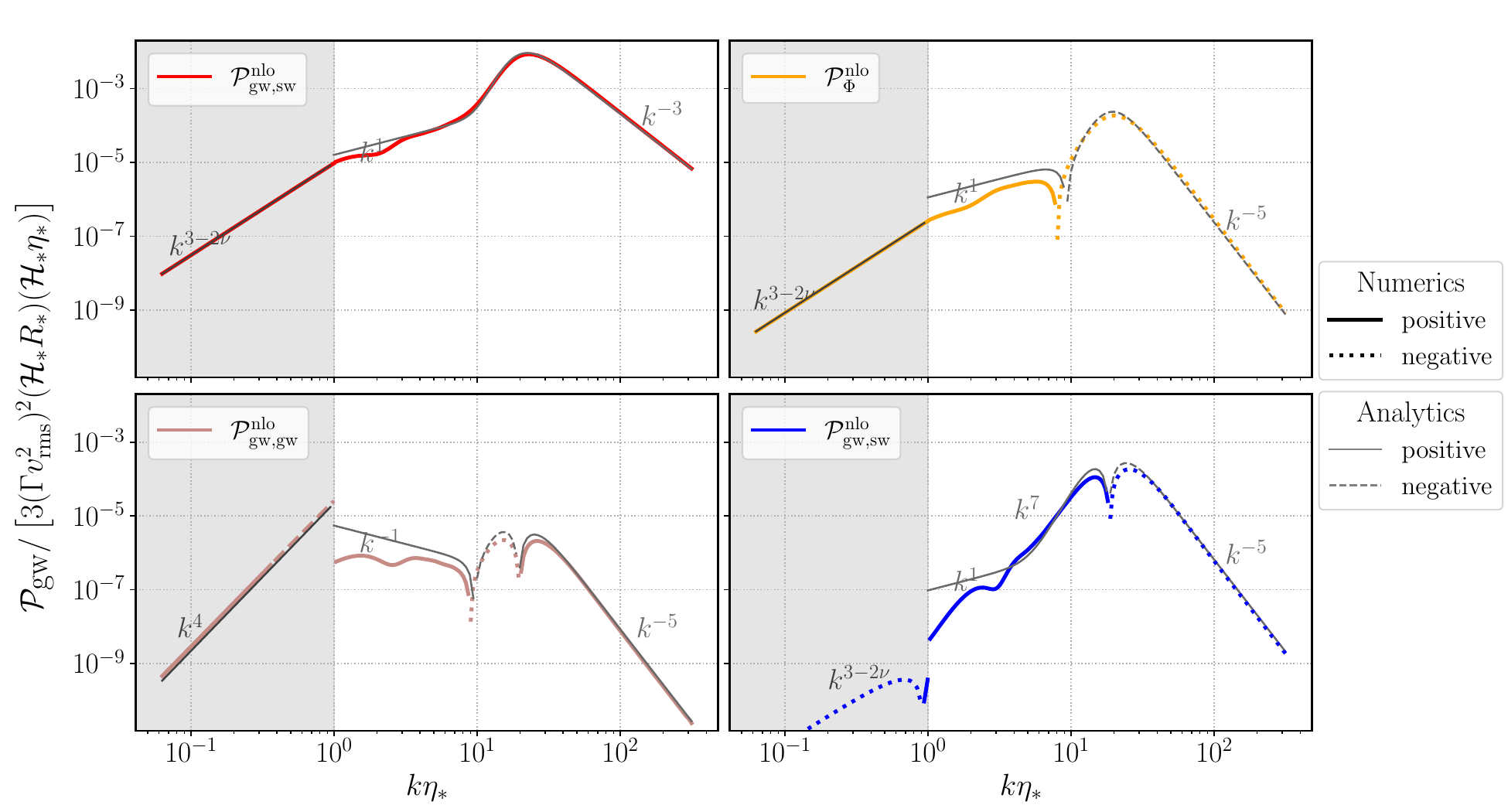}
    \caption{Power spectrum of gravitational waves from sound waves. Numerical results, obtained through the integration over the two sound wave wavenumbers in equation~\eqref{eq:pgw}, in coloured thick lines; semi-analytic approximations in thin gray lines. In the sub-horizon region $k\eta_*>1$, the semi-analytic approximations in the intermediate- and high-frequency regions are smoothly joined according to equation~\eqref{eq:join}. The leading-order contribution from sound waves and the NLO contribution from curvature perturbations are displayed in the top-left and top-right panels respectively. The contributions from the  dynamics of gravitational waves and sound waves are shown in the  bottom-left and bottom-right panels. Solid line are used for positive and dotted lines for negative values. Dashed lines are used in the bottom left panel in the low-frequency region when the contribution $\mathcal{P}_{\text{gw,gw}}^{\text{nlo}}$ in the super-horizon regime overcomes the value in the sub-horizon regime, that is for every $k$ such that  $\mathcal{P}_{\text{gw,gw}}^{\text{nlo, low}}(k\eta_*) > \mathcal{P}_{\text{gw,gw}}^{\text{nlo, int}}(k\eta_* =1)$. This is a region where we do not trust the low-frequency result, due to the fact that the regime $k\eta_{\text{end}}\sim 1$ breaks the perturbative expansion of the gravitational wave Green's function~\eqref{eq:gg_low_1}. The duration of the acoustic phase is set by $\eta_{\rm{end}} = N_{\rm{sh}}\eta_{\rm{sh}}$ and by choosing $N_{\rm{sh}}/v_{\rm{rms}} = 200$, which implies $\detav/\eta_* = 2.87$. We set the time of return to radiation to $a_{\text{r}} = a(\eta_{\rm{end}})$, the speed of sound to $c_s^2 = 0.2$, and the gravitational wave peak frequency to $k_p\eta_* = 20$, i.e. $\mathcal{H}_*R_* = 0.393$.} 
    \label{fig:approx}
\end{figure}

In Figure~\ref{fig:approx} we compare the numerical integrations with the semi-analytic approximations derived in Section~\ref{sec:analytic}.
The semi-analytic approximations describe quite accurately the shape of the gravitational wave power spectrum in different asymptotic regimes. We recall that the approximations in the high- and intermediate-frequency regimes are derived in the limit of fast oscillations, that is $k\detav \gg 1$.
We also remark that the discontinuities in the plot at $k\eta_* = 1$ are not physical discontinuities, but they arise from the fact that we are evaluating the gravitational wave power spectrum with a semi-analytical approach in two opposite regimes, $k\eta_* < k\eta_{\rm{end}} \ll 1$ in the shaded region with kernel functions~\eqref{eq:gg_low_1}, and $k\eta_{\rm{end}} >k\eta_* \gg 1$ in the blank region with kernel functions~\eqref{eq:kernel_sub}.   

In the large-bubble regime $R_*\mathcal{H}_* \lesssim \mathcal{O}(1)$, and thus $k_p\eta_* \gtrsim \mathcal{O}(10)$, the analytic estimation of the power spectrum in the intermediate-frequency range $1\ll k\eta_* \ll k_p \eta_*$ provides a good approximation only in a very narrow frequency window (see discussion in Section~\ref{subsec:intermediate}). 
Following the idea outlined in Ref.~\cite{Giombi:2024kju}, we can smoothly join the analytic approximations at high- and intermediate-frequency by inserting a complementary error function $\operatorname{erfc}(x)$ centred at $k_\star = 2c_s k_p$:
\begin{equation}\label{eq:join}
\begin{split}
    \mathcal{P}^{k\eta_*\gg 1}_{\text{gw}} \simeq &  3\left(\Gamma v_{\text{rms}}^2\right)^2 (\mathcal{H}_* R_*) (\mathcal{H}_* \eta_*) \left(\frac{a_*}{\bar{a}_{\text{r}}}\right)^{\frac{2\nu}{1+\nu}}  \frac{(kR_*)^3}{2\pi^2} \times  \qquad\qquad\qquad \\
    & \qquad\qquad\qquad\qquad\times\bigg\{\tilde{P}_{\text{gw}}^{\text{int}} (kR_*) \frac{1}{2}\operatorname{erfc}\left(2\pi \frac{(k-k_\star)}{k_\star}\right) + \tilde{P}_{\text{gw}}^{\text{high}} (kR_*)\bigg\}.
\end{split}
\end{equation}
We apply this recipe to all four contributions~\eqref{eq:kernel_sub}, plotted in gray lines in Figure~\ref{fig:approx} (white background region). 
This analytic method describes very accurately the transition from the shallow profile $\mathcal{P}^{\text{lo}}_{\text{gw, sw}} \propto k^1$ to the peak of the sound wave leading-order term, while it is less accurate in reproducing the other NLO contributions. Nevertheless, it is remarkably precise in locating and reproducing the zeros and the peaks of the contributions $\mathcal{P}^{\text{nlo}}_{\text{gw, gw}}$, $\mathcal{P}^{\text{nlo}}_{\text{gw, sw}}$ and $\mathcal{P}^{\text{nlo}}_{\text{gw,}\; \Phi}$ at $k\sim k_p$. 

The analytic approximation of the power spectrum around the peak works particularly well for the cases $\mathcal{P}^{\text{lo}}_{\text{gw, sw}}$ and $\mathcal{P}^{\text{nlo}}_{\text{gw, } \Phi}$, while it slightly overestimates the value of the other two contributions, $\mathcal{P}^{\text{nlo}}_{\text{gw, sw}}$ and $\mathcal{P}^{\text{nlo}}_{\text{gw, gw}}$. We remark however that our approximation, outlined in  Appendix~\ref{sec:appendix_high}, is able to reproduce the non-trivial structure of the latter terms, which at frequencies $k \gtrsim k_p$ present a single peak for $\mathcal{P}^{\text{lo}}_{\text{gw, sw}}$ and $\mathcal{P}^{\text{nlo}}_{\text{gw, } \Phi}$ and two different peaks with nearly opposite values for $\mathcal{P}^{\text{nlo}}_{\text{gw, sw}}$ and $\mathcal{P}^{\text{nlo}}_{\text{gw, gw}}$.  

The low-frequency tails $\mathcal{P}^{\text{lo, low}}_{\text{gw, sw}}, \mathcal{P}^{\text{nlo, low}}_{\text{gw, sw,}}, \mathcal{P}^{\text{nlo, low}}_{\text{gw, }\Phi}\propto k^{3-2\nu}$ and $\mathcal{P}^{\text{nlo, low}}_{\text{gw, gw}}\propto k^{4}$ manifest at gravitational wavelengths much larger than the typical correlation length of sound waves $kR_* \ll 1$. Our numerical results indicate that the low-frequency approximations, as estimated by the analytic expressions~\eqref{eq:P_gw_low}, provide a good approximation to the spectrum in the region $kR_* \lesssim \mathcal{O}(10^{-2})$. A smooth joining between the low-frequency and the intermediate-frequency profiles seems to be rather difficult to achieve with this analytical procedure due to the large separation of scale that divides the two regimes, the former valid for $k\eta_* \lesssim 0.01/(R_*\mathcal{H}_*)$, and the latter when $1\ll k\eta_*\ll k_p\eta_*$. Therefore, the larger the bubble-over-Hubble radius the larger the separation between the low- and high-frequency scales.  In our previous work (see Ref.~\cite{Giombi:2024kju}, Section 4.4) we were able to perform this smooth joining of the two profiles focusing on relatively smaller bubbles, whose size was set to $k_p\eta_* = 100$, that is $\mathcal{H}_*R_* \sim 0.1$.  In the large-bubble regime, one can reproduce correctly the transition between these two regimes only with an exact numerical integration. 

We finally remark that the transition between the low- and intermediate-frequency regimes is particularly tricky for the contribution $\mathcal{P}_{\text{gw, gw}}^{\text{nlo}}$. Indeed, this term takes into account the NLO correction to the Green's function of gravitational waves, where the expansion is performed in the super- or sub-horizon regimes, for which we have respectively $k\eta_* <k\eta_{\rm{end}} \ll 1 \ll k\eta$ and $k\eta_{\rm{end}} > k\eta_* \gg 1$. However, the NLO term in the Green's function expansion~\eqref{eq:gg_low_1} can grow larger than the leading-order term as $k\eta_*$ and $k\eta_{\rm{end}}$ approach unity from below, spoiling the perturbative expansion. For this reason, our results on $\mathcal{P}_{\text{gw, gw}}^{\text{nlo}}$ can only be trusted in the two opposite regimes $k\eta_* \ll 1$ and $k\eta_* \gg 1$. We leave a deeper analysis of the transition between these regimes for future work. 

\begin{figure}
    \centering
    \includegraphics[width=1.0\textwidth]{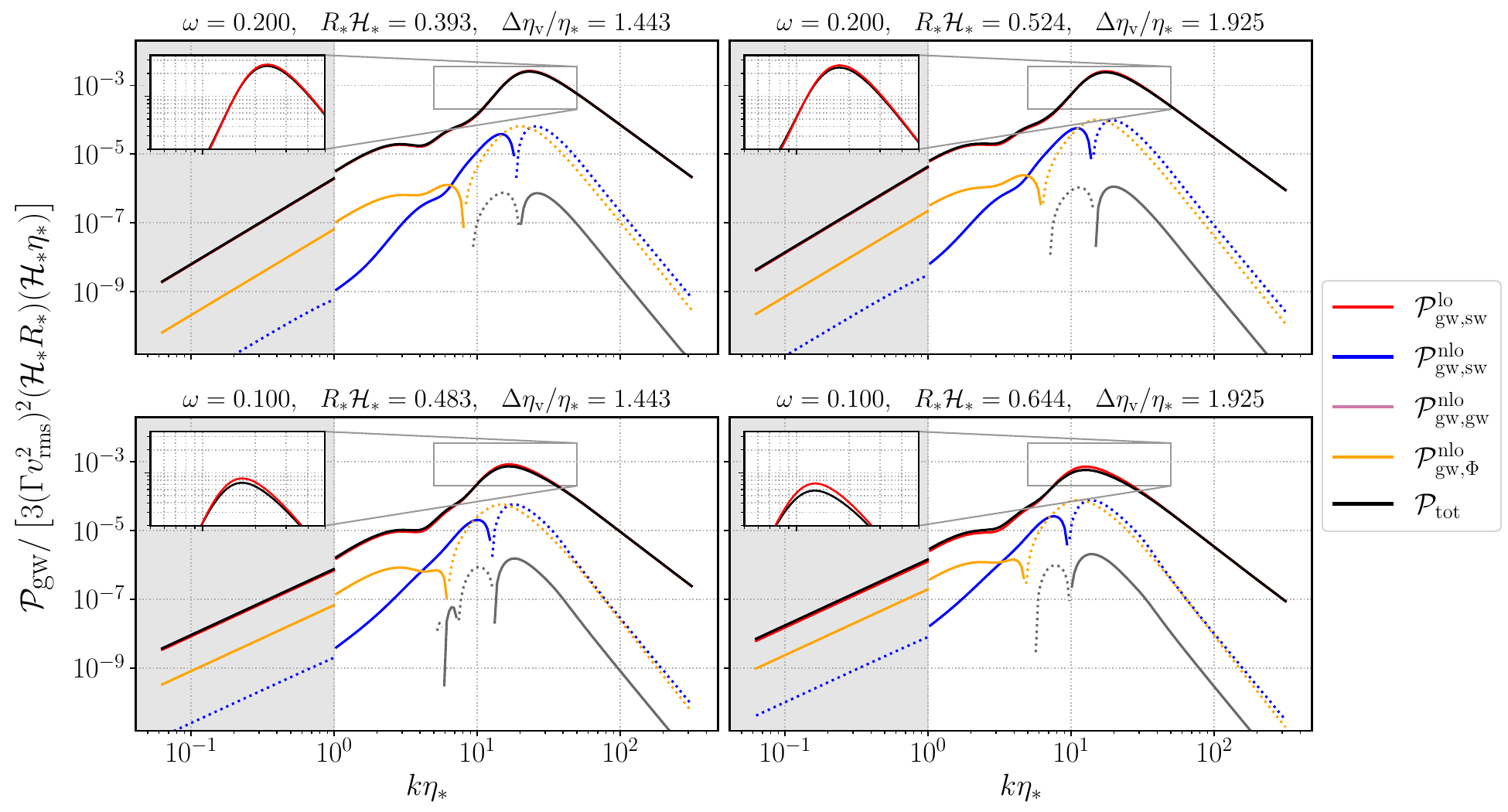}
    \caption{Power spectrum of gravitational waves from sound waves. Leading-order and NLO contributions are displayed in the same panels. Solid line are used for positive values, while dotted lines for negative values. The total power spectrum is obtained as the sum of each individual contribution, and is shown as a black line. The duration of the acoustic phase is set by $\eta_{\rm{end}} =  N_{\rm{sh}} \eta_{\rm{sh}}$ and by choosing $N_{\text{sh}}/v_{\rm{rms}} = 100$. The time of return to radiation is set to $a_{\text{r}} = a(\eta_{\text{end}})$. The peak wavenumber of the velocity field is set to $k_p\eta_* = 20$ in the upper panels, and $k_p\eta_* = 15$ in the bottom panels. The insets zoom in on the region around the peak of the gravitational wave power spectrum.}
    \label{fig:total}
\end{figure}

\paragraph{The total gravitational wave power spectrum:}
in Figure~\ref{fig:total} we compare the amplitude of each individual contributions and study their combination in the total gravitational wave power spectrum. In light of the discussion above, at the end of Section~\ref{sec:the_grav}, we cut the contribution from $\mathcal{P}^{\text{nlo}}_{\text{gw, gw}}$ to the region $k\eta_* \gtrsim \mathcal{O}(1)$. Due to the steep power scaling $\mathcal{P}^{\text{nlo, low}}_{\text{gw, gw}} \propto k^4$, this term brings a marginal contribution to the gravitational wave power spectrum at very small frequency $k\eta_* \ll 1$. We then neglect  $\mathcal{P}^{\text{nlo}}_{\text{gw, gw}}$ at small frequencies, leaving a more complete understanding of this contribution for future numerical analysis.

From Figure~\ref{fig:total} we first observe that the NLO terms slightly amplify the total power spectrum in the frequency regions $k < k_p$, i.e. on the left side of the peak. The most significant signature of the NLO contributions to the total gravitational wave power spectrum is found precisely at the peak. In particular, recalling the analytic estimates for the gravitational wave power spectrum at high frequency displayed in Section~\ref{sec:high}, we infer that the combination of the NLO terms suppresses the total peak amplitude by a factor
\begin{equation}\label{eq:A}
    \mathcal{P}_{\text{gw, tot}}(k_\star R_*) \sim \mathcal{P}^{\text{lo}}_{\text{gw, sw}}(k_\star R_*)  \left[1+\mathcal{A}\left(c_s, \frac{\detav}{\eta_*}, k_\star R_* \right)(R_*\mathcal{H}_*)^{2}\right], 
\end{equation}
where $k_\star$ denotes the location of the peak amplitude of the leading-order term $\mathcal{P}^{\text{lo}}_{\text{gw, sw}}$. From the expressions~\eqref{eq:kernel_sub} of the kernel contributions, we can state that the function $\mathcal{A}$ depends only on the speed of sound $c_s$, on the duration of the acoustic phase $\detav/\eta_*$, and on the sound wave correlation length $R_*$ through $z = kR_*$ (see for example the approximate expressions~\eqref{eq:P_gw_high}). However, since $k_\star$ is approximately set by the peak frequency $k_p$ as $k_\star \sim 2c_s k_p$, and since the peak frequency scale $k_p$ is related to the bubbles' mean separation by $k_p R_* = 2\pi$, we expect to find a weak dependence of $\mathcal{A}$ on the sound wave correlation length $R_*$. We remark that, given our model of the source in equation~\eqref{eq:Pv}, the coefficient $\mathcal{A}$ does not depend on the root mean squared fluid velocity $v_{\text{rms}}$. Indeed, in this model, $v_{\text{rms}}$ is just a constant multiplying factor that sets the amplitude of the spectral density $P_v$. Therefore, this factor cancels when taking the ratio of two gravitational wave power spectrum contributions, as implied by the definition of $\mathcal{A}$ in equation~\eqref{eq:A}.  
\begin{figure}[t]
    \centering
    \includegraphics[width=1.0\textwidth]{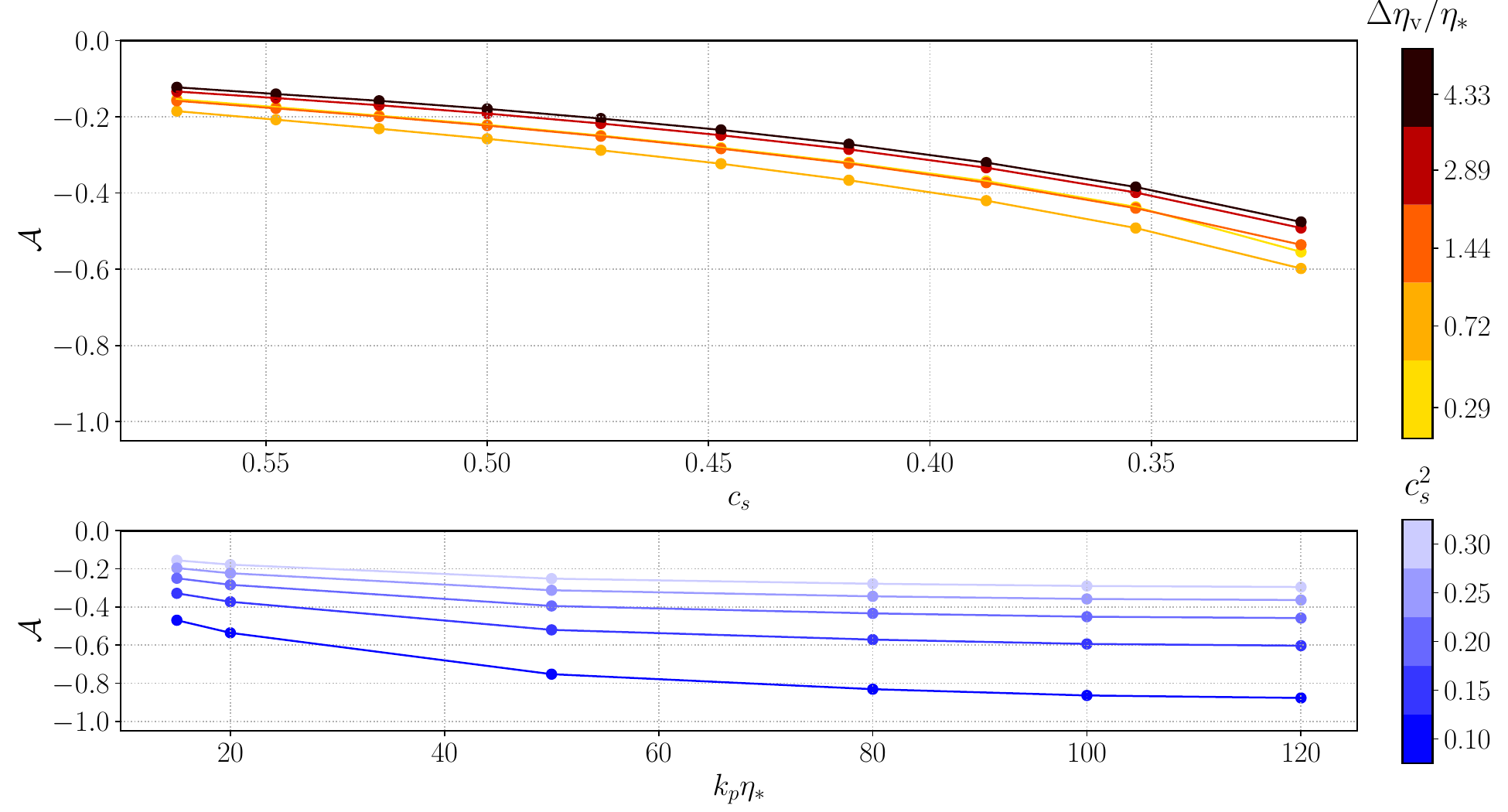}
    \caption{Suppression coefficient of the power spectrum of gravitational waves from the NLO contributions. Data obtained from the numerical integration of the gravitational wave power spectrum~\eqref{eq:Pgw} with kernel~\eqref{eq:kernel_sub}. In the upper panel the coefficient $\mathcal{A}$ is plotted for different source duration $\Delta\eta/\eta_*$ and speed of sound $c_s$, keeping the gravitational wave peak frequency fixed at $k_p\eta_* = 20$, i.e. $\mathcal{H}_*R_* \simeq (1+3c_s^2)^{-1}0.628$. In the lower panel the coefficient $\mathcal{A}$ is instead plotted for different speed of sound $c_s$ and gravitational wave peak frequency $k_p$, with the source duration fixed at $N_{\text{sh}} = 400$, that corresponds to $\detav/\eta_* \simeq 400/(2\sqrt{3}k_p\eta_*)$. }
    \label{fig:A_num}
\end{figure}

\begin{figure}[t]
    \centering
    \includegraphics[width=1.0\textwidth]{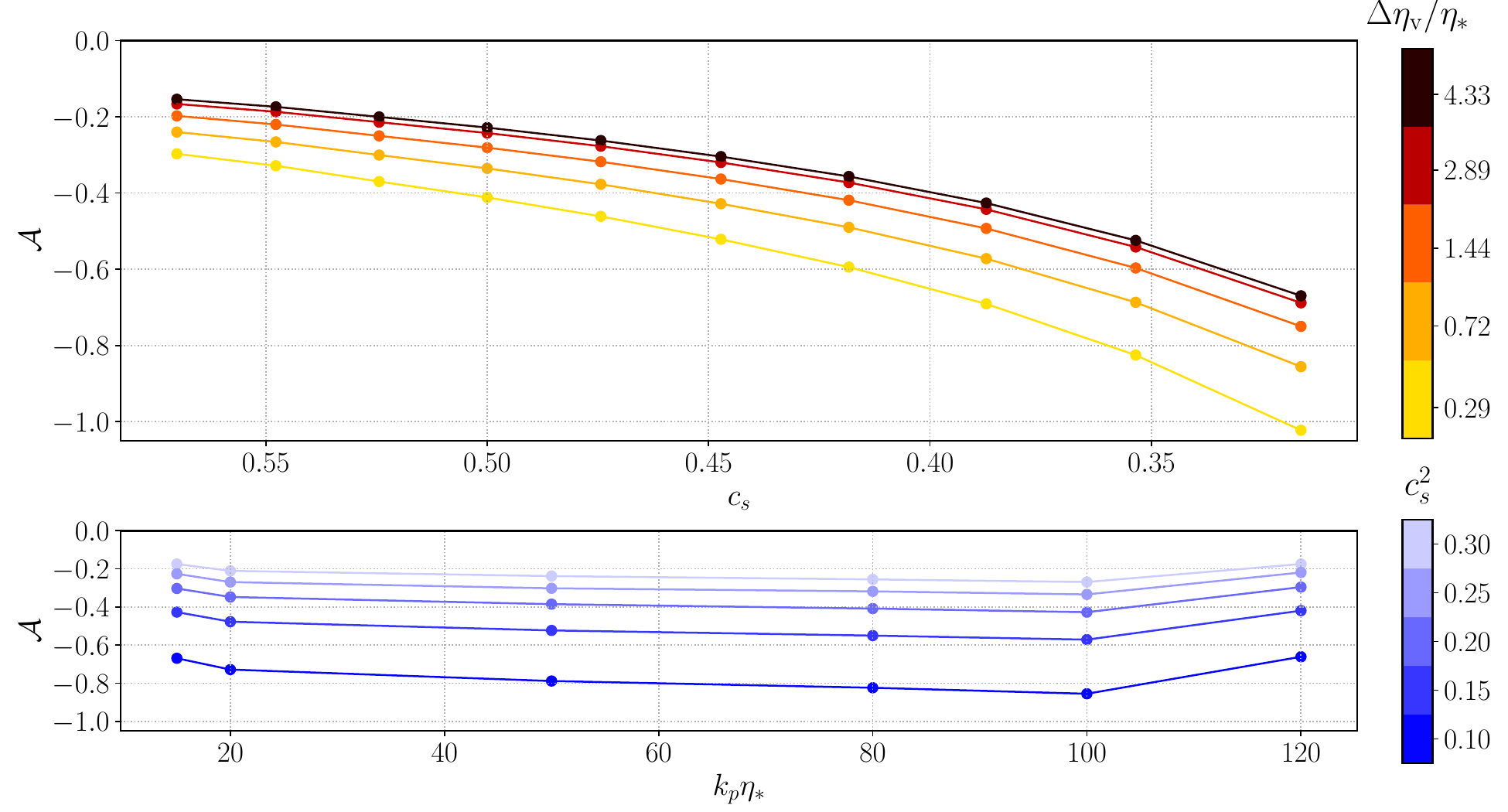}
    \caption{Suppression coefficient of the power spectrum of gravitational waves from the NLO contributions. Data obtained from the analytical approximations~\eqref{eq:P_gw_high} to the gravitational wave power spectrum. In the upper panel the coefficient $\mathcal{A}$ is plotted for different source duration $\Delta\eta/\eta_*$ and speed of sound $c_s$, keeping the gravitational wave peak frequency fixed at $k_p\eta_* = 20$, i.e. $\mathcal{H}_*R_* \simeq (1+3c_s^2)^{-1}0.628$. In the lower panel the coefficient $\mathcal{A}$ is instead plotted for different speed of sound $c_s$ and gravitational wave peak frequency $k_p$, with the source duration fixed at $N_{\text{sh}} = 400$, that corresponds to $\detav/\eta_* \simeq 400/(2\sqrt{3}k_p\eta_*)$. }
    \label{fig:A_approx}
\end{figure}

In Figure~\ref{fig:A_num} and~\ref{fig:A_approx} we display the value of $\mathcal{A}$ for different values of the speed of sound and source duration. Each data point in the plots is computed by finding the wavenumber $k_\star$ that corresponds to the location of the global maximum of $\mathcal{P}^{\text{lo}}_{\text{gw, sw}}$. This is obtained with the Nelder-Mead simplex algorithm encoded in the \texttt{scipy.optimize.fmin} function~\cite{2020SciPy-NMeth}, and then used to 
evaluate the different contributions to the gravitational wave power spectrum.  In Figure~\ref{fig:A_num} the profiles of the gravitational wave power spectrum that are used to compute $\mathcal{A}$ are obtained with the numerical integration of the kernel functions~\eqref{eq:kernel_sub} inside~\eqref{eq:pgw}. In Figure~\ref{fig:A_approx} instead, we use the semi-analytical approximations at high frequency from equations~\eqref{eq:P_gw_high}.

Figures~\ref{fig:A_num} and~\ref{fig:A_approx} show that $\mathcal{A}$ is always negative for every value of the parameters $c_s$,  $\detav/\eta_*$ and $k_p\eta_*$;  general relativistic corrections at the NLO always bring a suppression of the spectral peak amplitude.
The upper panels show that $\mathcal{A}$ decreases with the softening of the equation of state, that is reducing the speed of sound $c_s$. The bottom panels indicate that the coefficient $\mathcal{A}$ depends weakly on the peak frequency $k_p\eta_*$, as anticipated before. Equation~\eqref{eq:A} then implies that the effect of the suppression brought by NLO corrections grows quadratically with the bubble-over-Hubble radius parameter $R_*\mathcal{H}_*$. 

We also notice that, for sufficiently long-lasting sources $\detav/\eta_* \gtrsim \mathcal{O}(0.7)$ and for fixed $k_p\eta_*$ and $c_s$, the coefficient $\mathcal{A}$ is smaller for shorter source durations, and it increases with $\detav/\eta_*$ until converging to a limiting value. The dependence of $\mathcal{A}$ on the source duration reflects indeed that of the power spectrum investigated in Section~\ref{sec:time}. Since the NLO contributions in~\eqref{eq:P_gw_high}, which grow proportionally to $\Upsilon_{3+2\nu}(\eta_*/ \eta_{\text{end}})$, converge faster than the leading-order term~\eqref{eq:P_sw_high}, which grows as $\Upsilon_{1+2\nu}(\eta_*/ \eta_{\text{end}})$, they reach their limiting profile earlier and bring the largest suppression for shorter durations of the source.

Comparing Figure~\ref{fig:A_num}, which uses data computed through the numerical integration of the kernel~\eqref{eq:kernel_sub}, with Figure~\ref{fig:A_approx}, where the analytical approximations~\eqref{eq:P_gw_high} are used instead, we observe that our analytical approximations provide a good estimate of $\mathcal{A}$ only for sufficiently long lasting sources, $\detav/\eta_* \gtrsim \mathcal{O}(0.7)$. Indeed we found the analytical approximations~\eqref{eq:P_gw_high} by estimating the kernel function with a Dirac delta centered at the resonance between gravitational waves and sound waves, which is only justified when the gravitational wave modes around the peak $k\sim k_p$ oscillate several times during the activity of the source, that is $k\detav \gg 1$ (see Appendix~\ref{sec:appendix_high} for a rigorous discussion).

We remark that the analytical approximations always slightly overestimate the suppression factor $\mathcal{A}$ compared to the numerical results. A sample of this comparison, alongside the percentage suppression of the peak amplitude of the gravitational wave power spectrum, is given in Table~\ref{table}.

\begin{table}[t]
\begin{center}
\begin{tabular}{c c c c c | c c | c c}
$\bm{k_p\eta_*}$ & $\bm{c_s}$ & $\bm{\Delta c_s^2 (\%)}$& $\bm{\mathcal{H}_*R_*}$ & $\bm{\detav/\eta_*}$ & \multicolumn{2}{c}{$\bm{\mathcal{A}}$} & \multicolumn{2}{c}{$\bm{\Delta\mathcal{P}_{\text{gw}}(\%)}$} \\
 &  & &  &  & analytic & numeric & analytic & numeric\\
\hline
20 & 0.570 & 2.5 & 0.318 & 0.722 & $ -0.240 $ & $ -0.185$ & $ -2.429 $ & $ -1.874 $ \\
20 & 0.524 & 17.5 & 0.344 & 0.722 & $ -0.300 $ & $-0.231$ & $ -3.560 $ & $ -2.741 $ \\
20 & 0.447 & 40.0 & 0.393 & 0.722 & $ -0.428 $ & $-0.323$ & $ -6.601 $ & $ -4.983 $\\
20 & 0.387 & 55.0 & 0.433 & 0.722 & $ -0.572 $ & $-0.420$ & $ -10.743 $ & $ -7.887 $ \\
20 & 0.316 & 70.0 & 0.483 & 0.722 & $ -0.856 $ & $-0.598$ & $ -19.987 $ & $ -13.970 $ \\
20 & 0.500 & 25.0 & 0.359 & 1.443 & $ -0.281 $ & $ -0.223 $ & $ -3.620 $ & $ -2.875 $ \\
20 & 0.500 & 25.0 & 0.359 & 2.887 & $ -0.243 $ & $ -0.191 $ & $ -3.127 $ & $ -2.467 $ \\
30 & 0.447 & 40.0 & 0.262 & 0.962 & $ -0.399 $ & $ -0.332 $ &  $ -2.735 $ &  $ -2.279 $ \\
15 & 0.548 & 10.0 & 0.441 & 1.925 & $ -0.204 $ & $ -0.155 $ & $ -3.973 $ & $-3.019 $\\
15 & 0.447 & 40.0 & 0.524 & 1.925 & $ -0.342 $ & $ -0.249 $ & $ -9.385 $ & $ -6.836 $ \\
10 & 0.548 & 10.0 & 0.661 & 2.887 & $ -0.187 $ & $ -0.127 $ & $ -8.174 $ & $ -5.557 $ \\
10 & 0.447 & 40.0 & 0.785 & 2.887 & $ -0.320 $ & $ -0.205 $ & $ -19.727 $ & $ -12.660 $
\end{tabular}
\caption{Fractional suppression of the peak amplitude of the gravitational wave power spectrum for different values of the peak frequency $k_p$, sound speed $c_s$, and source duration $\detav/\eta_*$. The value of $v_{\text{rms}}$ does not affect the spectral peak suppression since, given the model of the source in equation~\eqref{eq:Pv}, this overall multiplying factor is canceled out when taking the ratio in the definition of $\mathcal{A}$ (\textit{cf.} Eq.~\eqref{eq:A}). The value of $\mathcal{A}$ is computed with a semi-analytic approach using the expressions~\eqref{eq:P_gw_high}, or by integrating the kernel functions~\eqref{eq:kernel_sub} inside the gravitational wave power spectrum~\eqref{eq:Pgw}.  The fractional suppression is computed as $\left[\mathcal{P}_{\text{gw, tot}} -\mathcal{P}^{\text{lo}}_{\text{gw, sw}}\right]/ \mathcal{P}^{\text{lo}}_{\text{gw, sw}}$ and displayed in percentage values. The quantity $\Delta c_s^2 = 1-3c_s^2$ measures the deviation from pure radiation.}\label{table}
\end{center}
\end{table}

\subsubsection{Comparison to LISA's sensitivity}\label{sec:comparison}
Finally, let us compare our predictions for some specific cases with the nominal sensitivity of LISA~\cite{Caprini:2019pxz}. 
Let us consider the gravitational wave power spectrum in the case $\omega = 0.2$, mean bubble size $R_*\mathcal{H}_* = 0.524$, source duration $\detav/\eta_* = 1.925$, and return to radiation time set to $a_{\rm{r}} = a(\eta_{\rm{end}})$. This case is shown in the top right panel of Figure~\ref{fig:total}. The amplitude of the spectrum is proportional to  $v_{\rm{rms}}^4$. Let us consider, for example $v_{\rm{rms}} = 0.08$, so that to satisfy the regime $v_{\rm{rms}} \ll R_*\mathcal{H}_*$. For the case at hand we find $\mathcal{P}_{\rm{gw, tot}} (k_\star R_*) \simeq 0.509 \times 10^{-6}$. The gravitational wave power spectrum is red-shifted to today and thus attenuated by a factor~\cite{Gowling:2021gcy, Caprini:2024hue}
\begin{equation}
    h^2 F_{\rm{gw, 0}} \simeq 1.64 \times 10^{-5} \left(\frac{100}{g_*} \right)^{1/3}, 
\end{equation}
where $g_*$ is the effective number of degrees of freedom at the end of the phase transition and $h = H_0/(100 \, \rm{km/s/Mpc})$, with $H_0$ the current Hubble rate. For a phase transition at the electroweak scale we now consider the reference value $g_* \approx100$~\cite{Kolb:1990vq}. 

This way, we estimate the spectral peak amplitude today of the gravitational wave power spectrum for the case at hand as
\begin{equation}\label{signal}
    h^2 \mathcal{P}^\star_{\rm{gw, tot, 0}} = h^2 F_{\rm{gw, 0}} \mathcal{P}_{\rm{gw, tot}} (k_\star R_*) \simeq 0.835\times 10^{-11} \left(\frac{100}{g_*} \right)^{1/3}.
\end{equation}
For reference, the reader can compare this number with the sensitivity curves of LISA, drawn for example in Ref.~\cite{Caprini:2024hue} Figure 1. LISA is maximally sensitive to gravitational wave where its amplitude spectral density (or for simplicity the sensitivity curve) has a minimum, which is approximately at $h^2\mathcal{P}_{\text{LISA}} \simeq 0.965 \times 10^{-11}$. This is however not a ``\textit{hard wall}'' discriminating observables signals, since the stochastic gravitational wave background is not a resolvable instantaneous signal, but instead a continuous signal over time. Therefore, even if the instantaneous strain of the signal is slightly below the sensitivity curve of LISA, by integrating over many waveform cycles the signal-to-noise ratio can grow above detectability.
A signal with the intensity of equation~\eqref{signal} could then be detected by LISA if the spectral peak amplitude is located near the frequency region where LISA is maximally sensitive. The position of the peak in the frequency space is determined by many parameters, such as the mean bubble separation $R_*$, the speed of sound $c_s$, and especially the transition temperature (see for example Ref.~\cite{Gowling:2021gcy} Figure 1.d).  

For the case considered by this example, we computed the relative suppression of the spectral peak amplitude $\Delta\mathcal{P}_{\rm{gw}} \simeq -6.836 \, \% $ (see Table~\ref{table}, line 10). Such difference is therefore potentially detectable by LISA.

\section{Conclusion}
In this work we presented a semi-analytical calculation of the power spectrum of gravitational waves from sound waves in a first-order phase transition. We focused on scenarios with large bubbles, meaning $R_*\mathcal{H}_* \lesssim \mathcal{O}(1)$, as these are the loudest and the first to be constrained in any developing dataset. 

The profiles of the gravitational wave power spectrum~\eqref{eq:Pgw} are obtained with a hybrid method in different asymptotic limits. The time integrations in the kernel function~\eqref{eq:Delta} are evaluated analytically into two different regimes: a) super-horizon modes $k\eta_* \ll 1 \ll k\eta$; and b) sub-horizon modes $k\eta_* \gg1$, with $\eta_*$ the conformal times at the beginning of the acoustic production and $\eta$ the time at which the gravitational wave power spectrum is evaluated, after the total dissipation of the acoustic source. The remaining two integrations over the sound wave wavenumbers in equation~\eqref{eq:pgw} are performed numerically. We refer to these results as \textit{numerical integrations}.

We improved the precision of existing model~\cite{Hindmarsh:2016lnk, Hindmarsh:2019phv, RoperPol:2023dzg} by adding the contributions at the next-to-leading-order (NLO) in the short sound wave wavelength expansion parameter $R_*\mathcal{H}_*$, where the peak wavenumber is at $k_p \simeq 2\pi/R_*$. The leading-order contribution to the gravitational wave power spectrum, $\mathcal{P}^{\text{lo}}_{\text{sw}}$, is the energy injection from sound waves, and was already studied in the context of the sound shell model with a pure radiation equation of state in Refs.~\cite{Hindmarsh:2016lnk, Hindmarsh:2019phv, RoperPol:2023dzg}, or with a softer equation of state in Ref.~\cite{Giombi:2024kju}. 

In addition to the effects considered in Ref.~\cite{Giombi:2024kju}, in this work we extended the approximation of the gravitational wave Green's function to super-horizon modes $k\eta_* <k\eta_{\rm{end}}\ll 1 \ll k\eta$. We found that, for modes outside the sound wave correlation length $kR_* \ll 1$, the background evolution of the Universe modifies the power law scaling of the spectrum to $\mathcal{P}_{\text{gw}} \propto k^{3-2\nu}$, with $\nu = (1-3c_s^2)/(1+3c_s^2)$. Gravitational wave modes on very low frequencies $k\eta \ll 1$ are not treated in this work. In this frequency region, the gravitational wave power spectrum is expected to scale as $\mathcal{P}_{\rm{gw}}\sim k^3$ due to causality arguments~\cite{Durrer:2003ja}, and is therefore less interesting observationally.

The NLO terms include, besides corrections to the dynamics of both gravitational and sound waves, a novel contribution from curvature perturbations. The most significant signature of the NLO corrections is the suppression of the spectral peak amplitude of the total gravitational wave power spectrum, located approximately at $k_\star \sim 2c_s k_p$, with respect to the leading-order contribution from sound waves. The extent of the suppression depends primarily on the average size of the bubbles, and more marginally on the speed of sound $c_s$ and on the duration of the source $\detav/\eta_*$. In our results, we find that this  suppression is of order $\Delta \mathcal{P}_{\text{gw}} \sim \mathcal{O}(5 \%)$ when $R_*\mathcal{H}_* \sim \mathcal{O}(0.4)$, and it grows quadratically with $R_*\mathcal{H}_*$ according to equation~\eqref{eq:A}. Some reference values of the suppression brought by the NLO contributions is given in Table~\ref{table}. Leading-order and NLO contributions become comparable in magnitude when $R_*\mathcal{H}_* \sim \mathcal{O}(1)$, showing the symptoms of perturbation theory breaking down; non-perturbative analysis should be used in this regime.

The numerical integrations are supported by semi-analytic approximations, where the additional integrations over the sound wave wavenumbers are performed analytically (entirely or partially)  by splitting the support of the gravitational wave power spectrum into three different regimes: (i) low-frequency regime $k\eta_* < k\eta_{\rm{end}}  \ll 1 \ll k\eta$,  (ii) intermediate-frequency regime $1 \ll k \eta_*  \ll k_p\eta_*$, and (iii) high-frequency regime $(k\sim k_p) \cup (k>k_p)$. We refer to Section~\ref{sec:table} and Table~\ref{table_approx} for a detailed explanation of the different regimes. To obtain analytical approximations in both the intermediate- and high-frequency regimes we also assume that these modes oscillate several times within the duration of the source, that is $k(\eta_{\rm{end}} - \eta_*) \gg 1$. 

The semi-analytical results provide good approximations of the numerical integrations for long-lasting sources $\eta_{\rm{end}} - \eta_* > \eta_*$ over most of the frequency range. We provided in equation~\eqref{eq:join} a useful analytic recipe to smoothly join the semi-analytical results in the high-frequency regime $(k\sim k_p)\cup (k> k_p)$  to the intermediate-frequency regime $1\ll k\eta_*\ll k_p\eta_*$. 
This method does not include, for now, the low-frequency region of the spectrum $k\eta_*\ll1$.  In the large-bubble regime $R_*\mathcal{H}_*\lesssim\mathcal{O}(1)$, the significant separation between the sub-horizon region $k\eta_* \gg1$ and the frequency region $kR_* \ll 1$, where analytic profiles at small wavenumbers provide a good approximation, causes the standard matching procedure~\eqref{eq:join} to fail in connecting these spectral regions. In order to further confirm the validity of our results in the asymptotic limit considered in this work, it might be useful to perform a full four-dimensional numerical integration of the gravitational wave power spectrum~\eqref{eq:pgw}. We leave this check for the future.

Our results are limited to non-relativistic compressional modes with $v_{\rm{rms}} \ll R_*\mathcal{H}_*$. In addition, we assumed fluid perturbations to be linear, stochastic, and Gaussian, approximating the source of sound waves as stationary throughout the acoustic phase. In a realistic phase transition, however, we expect the energy of the source to decay in time due to the formation of shocks, until sound waves eventually decorrelate. In our framework, we approximated the non-linearities as a sharp cut-off of the source at a time proportional to the shock formation time $\eta_{\rm{sh}} \sim R_*/v_{\rm{rms}}$. This corresponds to a scenario where non-linearities remain negligible until $\eta_{\rm{end}}$, when they suddenly form and instantaneously dissipate all the energy of sound waves into shocks. Proper considerations of the dynamical effect of non-linearities were introduced, for the leading sound wave contribution $\mathcal{P}^{\text{lo}}_{\text{sw}}$ in Refs.~\cite{Dahl:2021wyk, Dahl:2024eup, Caprini:2024gyk},  leading to a time-dependent spectral density of fluid perturbation and time-dependent peak-frequency $k_p$. The same analysis can be applied, with new numerical methods, to the next-to-leading-order terms. We leave this study for future work.   

In order to pursue a semi-analytic calculation of the gravitational wave power spectrum, we approximated the source of shear stress with the analytic function in equation~\eqref{eq:Pv}. A more realistic profile for the source can be obtained numerically with the sound shell model from the linear superposition of the fluid perturbations carried by each individual bubble~\cite{Hindmarsh:2016lnk, Hindmarsh:2019phv}. This model has so far been applied only to transitions with small bubbles, $R_*\mathcal{H}_* \ll 1$, whose dynamics can be approximated on a static and flat background~\cite{Hindmarsh:2016lnk, Hindmarsh:2019phv, RoperPol:2023dzg, Espinosa:2010hh, Giese:2020znk}. 
The study on the effects of the Universe expansion and self-gravitation of the fluid, relevant for large bubbles $R_*\mathcal{H}_* \lesssim \mathcal{O}(1)$, only started recently with Refs.~\cite{Cai:2018teh, Giombi:2023jqq, Jinno:2024nwb}, but it has not been applied to the sound shell model yet. We leave this study for the future.

\section*{Acknowledgements}
LG would like to thank Chiara Caprini, Alberto Roper Pol, and Simona Procacci for valuable discussions that contributed to the understanding of the gravitational wave power spectrum in the super-horizon regime.
The work of LG has been funded by the Alfred Kordelin Foundation, grant number 240111. MH was supported by the Research Council of Finland grant numbers 333609 and 363676. JD was supported by the Research Council of Finland grants numbers 354572 and 353131, and European research council grant CoCoS, 101142449.

\appendix
\section{Green's function method}\label{app:Green_function}
The Green's function for a linear differential operator $\mathcal{L}(k,\eta)$ is obtained by combining 
two independent solutions $g_{1}(k \eta)$ and $g_{2}(k \eta)$ of the homogeneous equation $\mathcal{L}(k,\eta)g_{j}(k \eta) =0$. For an initial value problem, the Green's function is
\begin{equation}
    G_k(\eta, \tilde{\eta}) = \frac{1}{\mathcal{N}_k} \left[g_{1}(k \eta)g_{2}(k \tilde{\eta}) - g_{1}(k\eta)g_{2}(k\tilde{\eta})\right] \Theta(\eta - \tilde{\eta}),
\end{equation}
with $\mathcal{N}_k = g_{1}^\prime(k\tilde{\eta})g_{2}(k\tilde{\eta}) - g_{1}(k\tilde{\eta}) g_{2}^\prime(k\tilde{\eta})$ 
a normalization that only depends on the time $\tilde{\eta}$, and $\Theta(\eta -\tilde{\eta})$ the Heaviside step function. Time derivatives in $\mathcal{N}_k$ are computed with respect to $\tilde{\eta}$. One can show that two homogeneous solutions of 
the wave equation~\eqref{eq:ell} are $g_{1}(k\eta) = k\eta\,j_{\nu}(k\eta)$ and $g_{2}(k\eta) = k\eta\, y_{\nu}(k\eta)$, 
with $y_\nu$ and $j_\nu$ the spherical Bessel functions of first and second kind respectively of order $\nu = (1-3\omega)/(1+3\omega)$. Then
\begin{equation}
    \mathcal{N}_k = - k(k\tilde{\eta})^2 \mathcal{W}\big(j_\nu(k\tilde{\eta}), y_\nu (k\tilde{\eta})\big) = - k
\end{equation}
with $\mathcal{W}\big(j_\nu(\tilde{z}), y_\nu (\tilde{z})\big) \equiv \partial_z j_\nu(\tilde{z})y_\nu (\tilde{z}) - j_\nu (\tilde{z}) \partial_z y_\nu (\tilde{z}) = \tilde{z}^{-2}$ the Wronskian of spherical Bessel functions~\cite[\href{http://dlmf.nist.gov/10.50.E1}{(10.50.1)}]{NIST:DLMF}, and
\begin{equation}\label{eq:green}
    G_k(\eta, \tilde{\eta}) = -k   \eta\Tilde{\eta}\left[j_{\nu}(k\eta)y_{\nu}(k\tilde{\eta}) - j_{\nu}(k\tilde{\eta})y_{\nu} (k\eta) \right], \qquad \nu =  \frac{1-3\omega}{1+3\omega}
\end{equation}
In the expression of the gravitational wave power spectrum~\eqref{P_hp}, the time variable $\tilde{\eta}$ is integrated over the time duration of the transition, while $\eta$ is evaluated long after the end of the phase transition. We can then consider $\eta \gg \tilde{\eta}$ and $k\eta \gg 1$. In the limit of large argument $z\equiv k\eta$, the spherical Bessel functions approximate to~\cite{ARFKEN2013643}
\begin{subequations}\label{eq:Bessel_limit}
\begin{eqnarray}\label{jy}
    j_{\nu}(z) &\simeq& \frac{1}{z} \bigg\{P_{\nu+\frac{1}{2}}(z)\cos\left(z - \frac{\pi}{2} (\nu+1)\right)  - Q_{\nu+\frac{1}{2}}(z)\sin\left(z - \frac{\pi}{2} (\nu+1)\right) \bigg\},\\
    y_{\nu}(z) &\simeq& \frac{1}{z} \bigg\{P_{\nu+\frac{1}{2}}(z)\sin\left(z - \frac{\pi}{2} (\nu+1)\right)  + Q_{\nu+\frac{1}{2}}(z)\sin\left(z - \frac{\pi}{2} (\nu+1)\right)\bigg\}.
\end{eqnarray}
\end{subequations}
with
\begin{equation}\label{eq:PQ}
    P_{\nu+\frac{1}{2}}(z) = \sum_{m=0}^\infty (-1)^m \frac{c_{2m}(\nu+\frac{1}{2})}{(2z)^{2m}}, \qquad Q_{\nu+\frac{1}{2}}(z) = \sum_{m=0}^\infty (-1)^m \frac{c_{2m+1}(\nu+\frac{1}{2})}{(2z)^{2m+1}},
\end{equation}
and 
\begin{equation}
    c_m(\nu+\frac{1}{2}) = \frac{\Gamma\left(\nu + m +1\right)}{m!\, \Gamma\left(\nu - m +1\right)}.
\end{equation} 
Therefore, considering only leading-order contributions in $1/z$, we can write the Green's function as
\begin{equation}
    \begin{split}
        G_k(\eta, \tilde{\eta}) = &-\tilde{\eta} \bigg\{ \bigg[\cos\left(k\eta - \frac{\pi}{2}(1+\nu)\right) - \frac{\nu(1+\nu)}{2k\eta} \sin\left(k\eta - \frac{\pi}{2}(1+\nu)\right)\bigg] y_\nu (k\tilde{\eta}) - \\
        & \qquad - j_\nu (k\tilde{\eta})\bigg[\sin\left(k\eta - \frac{\pi}{2}(1+\nu)\right) + \frac{\nu(1+\nu)}{2k\eta} \cos\left(k\eta - \frac{\pi}{2}(1+\nu)\right)\bigg]  \bigg\}.
    \end{split}
\end{equation}
Since the power spectrum of gravitational waves depends on the time derivative of tensor modes~\eqref{P_hp}, we consider the derivative of the Green's function $G_k(\eta, \tilde{\eta})$ with respect to the evaluation time $\eta$, yielding
\begin{equation}
    \begin{split}
        G^\prime_k(\eta, \tilde{\eta}) = & k\tilde{\eta} \bigg\{ \bigg[\sin\left(k\eta - \frac{\pi}{2}(1+\nu)\right) + \frac{\nu(1+\nu)}{2k\eta} \cos\left(k\eta - \frac{\pi}{2}(1+\nu)\right)\bigg] y_\nu (k\tilde{\eta}) + \\
        & \qquad + j_\nu (k\tilde{\eta})\bigg[\cos\left(k\eta - \frac{\pi}{2}(1+\nu)\right) - \frac{\nu(1+\nu)}{2k\eta} \sin\left(k\eta - \frac{\pi}{2}(1+\nu)\right)\bigg]  \bigg\}. 
    \end{split}
\end{equation}
Since the trigonometric periodic functions oscillate with a period $2\pi/k$, for wavenumbers $k$ such that $k\eta\gg 1$ we can perform the average over a large number of oscillations. For the product of two Green's functions, we finally obtain
\begin{equation}\label{GpGp}
        G^\prime_k(\eta, \eta_1)G^\prime_k(\eta, \eta_2) \overset{k\eta\gg 1}{\longrightarrow}  \frac{k\eta_1 k\eta_2}{2} \Big[ y_\nu(k\eta_1) y_\nu(k\eta_2) + j_\nu(k\eta_1) j_\nu(k\eta_2) \Big].
\end{equation}

A special case of equation~\eqref{GpGp} that serves as a comparison to the previous literature~\cite{Sharma:2023mao, RoperPol:2023dzg, Giombi:2024kju} is the case of pure radiation equation of state, $\omega = 1/3$ and $\nu = 0$. The spherical Bessel functions of integer order can be then represented in terms of trigonometric functions~\cite{ARFKEN2013643}, leading to 
\begin{equation}\label{eq:GG_rad}
        G^\prime_k(\eta, \eta_1)G^\prime_k(\eta, \eta_2)\bigg\vert_{\nu= 0} \overset{k\eta\gg 1}{\longrightarrow}  \frac{1}{2}  \cos\left[k(\eta_1-\eta_2) \right].
\end{equation}

In the following we will consider the general expression~\eqref{GpGp} in two different regimes: (i) the super-horizon regime $k\eta_*<k\eta_{\rm{end}} \ll 1 \ll k\eta$, where generally gravitational waves modes do not have time to complete one oscillation within the duration of the source, i.e. $k(\eta_{\rm{end}} - \eta_*) \ll1$; and (ii) the sub-horizon regime $k\eta_{\rm{end}} > k\eta_* \gg 1$.

\subsection{Green's function of gravitational waves on sub-horizon scales}
We first study the sub-horizon regime $k\eta_{\rm{end}} > k\eta_* \gg 1$. Since $\eta_1, \eta_2 \in [\eta_*, \eta_{\rm{end}}]$, we can take $k\eta_1, k\eta_2 > k\eta_* \gg 1$, and use again the expansion of the spherical Bessel functions for large arguments~\eqref{jy}, resulting in, at next-to-leading-order  in $1/k\eta_*$,
\begin{eqnarray}\label{eq:gpgp_result}
    G_k^\prime(\eta, \eta_1)G_k^\prime(\eta, \eta_2) &\overset{k\eta_*\gg1 }{=}& \frac{1}{2}\bigg[ \cos\left(k\eta_-\right) - \frac{\nu(1+\nu)}{2}\sin\left(k\eta_-\right) \left(\frac{1}{k\eta_1} -\frac{1}{k\eta_2}\right) \bigg],
\end{eqnarray}
with $\eta_- = \eta_1 - \eta_2$.

\subsection{Green's function of gravitational waves on super-horizon scales}
In the opposite regime, we consider gravitational wave modes such that $k\eta_1 , k\eta_2 < k\eta_{\rm{end}} \ll 1 \ll k\eta$. These are gravitational wave modes that do not have time to oscillate within the duration of the source, but still have enough time to oscillate within the time $\eta$ at which the gravitational wave power spectrum is computed (recall equation~\eqref{eq:pgw}). We then estimate the expression~\eqref{GpGp} by using the approximation of the spherical Bessel functions for small arguments.  To this end, we consider the limit of small argument starting from the power series representation of the Bessel function of the first kind of order $\nu\in \mathbb{R}$~\cite[\href{http://dlmf.nist.gov/10.2.E2}{(10.2.2)}]{NIST:DLMF}
\begin{equation}
    J_{\nu}(z) = \sum_{m= 0}^\infty \frac{(-1)^m}{m! \Gamma(m+\nu+1)} \left(\frac{z}{2} \right)^{2m+\nu} \underset{z\ll 1}{\longrightarrow} \frac{1}{\Gamma(1+\nu)}\left( \frac{z}{2} \right)^{\nu} + \mathcal{O}(z^{2+\nu}),
\end{equation}
and, for every index $\nu \not\in \mathbb{N}^0$, we evaluate the Bessel function of the second kind as
\begin{equation}\label{Yn_noint}
    Y_{\nu \not\in \mathbb{N}^0}(z) = \frac{J_{\nu}(z) \cos(\nu\pi) - J_{-\nu}(z)}{\sin(\nu\pi)} \underset{z\ll1}{\longrightarrow} -\frac{\Gamma(\nu)}{\pi} \left( \frac{z}{2}\right)^{-\nu} - \frac{\Gamma(-\nu)}{\pi}\cos(\nu\pi) \left( \frac{z}{2}\right)^{\nu} + \mathcal{O}(z^{2-\nu}).
\end{equation}
Here we used the reflection formula of the Euler Gamma function $\Gamma(\nu)\Gamma(1-\nu) = \pi/\sin(\nu\pi)$~\cite[\href{http://dlmf.nist.gov/5.5.E3}{(5.5.3)}]{NIST:DLMF}. 
The expression~\eqref{Yn_noint} is ill-defined for every integer index $\nu \in \mathbb{N}^0$, since in this case $\sin(\nu\pi) =0$. In such cases we can find instead an alternative expression. Let $\nu$ be an arbitrary non-negative integer $\nu\in \mathbb{N}^0$. For the Bessel function of the second kind of order $\nu\in \mathbb{N}^0$ one can use the following power series representation ~\cite[\href{https://dlmf.nist.gov/10.8.E1}{(10.8.1)}]{NIST:DLMF}
\begin{equation}
\begin{split}
    Y_{\nu\in \mathbb{N}^0}\left(z\right)=-\frac{(\tfrac{1}{2}z)^{-\nu}}{\pi}\sum_{\ell=0}^{\nu-1}\frac{(\nu-\ell-1)!}{\ell!} & \left(\tfrac{1}{4}z^{2}\right)^{\ell}+\frac{2}{\pi}\ln\left(\tfrac{1}{2}z\right)J_{n}\left(z\right)-\\
    &-\frac{(\tfrac{1}{2}z)^{\nu}}{\pi}\sum_{\ell=0}^{\infty}(\psi\left(\ell+1\right)+\psi\left(\nu+\ell+1\right))\frac{(-\tfrac{1}{4}z^{2})^{\ell}}{\ell!(\nu+\ell)!}
\end{split}
\end{equation}
with $\psi(x) \equiv \Gamma^\prime(x)/\Gamma(x)$ the digamma function. In the limit of small arguments (remember however that this asymptotic limit is bounded by the requirement $k\eta \gg 1$) so that 
\begin{equation}
    Y_{\nu\in \mathbb{N}^0}\left(z\right) \underset{z\ll 1 }{\longrightarrow} -\left(\frac{z}{2}\right)^{-\nu}\frac{\Gamma(\nu)}{\pi} +\frac{1}{\pi\Gamma(\nu+1)}\left(\frac{z}{2}\right)^{\nu} \left[ 2\gamma + 2\ln(z) - 2\ln(2) - H_{\nu} \right] + \mathcal{O}(z^{2-\nu}),
\end{equation}
with $\gamma \simeq 0.57721...$ the Euler–Mascheroni constant and $H_\nu = \sum_{\ell =1}^\nu \ell^{-1}$. 
Using the relation between the Bessel functions $J_\nu(z), Y_\nu(z)$ and the spherical Bessel functions 
\begin{equation}
    j_{\nu}(z) = \sqrt{\frac{\pi} {2z}} J_{\nu+\frac{1}{2}}(z), \qquad\quad y_{\nu}(z) = \sqrt{\frac{\pi} {2z}} Y_{\nu+\frac{1}{2}}(z),
\end{equation}
we can now write, up to terms of order $\mathcal{O}(z^2)$,
\begin{eqnarray}
    j_\nu(z) & \simeq & \frac{\sqrt{\pi}}{2\Gamma\left(\frac{3}{2}+\nu\right)} \left(\frac{z}{2} \right)^\nu,\\
    y_\nu(z) &\simeq & -\frac{\Gamma\left(\frac{1}{2}+\nu\right)}{2\sqrt{\pi}} \left(\frac{z}{2} \right)^{-1-\nu} + \left\lbrace\begin{aligned}
    &  \sin(\pi\nu) \frac{\Gamma\left(-\frac{1}{2}-\nu\right)}{2\sqrt{\pi}} \left(\frac{z}{2}\right)^{\nu},    \quad \nu \not\in \frac{\mathbb{N}^0}{2} \\
    & \frac{\left( \frac{z}{2} \right)^{\nu}}{2\sqrt{\pi} \Gamma\left(\nu +\frac{3}{2} \right)} \left[2\gamma + 2\ln\left(\frac{z}{2}\right) - H_{\nu+\tfrac{1}{2}}\right],    \quad \nu \in \frac{\mathbb{N}^0}{2}\qquad\;
  \end{aligned}\right. 
\end{eqnarray}
Finally, remembering the expression~\eqref{GpGp}, this leads to
\begin{equation}\label{eq:gg_long}
\begin{split}
        G^\prime_k & (\eta, \eta_1) G^\prime (\eta, \eta_2) \overset{k\eta_*\ll 1}{\simeq} \frac{\Gamma^2 \left(\frac{1}{2} +\nu \right)}{2\pi} \left(\frac{k\eta_1}{2} \right)^{-\nu} \left(\frac{k\eta_2}{2} \right)^{-\nu} \times \qquad\qquad\qquad\qquad\qquad \\
        & \times \left\lbrace\begin{aligned}
            & \Bigg\{ 1- \sin\left(\pi\nu\right)\frac{\Gamma \left(-\frac{1}{2} -\nu \right)}{\Gamma \left(\frac{1}{2} +\nu \right)} \left[ \left(\frac{k\eta_1}{2} \right)^{1+2\nu} + \left(\frac{k\eta_2}{2} \right)^{1+2\nu} \right] \Bigg\},    \quad \nu \not\in \frac{\mathbb{N}^0}{2} \\
            & \Bigg\{ 1- \frac{1}{\Gamma\left(\nu + \tfrac{3}{2}\right)\Gamma\left(\nu + \tfrac{1}{2}\right)}\left[ \left(\frac{k\eta_1}{2} \right)^{1+2\nu} f_\nu(k\eta_1) + \left(\frac{k\eta_2}{2} \right)^{1+2\nu} f_\nu(k\eta_2) \right] \Bigg\},    \quad \nu \in \frac{\mathbb{N}^0}{2}
        \end{aligned}\right.
        \end{split}
\end{equation}
where for brevity we defined $f_\nu(z) = 2\gamma + 2\ln\left(\frac{z}{2}\right) - H_{\nu+\tfrac{1}{2}}$. We again recall that this expression is obtained in the limit $k\eta_*< k\eta_{\rm{end}} \ll 1$ and $k\eta \gg 1$. 
Very long-wavelength modes with $k\eta \ll 1$ do not have time to complete one oscillation in the time $\eta$. These modes are constrained by causality~\cite{Durrer:2003ja}, and are not discussed in this work.

\section{Analytic solutions to the sound wave equation}\label{app:sw}
The dynamic equations for the fluid variables are given by the continuity equations $\nabla_\nu T^{\mu\nu} = 0$, which impose local conservation of energy and momentum. Given the metric~\eqref{metric}, the continuity equations for a barotropic perfect fluid with energy momentum tensor~\eqref{eq:T_munu} are
\begin{subequations}
    \begin{align}
         \nabla_\mu T^{\mu 0} = 0 & \rightarrow  \lambda^\prime + \partial_i v^i + 3\omega \mathcal{H}\lambda - 3  \Psi^\prime = 0,\\
    \nabla_\mu T^{\mu i} = 0 & \rightarrow v^{i\prime} + \mathcal{H}v^i + \partial^i\Phi + v^i\frac{\bar{p}^\prime}{\bar{w}} + \frac{\partial^i p}{\bar{w}} = 0.
    \end{align}
\end{subequations}
In Fourier space, neglecting the time dependence of the equation of state parameter $\omega = p/e$, we have 
\begin{subequations}\label{eqs:sw}
    \begin{align}
    \tilde{\lambda}_{\bm{p}}^\prime + p\tilde{v}_{\bm{p}}  - 3\Psi^\prime_{\bm{p}} = 0,\label{contunuity_0}\\
    \tilde{v}^\prime_{\bm{p}} + (1-3\omega) \mathcal{H} \tilde{v}_{\bm{p}}  - c_s^2p\tilde{\lambda}_{\bm{p}} - p\tilde{\Phi}_{\bm{p}} = 0.\label{contunuity_i}
    \end{align}
\end{subequations}
Notice that neglecting curvature perturbations, that is if we set $\Psi = 0 $ and $ \Phi = 0$,  these equations agree with equations (2.26) of Ref.~\cite{Giombi:2024kju}.
In our approximation we also neglect anisotropic pressure, as this identically vanishes for a perfect fluid. Einstein equations allow us then to set $\Phi = \Psi$~\cite{Durrer:2004fx}.  Moreover, we can use the linearized Einstein equations
\begin{subequations}
\begin{align}
    \tilde{\Psi}_{\bm{p}} = -\dfrac{3}{2} (1+\omega)\left(\dfrac{\mathcal{H}}{p}\right)^2 \left( \tlambda_{\bm{p}} + 3 \dfrac{\mathcal{H}}{p} \tv_{\bm{p}}\right),\\
    \tilde{\Psi}^\prime_{\bm{p}} + \mathcal{H} \tilde{\Psi}_{\bm{p}} = \frac{3}{2} (1+\omega)\left( \frac{\mathcal{H}}{p} \right)^2 p \tilde{v}_p ,
      \end{align}
\end{subequations}
to write the system of equations~\eqref{eqs:sw} in terms of $\tilde{v}_p$ and $\tilde{\lambda}_p$ only, leading to the system~\eqref{eq:eom_0}. For notational convenience we define $\zeta \equiv c_s p\eta$ and rewrite the system as
\begin{subequations}\label{subeqs:sw}
\begin{eqnarray}
    \frac{\partial\tilde{\lambda}_{\bm{p}}}{\partial \zeta} + \frac{1}{c_s}\left[1 - \dfrac{(2+\nu)(1-\nu)}{\zeta^2}\right]  \tv_{\bm{p}} = 0, \label{eq:lambda_zeta} \\
    \frac{\partial\tv_{\bm{p}}}{\partial \zeta} + \frac{2\nu}{\zeta} \tv_{\bm{p}} - c_s \left[1 - \dfrac{(2+\nu)(1+\nu)}{\zeta^2}\right] \tlambda_{\bm{p}}  = 0,
\end{eqnarray}
\end{subequations}
where we have also used the definition $\nu \equiv (1-3\omega)/(1+3\omega)$. In writing the equations~\eqref{subeqs:sw} we also used the solution to the background Friedmann equation for a barotropic fluid: $a(\eta) \sim \eta^{2/(1+3\omega)}\sim \eta^{1+\nu}$, which implies $\mathcal{H}(\eta) = (1+\nu)/\eta$. 

By differentiating equations~\eqref{subeqs:sw} with respect to $\zeta$ and neglecting terms of order $\mathcal{O}(1/\zeta^3)$, which we consider to be subdominant in the short sound wavelength expansion $R_*\mathcal{H}_* \lesssim \mathcal{O}(1)$, we can decouple the fluid energy fluctuations and peculiar velocity into 
\begin{subequations}\label{eqs:vl}
\begin{eqnarray}
    \frac{\partial^2\tv_{\bm{p}}}{\partial \zeta^2} + \frac{2\nu}{\zeta} \frac{\partial\tv_{\bm{p}}}{\partial \zeta} + \left[ 1 - 4\frac{1+\nu}{\zeta^2} \right] \tv_{\bm{p}} = 0,\\
    \frac{\partial^2 \tlambda_{\bm{p}}}{\partial\zeta^2} + \frac{2\nu}{\zeta} \frac{\partial  \tlambda_{\bm{p}}}{\partial\zeta} +  \left[ 1 - 2 \frac{2+\nu}{\zeta^2}  \right]  \tlambda_{\bm{p}} = 0.
\end{eqnarray}
\end{subequations} 
We now recognize that this is a system of Bessel equations. To make this clear, we perform a shift of the fluid perturbation fields
\begin{equation}\label{eq:field-redef}
     \tv_{\bm{p}} \equiv \zeta^{1-\nu} \tilde{V}_{\bm{p}}, \qquad \tilde{\lambda}_{\bm{p}} \equiv \zeta^{1-\nu} \tilde{\Lambda}_{\bm{p}}
\end{equation}
and simplify equations~\eqref{eqs:vl} as
\begin{subequations}\label{subeqs:bessel_sw}
\begin{eqnarray}
    \frac{\partial^2\tilde{V}_{\bm{p}}}{\partial \zeta^2} + \frac{2}{\zeta} \frac{\partial\tilde{V}_{\bm{p}}}{\partial \zeta} + \left[ 1 - \frac{\nu^2 +3\nu +4}{\zeta^2} \right] \tilde{V}_{\bm{p}} = 0, \label{sub:one_app}\\
    \frac{\partial^2 \tilde{\Lambda}_{\bm{p}}}{\partial\zeta^2} + \frac{2}{\zeta} \frac{\partial  \tilde{\Lambda}_{\bm{p}}}{\partial\zeta} +  \left[ 1 - \frac{\nu^2 +\nu +4}{\zeta^2} \right]  \tilde{\Lambda}_{\bm{p}} = 0.\label{eq:v_diff_app}
\end{eqnarray}
\end{subequations}
Equations~\eqref{sub:one_app} and~\eqref{eq:v_diff_app} are now manifestly written as Bessel equations. 
Their solutions can be written as a linear combination of spherical Bessel functions of order $n$ and $m$ respectively, with $n(n+1) = \nu^2 + 3\nu +4$ and $m(m+1) = \nu^2 + \nu +4$. Finally we can write the solutions for the fluid perturbations as
\begin{subequations}\label{sol_sw}
    \begin{eqnarray}
        \tilde{v}_{\bm{p}} &=& \left(\frac{\eta}{\eta_*}\right)^{1-\nu} \left[ \hat{c}_1 j_n(c_s p\eta)  + \hat{c}_2 y_n(c_s p\eta) \right], \label{eq:v_true}\\
        \tilde{\lambda}_{\bm{p}} &=& \left(\frac{\eta}{\eta_*}\right)^{1-\nu} \left[ \hat{c}_3 j_m(c_s p\eta)  + \hat{c}_4 y_m(c_s p\eta) \right],\label{eq:delta_true}
    \end{eqnarray}
\end{subequations}
with $\hat{c}_i$ ($i\in [1,2,3,4]$) real constants and $j_n, y_n$ spherical Bessel functions of the first and second kind of orders 
\begin{equation}
    n = \frac{1}{2}\sqrt{17 + 12\nu + 4\nu^2} - \frac{1}{2}, \qquad m = \frac{1}{2}\sqrt{17 + 4\nu + 4\nu^2} - \frac{1}{2}.
\end{equation}
We notice that in the original system of equations~\eqref{subeqs:sw} there can be only two independent solutions, and thus two independent real constants $\hat{c}_i$. This means that the constants $\hat{c}_i$ ($i\in [1,2,3,4]$) are not all independent from each other, but implicitly related through the sound wave equations~\eqref{subeqs:sw}.

\paragraph{Comparison to previous literature:} the contributions carried by curvature perturbations to the sound wave dynamics appear in the square brackets of equations~\eqref{subeqs:sw} as the terms that scale as $\zeta^{-2}$. Without these terms, equations~\eqref{subeqs:sw} describe the dynamics of sound waves in a simpler scenario, that is an expanding Universe with constant equation of state $p = \omega e$ without curvature perturbations. In this simpler situation, performing the same steps that led to equations~\eqref{eqs:vl}, we obtain
\begin{subequations}\label{subeqs:comparison_Midiri}
\begin{eqnarray}
    \frac{\partial^2\tv_{\bm{p}}}{\partial \zeta^2} + \frac{2\nu}{\zeta} \frac{\partial\tv_{\bm{p}}}{\partial \zeta} + \left[ 1 - \frac{2\nu}{\zeta^2} \right] \tv_{\bm{p}} = 0,\\
    \frac{\partial^2 \tlambda_{\bm{p}}}{\partial\zeta^2} + \frac{2\nu}{\zeta} \frac{\partial  \tlambda_{\bm{p}}}{\partial\zeta} + \tlambda_{\bm{p}} = 0.
\end{eqnarray}
\end{subequations} 
These equations agree with the results of Ref.~\cite{RoperPol:2025lgc}, equations (3.83).  This case also correspond to the scenario studied in Ref.~\cite{Giombi:2024kju}, as one can understand by comparing equations (2.26) therein with equations~\eqref{subeqs:sw}. Using the same field redefinition in equation~\eqref{eq:field-redef}, one can then write 
\begin{subequations}\label{subeqs:bessel_sw_nophi}
\begin{eqnarray}
    \frac{\partial^2\tilde{V}_{\bm{p}}}{\partial \zeta^2} + \frac{2}{\zeta} \frac{\partial\tilde{V}_{\bm{p}}}{\partial \zeta} + \left[ 1 - \frac{\nu (1+\nu)}{\zeta^2} \right] \tilde{V}_{\bm{p}} = 0,\\
    \frac{\partial^2 \tilde{\Lambda}_{\bm{p}}}{\partial\zeta^2} + \frac{2}{\zeta} \frac{\partial  \tilde{\Lambda}_{\bm{p}}}{\partial\zeta} +  \left[ 1 + \frac{\nu (1-\nu)}{\zeta^2} \right]  \tilde{\Lambda}_{\bm{p}} = 0.
\end{eqnarray}
\end{subequations}
Solutions to these second order differential equations are again spherical Bessel function. Therefore one can write a solutions in a way analogous to equation~\eqref{sol_sw}, where the order of the Bessel functions for the case without curvature perturbations are $n = \nu$ and $m = \nu- 1$, as obtained in Ref.~\cite{Giombi:2024kju} (see equation 2.27). 

Moreover, in absence of curvature perturbations and in the case $\nu = 0$ (which means $c_s = 1/\sqrt{3}$), our expressions~\eqref{subeqs:sw} recover the sound wave equations for a radiation fluid (see for example equations (24-25) of Ref.~\cite{RoperPol:2023dzg}) and the solutions of equations~\eqref{subeqs:comparison_Midiri} are sine and cosine functions, as in Ref.~\cite{RoperPol:2023dzg}.

\subsection{Asymptotic expansion at large wavenumbers}
The mean bubble spacing $R_*$ sets the largest typical size of correlated fluid regions. On length scales larger than $R_*$, fluid perturbations are not coherent. The relevant length scales for the source are therefore smaller than $R_*$, that is wavenumbers $p>1/R_*$. 
Since in this paper we consider that the characteristic wavelength of the fluid is shorter than the Hubble length, meaning that $R_*\mathcal{H}_* \lesssim \mathcal{O}(1)$, the relevant wavenumbers satisfy $p\eta > p\eta_*> (R_*\mathcal{H}_*)^{-1} \gtrsim \mathcal{O}(1)$. With the same level of approximation in the bubble mean spacing expansion $R_*\mathcal{H}_* \lesssim \mathcal{O}(1)$ used to derived the sound wave solutions~\eqref{sol_sw}, we can expand the spherical Bessel functions for large arguments $\zeta\gg 1$. At any order one has~\cite{ARFKEN2013643} 
\begin{subequations}
\begin{eqnarray}
    j_{m}\left(\zeta\right) &\simeq& \frac{1}{\zeta} \bigg\{P_{m+\frac{1}{2}}(\zeta)\cos\left[\zeta - (m+1)\frac{\pi}{2}\right]  - Q_{m+\frac{1}{2}}(\zeta)\sin\left[\zeta - (n+1) \frac{\pi}{2}\right] \bigg\},\\
    y_{m}(\zeta) &\simeq& \frac{1}{\zeta} \bigg\{P_{m+\frac{1}{2}}(\zeta)\sin\left[\zeta - (m+1)\frac{\pi}{2}\right]  + Q_{m+\frac{1}{2}}(\zeta)\cos\left[\zeta - (m+1)\frac{\pi}{2}\right]\bigg\},
\end{eqnarray}
\end{subequations}
with the expansion polynomials
\begin{equation}
    P_{m+\frac{1}{2}}(\zeta) = \sum_{j=0}^\infty (-1)^j \frac{c_{2j}({m+\frac{1}{2}})}{(2\zeta)^{2j}}, \qquad Q_{m+\frac{1}{2}}(\zeta) = \sum_{j=0}^\infty (-1)^j \frac{c_{2j+1}({m+\frac{1}{2}})}{(2\zeta)^{2j+1}},
\end{equation}
and coefficients
\begin{equation}
    c_j(m) = \frac{\Gamma\left(m + j +\frac{1}{2}\right)}{j!\, \Gamma\left(m - j +\frac{1}{2}\right)}.
\end{equation} 
At the next-to-leading-order in the bubble-over-Hubble radius expansion $R_*\mathcal{H}_* \lesssim \mathcal{O}(1)$, we have $P_{m+\frac{1}{2}}(\zeta) = 1 + o(\zeta^{-2})$ and $Q_{m+\frac{1}{2}}(\zeta) = m(m+1)/2\zeta  + o(\zeta^{-3})$. Thus, from equation~\eqref{eq:delta_true}, and writing the trigonometric functions as sum of exponential functions, we obtain
\begin{equation}\label{eq:lambda_app}
    \tilde{\lambda}_{\bm{p}} = -\frac{i}{c_s}\left(\frac{\eta}{\eta_*}\right)^{-\nu} \left[ c_1 \left( 1-i \frac{m(m+1)}{2\zeta} \right) e^{-i\zeta} + c_2 \left( 1+ i \frac{m(m+1)}{2\zeta} \right) e^{i\zeta} \right],
\end{equation}
where we defined the constants $c_1$ and $c_2$ as
\begin{equation}
    -i p \eta_* c_1 \equiv\frac{\hat{c}_3+i\hat{c}_4}{2} e^{i (m+1)\frac{\pi}{2}} , \qquad -i p \eta_* c_2 \equiv \frac{\hat{c}_3-i\hat{c}_4}{2} e^{-i (m+1)\frac{\pi}{2}} .
\end{equation}
The redefinition of the constants is done in such a way that the solution fluid density fluctuation field~\eqref{eq:lambda_app} recovers the known solution in flat Minkowski spacetime (see for example equation 12 of Ref.~\cite{Hindmarsh:2016lnk}) at the initial time $\eta_*$ and in the asymptotic limit $\zeta \rightarrow \infty$.

We remind the reader that, in order to satisfy the system of equations~\eqref{subeqs:sw}, also the real constants $\hat{c_1}$ and $\hat{c}_2$ that specify the solution~\eqref{sol_sw} must be related to $c_1$ and $c_2$, as there are only two independent constants.
To obtain the fluid peculiar velocity we now insert this expression back into the starting differential equation~\eqref{eq:lambda_zeta}. 
At the next-to-leading-order in $1/\zeta$ we then have
\begin{equation}\label{eq:v_p_app}
    \tilde{v}_{\bm{p}} = -\frac{i}{c_s}\left(\frac{\eta}{\eta_*}\right)^{-\nu} \left[ c_1 \left( 1-i \frac{m(m+1) +2\nu}{2\zeta} \right) e^{-i\zeta} - c_2 \left( 1+ i \frac{m(m+1)+2\nu}{2\zeta} \right) e^{i\zeta} \right],
\end{equation}
Recalling equation~\eqref{eq:v_true}, the solution~\eqref{eq:v_p_app} implicitly defines the relation between the real constants $\hat{c}_{1,2}$ and $c_{1,2}$. 
Finally, we can connect the complex constant $c_1$ and $c_2$ to the plane wave amplitudes by defining $c_1 \equiv \hat{p}_i v_{\bm{p}}^i$ and $c_2 = c_1^* = - \hat{p}_i v_{-\bm{p}}^i$. Remembering further that $m(m+1) = \nu^2 +\nu +4$, we recover the result of equation~\eqref{eq:sound_waves} anticipated above.

\section{Additional material for the calculation of the kernel functions and the gravitational wave power spectrum}\label{app:kernel}
In this section of the Appendix we provide additional material that leads to the analytical integration of the gravitational wave power spectrum~\eqref{eq:Pgw}. Let us first remember the kernel for the gravitational wave power spectrum from equation~\eqref{eq:Delta},
\begin{equation}\label{eq:kern_app}
    \begin{split}
        \Delta &\left(z, x, y, \uptau_*, \uptau_{\text{end}}\right) =  \iint_{\uptau_*}^{\uptau_{\text{end}}} \frac{d\uptau_1 d\uptau_2}{\uptau_*^2} \left(\frac{\uptau_*^2}{\uptau_1\uptau_2} \right)^{1+\nu} G^\prime_z(\uptau, \uptau_1)G^\prime_z(\uptau, \uptau_2) \times \\
        & \times \bigg\{ \cos(c_s x \uptau_-) \cos(c_s y\uptau_-)\\
        &\qquad - \frac{4+\nu(3+\nu)}{2c_s} \left(\frac{1}{\uptau_1} - \frac{1}{\uptau_2}\right) \Big[ \frac{\sin(c_s x \uptau_-)\cos(c_s y \uptau_-) }{x} + \frac{\sin(c_s y \uptau_-)\cos(c_s x \uptau_-) }{y}\Big] \\
        & \qquad + \frac{(1+\nu)(2+\nu)}{c_s^2} \bigg(\frac{1}{\uptau_2^2} + \frac{1}{\uptau_1^2}\bigg)\frac{\sin(c_s x \uptau_-)\sin(c_s y \uptau_-)}{xy} \bigg\}.
    \end{split}
\end{equation}
As done in the main text of the paper, we divide this calculation in super- and sub-horizon modes.

\subsection{Approximation for super-horizon gravitational wave modes}\label{app_low}
On super-horizon scales $k\eta_*<k\eta_{\rm{end}} \ll1 \ll k\eta$, the product of Green's functions of gravitational waves is well approximated by equation~\eqref{eq:gg_long}, so that the individual terms of the kernel can be written as
\begin{subequations}
    \begin{eqnarray}
    \Delta^{\text{lo}}_{\text{sw}} &\underset{k\eta_{\rm{end}}\ll 1}{\simeq}& \left(\frac{k\eta_*}{2} \right)^{-2\nu}  \frac{\Gamma^2 \left(\frac{1}{2}+\nu\right)}{4\pi} \sum_{m=\pm} \iint_{\uptau_*}^{\uptau_{\text{end}}} \frac{d\uptau_1 d\uptau_2}{\uptau_*^2}  \left(\frac{\uptau_*^2}{\uptau_1\uptau_2}\right)^{1+2\nu} \cos(\omega_m \uptau_-),\\
    \Delta^{\text{nlo}}_{\text{gw}} &\underset{k\eta_{\rm{end}}\ll 1}{\simeq}& -\left(\frac{k\eta_*}{2} \right)\sin(\pi\nu)  \frac{\Gamma\left(-\frac{1}{2}-\nu\right)\Gamma \left(\frac{1}{2}+\nu\right)}{4\pi} \sum_{m= \pm} \iint_{\uptau_*}^{\uptau_{\text{end}}} \frac{d\uptau_1 d\uptau_2}{\uptau_*^2}     \times\nonumber\\
    &&\qquad\qquad\qquad\qquad\qquad\qquad \times \left[\left(\frac{\uptau_*}{\uptau_1}\right)^{1+2\nu} + \left(\frac{\uptau_*}{\uptau_2}\right)^{1+2\nu} \right] \cos(\omega_m\uptau_-), \label{eq:delta_gw} \\
    \Delta^{\text{nlo}}_{\text{sw}} &\underset{k\eta_{\rm{end}}\ll 1}{\simeq}& - \left(\frac{k\eta_*}{2} \right)^{-2\nu}  \frac{\Gamma^2 \left(\frac{1}{2}+\nu\right)}{4\pi}\frac{4+\nu(3+\nu)}{2 c_s^2 xy }\sum_{m=\pm} m\,  \times  \nonumber\\
    && \qquad\qquad\quad \times \iint_{\uptau_*}^{\uptau_{\text{end}}} \frac{d\uptau_1 d\uptau_2}{\uptau_*^2}   \left(\frac{\uptau_*^2}{\uptau_1\uptau_2}\right)^{1+2\nu}  \left(\frac{1}{\uptau_1} - \frac{1}{\uptau_2}\right) \omega_m\sin(\omega_m \uptau_-),\quad\\
    \Delta^{\text{nlo}}_\Phi &\underset{k\eta_{\rm{end}}\ll 1}{\simeq}& -\left(\frac{k\eta_*}{2} \right)^{-2\nu}  \frac{\Gamma^2 \left(\frac{1}{2}+\nu\right)}{4\pi} \frac{(1+\nu)(2+\nu)}{c_s^2 xy} \sum_{m=\pm}\iint_{\uptau_*}^{\uptau_{\text{end}}} \frac{d\uptau_1 d\uptau_2}{\uptau_*^2}   \left(\frac{\uptau_*^2}{\uptau_1\uptau_2}\right)^{1+2\nu} \times \nonumber\\
    && \qquad\qquad\qquad\qquad\qquad\qquad\qquad\qquad\qquad \times\bigg(\frac{1}{\uptau_2^2} + \frac{1}{\uptau_1^2}\bigg) m\cos(\omega_m\uptau_-)\big].
\end{eqnarray}
\end{subequations}
In writing the above expression, we also made use of the trigonometric properties
\begin{equation}
    \begin{split}
        \cos(A)\cos(B) = \frac{1}{2} \sum_\pm \cos(A\pm B), \qquad \sin(A)\sin(B) = -\frac{1}{2} \sum_\pm  \pm \cos(A\pm B), \\ \sin(A)\cos(B) = \frac{1}{2} \sum_\pm \sin(A\pm B).\qquad\qquad\qquad\qquad\quad
    \end{split}
\end{equation}
We notice that the NLO contribution $\Delta_{\text{gw}}^{\text{nlo}}$ grows linearly with $k$, and it is identically zero in pure radiation, when $\nu= 0$. On the opposite, all the other contributions decay with the gravitational wavenumber as $k^{-2\nu}$. Notice further that, in writing equation~\eqref{eq:delta_gw}, we used the expression of the product of Green's functions~\eqref{eq:gg_long} with $\nu\not\in \mathbb{N}^0/2$. This choice implies that $\Delta_{\text{gw}}^{\text{nlo}}$ is not defined for $\nu =1/2$. The case with $\nu = 1/2$ is more difficult to integrate analytically because of the presence of a logarithmic function in $f_\nu(z) = 2\gamma + 2\ln\left(z/2\right) - H_{\nu+1/2}$ in equation~\eqref{eq:gg_long}. We therefore prefer to use equation~\eqref{eq:delta_gw} to evaluate $\Delta_{\text{gw}}^{\text{nlo}}$ for all purposes and understand the case $\nu= 1/2$ as a limit, that is  $\Delta_{\text{gw}}^{\text{nlo}}(\nu= 1/2) \equiv \underset{\nu\rightarrow1/2}{\operatorname{lim}} \Delta_{\text{gw}}^{\text{nlo}}(\nu\not\in\mathbb{N}^0/2)$.

We can now use the properties of summation of trigonometric functions
to separate the integration variables $\uptau_1$ and $\uptau_2$. The evaluation of the kernel function is thereby reduced to the calculation of one dimensional integrals of the form
\begin{subequations}\label{eqs:sici}
\begin{eqnarray}
    \int_{\uptau_*}^{\uptau_{\text{end}}} d\uptau \, \frac{\cos(\omega_{m n}\uptau)}{\uptau^{j+\nu}} &=& \abs{\omega_{mn}}^{-1+j+\nu} \left[ \operatorname{ci}_{1-j-\nu}\left(\abs{\omega_{mn}}\uptau_*\right) - \operatorname{ci}_{1-j-\nu}\left(\abs{\omega_{mn}}\uptau_{\text{end}}\right)\right],\\
    \int_{\uptau_*}^{\uptau_{\text{end}}} d\uptau \, \frac{\sin(\omega_{m n}\uptau)}{\uptau^{j+\nu}} &=& \operatorname{sign}(\omega_{mn})\abs{\omega_{mn}}^{-1+j+\nu} \left[ \operatorname{si}_{1-j-\nu}\left(\abs{\omega_{mn}}\uptau_*\right) - \operatorname{si}_{1-j-\nu}\left(\abs{\omega_{mn}}\uptau_{\text{end}}\right)\right],\quad\nonumber\\
    &&
\end{eqnarray} 
\end{subequations}
with $j\in  \mathbb{N}^+$, and $\operatorname{si}_\nu(x)$, $\operatorname{ci}_\nu(x)$ the generalized sine and cosine integral functions defined in~\eqref{eqs:trigo}.
This way, we finally find the following expressions
\begin{subequations}
    \begin{eqnarray}
    \Delta^{\text{lo}}_{\text{sw}} &\overset{k\eta_{\rm{end}}\ll 1}{=}& \left(\frac{k\eta_*}{2}\right)^{-2\nu}\frac{\Gamma^2 \left( \frac{1}{2} +\nu\right)}{4\pi} \sum_{m = \pm}  \left\vert \omega_m\uptau_*\right\vert^{4\nu} \times \nonumber\\
    && \times \left[\left(\operatorname{ci}_{-2\nu}(\omega_m\uptau)\Big\vert^{\uptau_{\text{end}}}_{\uptau_*} \right)^2  + \left( \operatorname{si}_{-2\nu}(\omega_m\uptau)\Big\vert^{\uptau_{\text{end}}}_{\uptau_*} \right)^2 \right], \\ 
    \Delta^{\text{nlo}}_{\text{gw}} &\overset{k\eta_{\rm{end}}\ll 1}{=}& -\left(\frac{k\eta_*}{2} \right)\sin(\pi\nu)  \frac{\Gamma\left(-\frac{1}{2}-\nu\right)\Gamma \left(\frac{1}{2}+\nu\right)}{2\pi} \sum_{m= \pm} \vert \omega_m \uptau_* \vert^{-1+2\nu} \times \nonumber\\
    && \times\left[\operatorname{ci}_{-2\nu}(\omega_m\uptau)\Big\vert^{\uptau_{\text{end}}}_{\uptau_*} \sin(\omega_m\uptau)\Big\vert^{\uptau_{\text{end}}}_{\uptau_*}  - \operatorname{si}_{-2\nu}(\omega_m\uptau)\Big\vert^{\uptau_{\text{end}}}_{\uptau_*} \cos(\omega_m\uptau) \Big\vert^{\uptau_{\text{end}}}_{\uptau_*}\right], \\
    \Delta^{\text{nlo}}_{\text{sw}} &\overset{k\eta_{\rm{end}}\ll 1}{=}& -\left(\frac{k\eta_*}{2}\right)^{-2\nu}\frac{\Gamma^2 \left( \frac{1}{2} +\nu\right)}{2\pi} \frac{4+\nu(3+\nu)}{2c_s^2 \uptau_*^2 xy } \sum_{m =\pm} m    \left\vert \omega_m \uptau_*\right\vert^{2+4\nu} \times \nonumber\\
    && \times\Big[\operatorname{si}_{-1-2\nu}(\omega_m\uptau)\Big\vert^{\uptau_{\text{end}}}_{\uptau_*} \operatorname{ci}_{-2\nu}(\omega_m\uptau)\Big\vert^{\uptau_{\text{end}}}_{\uptau_*} - \operatorname{ci}_{-1-2\nu}(\omega_m\uptau)\Big\vert^{\uptau_{\text{end}}}_{\uptau_*} \operatorname{si}_{-2\nu}(\omega_m\uptau)\Big\vert^{\uptau_{\text{end}}}_{\uptau_*} \Big], \\
    \Delta^{\text{nlo}}_\Phi &\overset{k\eta_{\rm{end}}\ll 1}{=}& -\left(\frac{k\eta_*}{2}\right)^{-2\nu}\frac{\Gamma^2 \left( \frac{1}{2} +\nu\right)}{2\pi}  \frac{(1+\nu)(2+\nu)}{c_s^2 xy\uptau_*^2} \sum_{m = \pm} m \left\vert \omega_m \uptau_* \right\vert^{2+4\nu} \times  \nonumber\\
    && \times \Big[\operatorname{ci}_{-2-2\nu}(\omega_m\uptau)\Big\vert^{\uptau_{\text{end}}}_{\uptau_*} \operatorname{ci}_{-2\nu}(\omega_m\uptau)\Big\vert^{\uptau_{\text{end}}}_{\uptau_*} + \operatorname{si}_{-2-2\nu}(\omega_m\uptau)\Big\vert^{\uptau_{\text{end}}}_{\uptau_*} \operatorname{si}_{-2\nu}(\omega_m\uptau)\Big\vert^{\uptau_{\text{end}}}_{\uptau_*} \Big] ,\qquad\quad
\end{eqnarray}
\end{subequations}
with $\omega_\pm = c_s (x\pm y)$.

On super-horizon scales $k\eta_*<k\eta_{\rm{end}} \ll 1 \ll k\eta$ we can further consider the limit $kR_*\sim (R_*\mathcal{H}_*) k\eta_* \ll 1$. We recall that, since the fluid flow has support in $R_*$, the relevant wavenumbers of sound waves satisfy $p\gtrsim 1/R_*$. This means that we can consider $k/p \ll 1$. In this limit, the incoming sound wave wavenumbers tend to align
$\bm{q} \approx \bm{p}$, and we consider $y\approx  x$. This approximation simplifies the above system of equations to 
\begin{subequations}\label{kenerl_super}
    \begin{align}
    \Delta^{\text{lo}}_{\text{sw}}  &\underset{z\ll 1}{\simeq}  \left(\frac{k\eta_*}{2}\right)^{-2\nu} \frac{\Gamma^2\left(\frac{1}{2}+\nu \right)}{4\pi} \Bigg\{ \frac{1}{4\nu^2}\left[1 - \left(\frac{\uptau_*}{\uptau_{\text{end}}}\right)^{2\nu}\right]^2 + \nonumber \\
    & \qquad\qquad\qquad + (2c_sx\uptau_*)^{4\nu} \left[\left(\operatorname{ci}_{-2\nu}(2c_s x\uptau)\Big\vert^{\uptau_{\text{end}}}_{\uptau_*} \right)^2  + \left( \operatorname{si}_{-2\nu}(2c_s x\uptau)\Big\vert^{\uptau_{\text{end}}}_{\uptau_*} \right)^2 \right]\Bigg\}, \\
    \Delta^{\text{nlo}}_{\text{gw}} &\underset{z\ll 1}{\simeq} -\left(\frac{k\eta_*}{2} \right)\sin(\pi\nu)  \frac{\Gamma\left(-\frac{1}{2}-\nu\right)\Gamma \left(\frac{1}{2}+\nu\right)}{2\pi} \Bigg\{ \frac{1}{2\nu} \left[\frac{\uptau_{\text{end}}}{\uptau_*} -1\right] \left[1- \left(\frac{\uptau_*}{\uptau_{\text{end}}}\right)^{2\nu} \right] \nonumber\\
    & + (2c_sx\uptau_*)^{-1+2\nu}\left[\operatorname{ci}_{-2\nu}(\omega_+\uptau)\Big\vert^{\uptau_{\text{end}}}_{\uptau_*} \sin(\omega_+\uptau)\Big\vert^{\uptau_{\text{end}}}_{\uptau_*}  - \operatorname{si}_{-2\nu}(\omega_+\uptau)\Big\vert^{\uptau_{\text{end}}}_{\uptau_*} \cos(\omega_+\uptau) \Big\vert^{\uptau_{\text{end}}}_{\uptau_*}\right] \Bigg\}, \quad \\
    \Delta^{\text{nlo}}_{\text{sw}}  &\underset{z\ll 1}{\simeq}  - \left(\frac{k\eta_*}{2}\right)^{-2\nu} \frac{\Gamma^2\left(\frac{1}{2}+\nu \right)}{2\pi}\left[4+\nu(3+\nu)\right] 2(2c_s x\uptau_*)^{4\nu} \times\nonumber \\
    & \times \left[\operatorname{si}_{-1-2\nu} (2c_s x\uptau) \Big\vert^{\uptau_{\text{end}}}_{\uptau_{*}} \operatorname{ci}_{-2\nu} (2c_s x\uptau) \Big\vert^{\uptau_{\text{end}}}_{\uptau_{*}}  -  \operatorname{ci}_{-1-2\nu} (2c_s x\uptau) \Big\vert^{\uptau_{\text{end}}}_{\uptau_{*}} \operatorname{si}_{-2\nu} (2c_s x\uptau) \Big\vert^{\uptau_{\text{end}}}_{\uptau_{*}}\right], \\
    \Delta^{\text{nlo}}_{\Phi}  &\underset{z\ll 1}{\simeq}  \left(\frac{k\eta_*}{2}\right)^{-2\nu} \frac{\Gamma^2\left(\frac{1}{2}+\nu \right)}{2\pi}\frac{(2+\nu)}{\nu (2 c_s x \uptau_*)^2} \Bigg\{\left[1-\left(\frac{\uptau_*}{\uptau_{\text{end}}}\right)^{2\nu} \right]\left[1-\left(\frac{\uptau_*}{\uptau_{\text{end}}}\right)^{2+2\nu} \right] +\nonumber\\
    & \qquad\qquad\qquad\qquad \times (2c_s x\uptau_*)^{2+4\nu} \Big[\operatorname{ci}_{-2-2\nu}(\omega_m\uptau)\Big\vert^{\uptau_{\text{end}}}_{\uptau_*} \operatorname{ci}_{-2\nu}(\omega_m\uptau)\Big\vert^{\uptau_{\text{end}}}_{\uptau_*} + \nonumber\\
    & \qquad\qquad\qquad\qquad\qquad\qquad\qquad\qquad + \operatorname{si}_{-2-2\nu}(\omega_m\uptau)\Big\vert^{\uptau_{\text{end}}}_{\uptau_*} \operatorname{si}_{-2\nu}(\omega_m\uptau)\Big\vert^{\uptau_{\text{end}}}_{\uptau_*} \Big]\Bigg\} .
    \end{align}
\end{subequations}
The generalized sine and cosine functions provide oscillatory and decaying contributions that become subdominant for large arguments $x\uptau >x\uptau_* \gtrsim (R_*\mathcal{H}_*)^{-1}\gtrsim \mathcal{O}(1)$. In providing an estimation for the gravitational wave power spectrum, carried out in Section~\ref{sec:low}, we will therefore neglect these terms.

Since the gravitational wave power spectrum~\eqref{eq:Pgw} scales as $\Pgw \propto (k R_*)^3 \tilde{P}_{\text{gw}}(kR_*)$, this estimation of the kernel at low frequency predicts, for all wavenumbers $k\eta_* \ll 1$, a scaling $\Pgw \propto k^{3-2\nu}$ for the contributions from $\Delta_{\text{sw}}^{\text{lo}}, \Delta_{\text{sw}}^{\text{nlo}}, \Delta^{\text{nlo}}_{\Phi}$, while a steeper scaling $\Pgw\propto k^4$ for the contribution from $\Delta_{\text{gw}}^{\text{nlo}}$.

We recall however that the above derivation is done in the limit $k\eta \gg 1$. At gravitational wave wavenumbers $k\eta\ll 1$ the above arguments are not valid, and the gravitational wave power spectrum is bound instead by causality to scale as $\Pgw \propto k^{3}$~\cite{Durrer:2003ja}.

\subsection{Approximation for sub-horizon gravitational wave modes}\label{app_large}
Gravitational wave modes with $k\eta_* \gg 1$ typically oscillate several times within the duration of the source. The product of Green's functions inside the kernel~\eqref{eq:Delta} is well approximated by equation~\eqref{eq:gpgp_result}. The four contributions are then
\begin{subequations}\label{eqs:k_sub}
    \begin{eqnarray}
    \Delta^{\text{lo}}_{\text{sw}} &\overset{k\eta_*\gg 1}{=}& \frac{1}{2} \int_{\uptau_*}^{\uptau_{\text{end}}} \frac{d\uptau_1 d\uptau_2}{\uptau_*^2}  \left(\frac{\uptau_*^2}{\uptau_1\uptau_2}\right)^{1+\nu}  \cos(z\uptau_-) \cos(c_s x\uptau_-)\cos(c_s y\uptau_-),\\
    \Delta^{\text{nlo}}_{\text{gw}} &\overset{k\eta_*\gg 1}{=}& -\frac{1}{2} \frac{\nu(1+\nu)}{2} \int_{\uptau_*}^{\uptau_{\text{end}}} \frac{d\uptau_1 d\uptau_2}{\uptau_*^2}   \left(\frac{\uptau_*^2}{\uptau_1\uptau_2}\right)^{1+\nu}  \left(\frac{1}{\uptau_1} - \frac{1}{\uptau_2}\right) \times \nonumber \\
    && \qquad\qquad\qquad\qquad\qquad\qquad\qquad\qquad\times \; \frac{\sin(z\uptau_-)\cos(c_s x\uptau_-)\cos(c_s y\uptau_-)}{z},  \\
    \Delta^{\text{nlo}}_{\text{sw}} &\overset{k\eta_*\gg 1}{=}& -\frac{1}{2} \frac{4+\nu(3+\nu)}{2c_s}\int_{\uptau_*}^{\uptau_{\text{end}}} \frac{d\uptau_1 d\uptau_2}{\uptau_*^2}   \left(\frac{\uptau_*^2}{\uptau_1\uptau_2}\right)^{1+\nu}     \left(\frac{1}{\uptau_1} - \frac{1}{\uptau_2}\right)\cos(z\uptau_-)\times \nonumber\\
    && \qquad\qquad\qquad\qquad \times \left[\frac{\cos(c_s x \uptau_-)\sin(c_s y \uptau_-)}{y} + \frac{\sin(c_s x \uptau_-)\cos(c_s y \uptau_-)}{x}\right],\qquad\quad\\
    \Delta^{\text{nlo}}_\Phi &\overset{k\eta_*\gg 1}{=}& \frac{1}{2} \frac{(1+\nu)(2+\nu)}{c_s^2} \int_{\uptau_*}^{\uptau_{\text{end}}} \frac{d\uptau_1 d\uptau_2}{\uptau_*^2}   \left(\frac{\uptau_*^2}{\uptau_1\uptau_2}\right)^{1+\nu}  \bigg(\frac{1}{\uptau_2^2} + \frac{1}{\uptau_1^2}\bigg)\times\nonumber\\
    && \qquad\qquad\qquad\qquad\qquad\qquad\qquad\qquad\times \;\frac{\cos(z\uptau_-)\sin(c_s x \uptau_-)\sin(c_s y \uptau_-)}{xy}.
\end{eqnarray}
\end{subequations}
The kernel contributions~\eqref{eqs:k_sub} contain a double integration over dimensionless time variables $\uptau_1$ and $\uptau_2$. As done for the non-oscillatory modes $k\eta_* \ll 1$, we aim to split the integration variables and reduce the kernel terms to products of one-dimensional integrals. To this end, we make use of the properties of trigonometric functions
\begin{eqnarray}
&&    \cos(z\uptau_-) \cos(x c_s\uptau_-)\cos(y c_s\uptau_-) = \frac{1}{4}\sum_{m, n = \pm 1} \cos(\omega_{m n}\uptau_-),\\
&&    \cos(z\uptau_-) \sin(x c_s\uptau_-)\sin(y c_s\uptau_-) = -\frac{1}{4}\sum_{m, n = \pm 1} mn\cos(\omega_{m n}\uptau_-),\\
&&    \cos(z\uptau_-) \cos(x c_s\uptau_-)\sin(y c_s\uptau_-) = \frac{1}{4}\sum_{m, n = \pm 1} n\sin(\omega_{m n}\uptau_-),\\
&&    \sin(z\uptau_-) \cos(x c_s\uptau_-)\cos(y c_s\uptau_-) = \frac{1}{4}\sum_{m, n = \pm 1} \sin(\omega_{m n}\uptau_-),
\end{eqnarray}
with $\omega_{mn} = z + c_s(mx + ny)$, and write the kernel terms as 
\begin{subequations}\label{eqs:delta_appendix}
    \begin{eqnarray}
    \Delta^{\text{lo}}_{\text{sw}} &=& \frac{1}{8}\sum_{m, n = \pm 1} \int_{\uptau_*}^{\uptau_{\text{end}}} \frac{d\uptau_1 d\uptau_2}{\uptau_*^2}  \left(\frac{\uptau_*^2}{\uptau_1\uptau_2}\right)^{1+\nu}  \times \nonumber \\
    && \qquad\qquad\qquad\qquad \times \Big\{\cos(\omega_{m n}\uptau_1) \cos(\omega_{m n}\uptau_2) +\sin(\omega_{m n}\uptau_1) \sin(\omega_{m n}\uptau_2)\Big\},\\
    \Delta^{\text{nlo}}_{\text{gw}} &=& -\frac{1}{8} \frac{\nu(1+\nu)}{2z} \sum_{m, n = \pm 1} \int_{\uptau_*}^{\uptau_{\text{end}}} \frac{d\uptau_1 d\uptau_2}{\uptau_*^2}   \left(\frac{\uptau_*^2}{\uptau_1\uptau_2}\right)^{1+\nu}  \left(\frac{1}{\uptau_1} - \frac{1}{\uptau_2}\right) \times \nonumber\\
    && \qquad\qquad\qquad\qquad\times \Big\{\sin(\omega_{m n}\uptau_1) \cos(\omega_{m n}\uptau_2) - \cos(\omega_{m n}\uptau_1) \sin(\omega_{m n}\uptau_2) \Big\}, \\
    \Delta^{\text{nlo}}_{\text{sw}} &=& -\frac{1}{8} \frac{4+\nu(3+\nu)}{2c_s}\sum_{m, n = \pm 1}  \left( \frac{n}{y} +\frac{m}{x}\right)\int_{\uptau_*}^{\uptau_{\text{end}}} \frac{d\uptau_1 d\uptau_2}{\uptau_*^2}   \left(\frac{\uptau_*^2}{\uptau_1\uptau_2}\right)^{1+\nu}     \left(\frac{1}{\uptau_1} - \frac{1}{\uptau_2}\right) \times \nonumber\\
    && \qquad\qquad\qquad\qquad \times \Big\{\sin(\omega_{m n}\uptau_1) \cos(\omega_{m n}\uptau_2) - \cos(\omega_{m n}\uptau_1) \sin(\omega_{m n}\uptau_2) \Big\},\qquad\\
    \Delta^{\text{nlo}}_\Phi &=& -\frac{1}{8} \frac{(1+\nu)(2+\nu)}{c_s^2} \sum_{m, n = \pm 1} \frac{mn}{xy} \int_{\uptau_*}^{\uptau_{\text{end}}} \frac{d\uptau_1 d\uptau_2}{\uptau_*^2}   \left(\frac{\uptau_*^2}{\uptau_1\uptau_2}\right)^{1+\nu}  \bigg(\frac{1}{\uptau_2^2} + \frac{1}{\uptau_1^2}\bigg)\times\nonumber\\
    && \qquad\qquad\qquad\qquad\times \Big\{\cos(\omega_{m n}\uptau_1) \cos(\omega_{m n}\uptau_2) + \sin(\omega_{m n}\uptau_1) \sin(\omega_{m n}\uptau_2) \Big\}.
\end{eqnarray}
\end{subequations}
 With equations~\eqref{eqs:sici}, it is straightforward now to recover the expressions of the kernel contributions~\eqref{eq:kernel_sub}.

\subsubsection{Approximation at intermediate frequency}\label{sec:append_int}
Let us now consider the gravitational wave power spectrum in the intermediate-frequency range $1\ll k\eta_*\ll k_p\eta_*$. In the large-bubble regime with $R_*\mathcal{H}_* \lesssim\mathcal{O}(1)$ and $k_p\eta_* \sim 2\pi/(R_*\mathcal{H}_*)$, this asymptotic regime might be very narrow and thus not very relevant. Nonetheless it can at least help us understand the overall shape of the power spectrum and in particular how to join the high- and low-frequency regimes. In the intermediate-frequency range  the gravitational wave power spectrum is dominated by the odd terms of the kernel sum~\eqref{eq:kernel_sub}, when $\omega_{mn} = \omega_m \equiv z + m c_s(x-y)$~\cite{Sharma:2023mao}. Since $k\ll k_p$, it follows that $kR_* \sim(R_*\mathcal{H}_*)k\eta_* \ll  2\pi$. Moreover, since $k\eta_* \gg1$, it follows that $kR_* \gtrsim 1$. We can then expand $\bm{y} = \bm{x} - \bm{z}$ for small $z$ and approximate $y \simeq x-\mu z$, with $\mu = \hat{\bm{x}}\cdot\hat{\bm{z}}$, so that $\omega_m \simeq z(1 + m c_s\mu)$. The arguments of the generalized trigonometric integrals have typically very large values in the intermediate-frequency regime, so that we are motivated to consider the expansion 
\begin{subequations}\label{eq:trigo_large}
\begin{eqnarray}
    \operatorname{si}_{-\nu} (x) & \underset{x\gg 1}{=} & \frac{\cos(x)}{x^{1+\nu}} +(1+\nu)\frac{\sin(x)}{x^{2+\nu}} + o\left(\frac{1}{x^{3+\nu}}\right), \\
    \operatorname{ci}_{-\nu} (x) &\underset{x\gg 1}{=}& - \frac{\sin(x)}{x^{1+\nu}}  +(1+\nu) \frac{\cos(x)}{x^{2+\nu}} + o\left(\frac{1}{x^{3+\nu}}\right).
\end{eqnarray}
\end{subequations}
Moreover, for the purpose of this analytic estimation of the power spectrum, we just focus on the case of long-lasting sources with $\eta_{\text{end}} \gg \eta_*$, where we can approximate
\begin{subequations}
    \begin{eqnarray}
    \operatorname{ci}_{-\nu}(\omega_{m}    \uptau)\Big\vert^{\uptau_{\text{end}}}_{\uptau_*} &\simeq& - \frac{\sin(\omega_m \uptau_*)}{(\omega_m \uptau_*)^{1+\nu}}  +(1+\nu) \frac{\cos(\omega_m \uptau_*)}{(\omega_m \uptau_*)^{2+\nu}}, \\
    \operatorname{si}_{-\nu}(\omega_{m}    \uptau)\Big\vert^{\uptau_{\text{end}}}_{\uptau_*} &\simeq& \frac{\cos(\omega_m \uptau_*)}{(\omega_m \uptau_*)^{1+\nu}}  +(1+\nu) \frac{\sin(\omega_m \uptau_*)}{(\omega_m \uptau_*)^{2+\nu}}.
    \end{eqnarray}
\end{subequations}
At leading-order in $1/k\eta_*$ the kernel terms~\eqref{eq:kernel_sub} approximate to
\begin{subequations}
    \begin{eqnarray}
        \Delta^{\text{lo}}_{\text{sw}}(1\ll k\eta_* \ll k_p\eta_*) &\simeq& \frac{1}{8}  \sum_{m = \pm 1}  \left\vert\omega_{m} \uptau_*\right\vert^{-2},\\
        \Delta^{\text{nlo}}_{\text{gw}}(1\ll k\eta_* \ll k_p\eta_*) &\simeq& \frac{\nu (1+\nu)}{8}  \sum_{m = \pm 1}  \frac{\left\vert\omega_{m} \uptau_*\right\vert^{-3}}{z\uptau_*}, \\
        \Delta^{\text{nlo}}_{\text{sw}}(1\ll k\eta_* \ll k_p\eta_*) &\simeq& -\frac{1}{2} \frac{4+\nu(3+\nu)}{(2c_s x\uptau_*)^2} c_s \mu z\uptau_*\sum_{m= \pm 1} m \left\vert\omega_{m} \uptau_*\right\vert^{-3}, \\ 
        \Delta^{\text{nlo}}_{\Phi}(1\ll k\eta_* \ll k_p\eta_*) &\simeq& \frac{(1+\nu) (2+\nu)}{(2 c_s x \uptau_*)^2}  \sum_{m = \pm 1}  \left\vert\omega_{m} \uptau_*\right\vert^{-2}.
    \end{eqnarray}
\end{subequations}
Finally, we can expand the sums to get
\begin{subequations}\label{eq:kernel_sub_int}
    \begin{eqnarray}
        \Delta^{\text{lo}}_{\text{sw}}(1\ll k\eta_* \ll k_p\eta_*) &=& \frac{1}{4(k\eta_*)^2}  \frac{1+\mu^2 c_s^2}{(1-\mu^2c_s^2)^2} ,\\
        \Delta^{\text{nlo}}_{\text{gw}}(1\ll k\eta_* \ll k_p\eta_*) &=& \frac{\nu (1+\nu)}{4}  \frac{1}{(k\eta_*)^4}  \frac{1+3\mu^2 c_s^2}{(1-\mu^2c_s^2)^3}, \\
        \Delta^{\text{nlo}}_{\text{sw}}(1\ll k\eta_* \ll k_p\eta_*) &=&  \frac{4+\nu(3+\nu)}{(2c_s x\uptau_*)^2} \frac{c_s^2 \mu^2}{(k\eta_*)^2} \frac{3+\mu^2 c_s^2}{(1-\mu^2c_s^2)^3} ,\\ 
        \Delta^{\text{nlo}}_{\Phi}(1\ll k\eta_* \ll k_p\eta_*) &=& 2\frac{(1+\nu) (2+\nu)}{(2 c_s x \uptau_*)^2} \frac{1}{(k\eta_*)^2}  \frac{1+\mu^2 c_s^2}{(1-\mu^2c_s^2)^2}.
    \end{eqnarray}
\end{subequations}
We now find it convenient to rewrite the dimensionless spectral density~\eqref{eq:pgw} using $\rho(z, x, y) = (1-\mu^2)^2 x^3 z^2 /y$ and $y\,dy = - xz\,d\mu$. At leading-order in $z/x \ll 1$, we can further approximate $\tilde{P}_v(y) \simeq \tilde{P}_v(x)$ and write
\begin{equation}
    \tilde{P}_{\text{gw}}^{\text{int}} (kR_*) \simeq \frac{\uptau_*}{\pi^2} \int_0^\infty dx \, x^2 \tilde{P}^2_v(x) 
    \int_{-1}^{1} d\mu (1-\mu^2)^2 \Delta^{\text{int}} (z, \mu, x, \uptau_*, \uptau_{\text{end}}),
\end{equation}
where the superscript ``$\text{int}$'' specifies that the quantity is evaluated in the intermediate-frequency range $1\ll k\eta_*\ll k_p\eta_*$. 
The integration over the angle $\mu$ can now be performed on a case-by-case basis with the expressions~\eqref{eq:kernel_sub_int}, resulting in
\begin{subequations}
    \begin{eqnarray}
        \tilde{P}^{\text{lo,int}}_{\text{gw, sw}} (kR_*) &\simeq & \frac{4}{3c_s^4}  \left[ 3-2c_s^2 - \frac{3}{c_s}(1-c_s^2) \operatorname{arctanh}(c_s) \right]   \frac{\mathcal{I}_v}{\uptau_* z^2},\\
        \tilde{P}^{\text{nlo, int}}_{\text{gw, gw}} (kR_*) &\simeq &  -\frac{2\nu (1+\nu)}{c_s^4}    \left[ 3   - \frac{1}{c_s}(3-c_s^2) \operatorname{arctanh}(c_s) \right]   \frac{\mathcal{I}_v}{\uptau_*^3 z^4}, \\
        \tilde{P}^{\text{nlo, int}}_{\text{gw, sw}} (kR_*) &\simeq & -2\frac{4+\nu(3+\nu)}{3c_s^6}  \left[ 15 - 4c_s^2  - \frac{3}{c_s}(5-3c_s^2) \operatorname{arctanh}(c_s) \right]   \frac{\mathcal{J}_v}{\uptau_*^3 z^2}, \\
        \tilde{P}^{\text{nlo, int}}_{\text{gw}, \Phi} (kR_*) &\simeq & 8\frac{(1+\nu) (2+\nu)}{3 c_s^6 }   \left[ 3-2c_s^2 - \frac{3}{c_s}(1-c_s^2) \operatorname{arctanh}(c_s) \right]   \frac{\mathcal{J}_v}{\uptau_*^3 z^2},
    \end{eqnarray}
\end{subequations}
where we have defined
\begin{equation}
    \mathcal{I}_v \equiv \frac{1}{2\pi^2} \int_0^\infty dx x^2 \tilde{P}_v^2(x), \qquad \mathcal{J}_v \equiv \frac{1}{2\pi^2} \int_0^\infty dx \tilde{P}_v^2(x).
\end{equation}
Recalling the definition $\uptau_* \equiv \eta_*/R_* \sim (\mathcal{H}_*R_*)^{-1}$, we notice that the general relativistic corrections are suppressed by a factor $(\mathcal{H}_*R_*)^{2}$ with respect to the leading-order term. Moreover, we notice that the correction coming from the modified propagation of gravitational waves, $\tilde{P}^{\text{int, nlo}}_{\text{gw, gw}} (kR_*) $, decays faster with the gravitational wave frequency. At the level of the power spectrum~\eqref{eq:Pgw}, given that $\mathcal{P}_{\text{gw}} \sim (kR_*)^3 \tilde{P}_{\text{gw}}(kR_*) $, we find $\mathcal{P}^{\text{lo, int}}_{\text{gw, sw}} (k)\sim \mathcal{P}^{\text{nlo, int}}_{\text{gw, sw}}(k)\sim \mathcal{P}^{\text{nlo, int}}_{\text{gw}, \Phi} (k)\sim k^1$, while $\mathcal{P}^{\text{int, lo}}_{\text{gw, sw}}(k)\sim k^{-1}$.

\subsubsection{Approximation at high frequency}\label{sec:appendix_high}
Let us finally consider gravitational wave modes around or beyond the energy injection scale $(k\sim k_p) \cup (k> k_p)$, which implies $kR_*\gtrsim 2\pi$. For a sufficiently long-lasting source $ k(\eta_{\text{end}}- \eta_*) \gg 1$, these mode oscillate many times during the acoustic phase. In this regime, the damped oscillatory behaviour of the kernel allows us to apply some further simplifications. Following the analysis of Refs.~\cite{Hindmarsh:2016lnk, Hindmarsh:2019phv}, let us start with a change of integration variables $\uptau_+ = (\uptau_1 +\uptau_2)/2$ and $\uptau_- = \uptau_1-\uptau_2$, so that
\begin{subequations}\label{eqs:kernel_high_0}
    \begin{eqnarray}
    \Delta^{\text{lo}}_{\text{sw}} &=& \frac{\uptau_*^{2\nu}}{8}\sum_{m, n = \pm 1} \int_{\uptau_*}^{\uptau_{\text{end}}}d\uptau_+ \int_{-\bar{\uptau}_+}^{\bar{\uptau}_+}d\uptau_-   \frac{\cos(\omega_{mn}\uptau_-)}{\left(\uptau_+^2 -\uptau_-^2/4\right)^{1+\nu}}, \\
    \Delta^{\text{nlo}}_{\text{gw}} &=& \frac{\uptau_*^{2\nu}}{8} \frac{\nu(1+\nu)}{2z} \sum_{m, n = \pm 1} \int_{\uptau_*}^{\uptau_{\text{end}}}d\uptau_+ \int_{-\bar{\uptau}_+}^{\bar{\uptau}_+}d\uptau_-   \frac{\uptau_-}{\left(\uptau_+^2 -\uptau_-^2/4\right)^{2+\nu}} \sin(\omega_{mn}\uptau_-), \\
    \Delta^{\text{nlo}}_{\text{sw}} &=& \frac{\uptau_*^{2\nu}}{8} \frac{4+\nu(3+\nu)}{2c_s}\sum_{m, n = \pm 1}  \left( \frac{n}{y} +\frac{m}{x}\right) \int_{\uptau_*}^{\uptau_{\text{end}}}d\uptau_+ \int_{-\bar{\uptau}_+}^{\bar{\uptau}_+}d\uptau_-   \frac{\uptau_-}{\left(\uptau_+^2 -\uptau_-^2/4\right)^{2+\nu}} \sin(\omega_{mn}\uptau_-),  \nonumber\\
    && \\
    \Delta^{\text{nlo}}_\Phi &=& -\frac{1}{8} \frac{(1+\nu)(2+\nu)}{c_s^2} \sum_{m, n = \pm 1} \frac{mn}{xy} \int_{\uptau_*}^{\uptau_{\text{end}}}d\uptau_+ \int_{-\bar{\uptau}_+}^{\bar{\uptau}_+}d\uptau_-   \frac{\uptau_+^2 -\uptau_-^2/4}{\left(\uptau_+^2 -\uptau_-^2/4\right)^{3+\nu}} \sin(\omega_{mn}\uptau_-),\nonumber\\
    &&
\end{eqnarray}
\end{subequations}
where, as shown in Ref.~\cite{RoperPol:2023dzg}, the extrema of integration are set by
\begin{equation}
    \bar{\uptau}_+ = \left\lbrace \begin{array}{cc}
         2(\uptau_+ -\uptau_*) & \rm{if} \quad \uptau_+ < \uptau_{\rm{m}}  ,\\
         2(\uptau_{\rm{end}} -\uptau_+) & \rm{if} \quad \uptau_+ \geq \uptau_{\rm{m}},  
    \end{array}\right.
\end{equation}
with $\uptau_{\rm{m}} = (\uptau_{\rm{end}} - \uptau_*)/2$ the half source duration.

To pursue an analytical estimation, we now introduce some approximations.  
Due to the oscillatory behaviour of the Green's functions and thus of the integrands in equations~\eqref{eqs:kernel_high_0}, sound waves rapidly decorrelate after a few oscillations as $\omega_{mn}\uptau_-\rightarrow \infty$, so that gravitational waves are mostly sourced in the first Hubble time~\cite{Guo:2020grp}. In the limit of long-lasting source, modes around the energy injection scale oscillate several times within the duration of the source $k(\eta_{\rm{end}}- \eta_*) \gg 1$, so that the integrands in equations~\eqref{eqs:kernel_high_0} have enough time to perform several oscillations within the support of $\uptau_-$. For this reason, extending the support of $\uptau_-$ to $\uptau_-\in (-\infty, \infty)$ does not affect the final result much, as it adds to the integral regions of phase space where integrands oscillate very fast with zero average.
Similarly, in the limit $k(\eta_{\rm{end}}- \eta_*) \gg 1$, the largest contributions to the kernels come from the region $\uptau_- \ll \uptau_+$ (see for example Ref.~\cite{RoperPol:2023dzg} Figure 5). We then evaluate the integrand of the kernel contributions~\eqref{eqs:kernel_high_0} using a Taylor expansion in $\uptau_-/\uptau_+\ll 1$.  Neglecting contributions of order $(\uptau_-/\uptau_+)^2$, we perform the integration over $\uptau_+$ resulting in 
\begin{subequations}\label{eqs:kernel_high_1}
    \begin{eqnarray}
    \Delta^{\text{lo}}_{\text{sw}} &\simeq & \frac{\uptau_*^{-1}}{8}\frac{1}{1+2\nu} \left[1- \left(\frac{\uptau_*}{\uptau_{\text{end}}}\right)^{1+2\nu}\right] \sum_{m, n = \pm 1}  \int_{-\uptau_{\rm{m}}}^{\uptau_{\rm{m}}}d\uptau_-   \cos(\omega_{mn}\uptau_-), \\
    \Delta^{\text{nlo}}_{\text{gw}} &=& \frac{\nu(1+\nu)}{16z} \frac{\uptau_*^{-3}}{3+2\nu} \left[1- \left(\frac{\uptau_*}{\uptau_{\text{end}}}\right)^{3+2\nu}\right] \sum_{m, n = \pm 1}   \int_{-\uptau_{\rm{m}}}^{\uptau_{\rm{m}}}d\uptau_-   \uptau_- \sin(\omega_{mn}\uptau_-), \\
    \Delta^{\text{nlo}}_{\text{sw}} &=&  \frac{4+\nu(3+\nu)}{16 c_s}\frac{\uptau_*^{-3}}{3+2\nu} \left[1- \left(\frac{\uptau_*}{\uptau_{\text{end}}}\right)^{3+2\nu}\right] \sum_{m, n = \pm 1}  \left( \frac{n}{y} +\frac{m}{x}\right) \int_{-\uptau_{\rm{m}}}^{\uptau_{\rm{m}}}d\uptau_-   \uptau_- \sin(\omega_{mn}\uptau_-),  \nonumber\\
    && \\
    \Delta^{\text{nlo}}_\Phi &=& - \frac{(1+\nu)(2+\nu)}{4 c_s^2} \frac{\uptau_*^{-3}}{3+2\nu} \left[1- \left(\frac{\uptau_*}{\uptau_{\text{end}}}\right)^{3+2\nu}\right] \sum_{m, n = \pm 1} \frac{mn}{xy} \int_{-\uptau_{\rm{m}}}^{\uptau_{\rm{m}}}d\uptau_-  \cos(\omega_{mn}\uptau_-).\nonumber\\
    &&
\end{eqnarray}
\end{subequations}
We note that every term of the kernel is proportional to the suppression factor
\begin{equation}\label{eq:Upsilon}
    \Upsilon_\ell \left(\frac{\eta_*}{\eta_{\rm{end}}}\right) = \frac{1}{\ell(\nu)} \left[1- \left(\frac{\eta_*}{\eta_{\text{end}}}\right)^{\ell(\nu)}\right],
\end{equation}
where $\ell(\nu) = 1+2\nu$ for the leading-order term, and $\ell(\nu) = 3+2\nu$ for the NLO terms. This suppression arises from the finite lifetime of sound waves, and it was first studied for the leading-order term in Ref.~\cite{Guo:2020grp}. Our suppression factor in the leading-order term $\Upsilon_{1+2\nu}$ reduces to the known expression $\Upsilon_1$ in the case of pure radiation ($\omega = 1/3$ and $\nu = 0$)~\cite{Guo:2020grp, RoperPol:2023dzg} and  $\Upsilon_3$ in the case of pressureless matter ($\omega = 0$ and $\nu = 1$)~\cite{Guo:2020grp}. 

If the source lasts for many Hubble times $\uptau_{\text{end}} \gg \uptau_*$, we can consider the limit $\uptau_{\rm{m}} \rightarrow \infty$ and use the properties of the Dirac delta function $\delta(x)$
\begin{subequations}
    \begin{eqnarray}
        \int_{-\infty}^{\infty} d\uptau_-  \cos(\omega_{mn}\uptau_-) &=& 2\pi\, \delta(\omega_{mn}),\\
        \int_{-\infty}^{\infty} d\uptau_- \uptau_-  \sin(\omega_{mn}\uptau_-) &=& -\frac{n}{c_s}\partial_{y} \int_{-\infty}^{\infty} d\uptau_-  \cos(\omega_{mn}\uptau_-) = - \frac{2\pi n}{c_s} \partial_{y}\delta(\omega_{mn}).
    \end{eqnarray}
\end{subequations}
We notice at this point that the Dirac delta functions are all centred at $z +c_s(mx +ny) = 0$. Since in the gravitational wave power spectrum~\eqref{eq:Pgw} the variables $x$ and $y$ are weighted by the spectral densities $P_v(x)$ and $P_v(y)$ via equation~\eqref{eq:pgw}, in the high gravitational frequency regime $z\gtrsim \mathcal{O}(1)$ only the case $m=n=-1$ can realize the condition $\omega_{mn}=0$. This way we finally obtain 
\begin{subequations}\label{eqs:kernel_high}
    \begin{eqnarray}
    \Delta^{\text{lo}}_{\text{sw}} &\simeq & \frac{\uptau_*^{-1}}{8}\Upsilon_{1+2\nu} \left(\frac{\eta_*}{\eta_{\rm{end}}}\right) \, 2\pi\, \delta(z-c_s(x+y)), \label{delta_lo}\\
    \Delta^{\text{nlo}}_{\text{gw}} &\simeq& \uptau_*^{-3} \, \frac{\nu(1+\nu)}{16 c_s z} \Upsilon_{3+2\nu} \left(\frac{\eta_*}{\eta_{\rm{end}}}\right) \, 2\pi\, \partial_{y}\delta(z-c_s(x+y)), \label{delta_nlo_gw}\\
    \Delta^{\text{nlo}}_{\text{sw}} &\simeq&  - \uptau_*^{-3}\, \frac{4+\nu(3+\nu)}{16 c_s^2} \Upsilon_{3+2\nu} \left(\frac{\eta_*}{\eta_{\rm{end}}}\right)\,   \left( \frac{1}{y} +\frac{1}{x}\right) 2\pi\, \partial_{y}\delta(z-c_s(x+y)), \label{delta_nlo_sw} \qquad\\
    \Delta^{\text{nlo}}_\Phi &\simeq& - \uptau_*^{-3} \, \frac{(1+\nu)(2+\nu)}{4 c_s^2 xy} \Upsilon_{3+2\nu} \left(\frac{\eta_*}{\eta_{\rm{end}}}\right) \,   2\pi\, \delta(z-c_s(x+y)).\label{delta_phi}
\end{eqnarray}
\end{subequations}

\paragraph{High-frequency approximation of the gravitational wave spectral density function}
The Dirac delta function can be now used to perform one integration over the sound wave wavenumbers in the expression of the gravitational wave dimensionless spectral density function~\eqref{eq:pgw}
\begin{equation}
    \tilde{P}_{\text{gw}} (kR_*) = \frac{\uptau_*}{\pi^2 z^3}  \int_0^\infty dx \int_{\vert x - z\vert}^{x+z}dy \, \rho(z, x, y) \tilde{P}_v(x) \tilde{P}_v(y) \Delta \left(z, x, y, \uptau_*, \uptau_{\text{end}}\right).
\end{equation}
Studying each contribution individually, the integration of the gravitational wave spectral density function can be split into the evaluation of the terms 
\begin{subequations}
    \begin{eqnarray}
        \mathcal{I} &=&   \int_0^\infty dx \int_{\vert x - z\vert}^{x+z}dy \, \rho(z, x, y) f(x, y) \tilde{P}_v(x) \tilde{P}_v(y) \, \delta(z-c_s(x+y)),\\
        \mathcal{J} &=&   \int_0^\infty dx \int_{\vert x - z\vert}^{x+z}dy \, \rho(z, x, y) g(x, y) \tilde{P}_v(x) \tilde{P}_v(y) \, \partial_{y}\delta(z-c_s(x+y)).
    \end{eqnarray}
\end{subequations}
where $f(x, y) = 1$ for the leading-order contribution~\eqref{delta_lo}, and $f(x, y) = 1/(xy)$ for the  contribution from curvature perturbations~\eqref{delta_phi}; $g(x, y) = 1$ for the NLO contribution from gravitational wave propagation~\eqref{delta_nlo_gw}, and $g(x, y) = (x+y)/(xy)$ for the  NLO contribution from sound wave propagation~\eqref{delta_nlo_sw}.
The $y$-integration in $\mathcal{I}$ is now trivial, and gives
\begin{equation}
    \mathcal{I} =   \frac{z^2}{c_s}\left(\frac{1-c_s^2}{c_s^2}\right)^2\int_{x_-}^{x_+} dx \,  \frac{(x-x_+)^2(x-x_-)^2}{x(x_+ + x_- - x)} f(x, x_++x_--x) \tilde{P}_v(x) \tilde{P}_v(x_++x_--x) ,
\end{equation}
with $x_\pm = z(1\pm c_s)/(2c_s)$. Therefore
\begin{equation}
    \begin{split}
        \tilde{P}^{\text{lo, high}}_{\text{gw, sw}} (kR_*) = \frac{1}{4\pi c_s z} & \Upsilon_{1+2\nu}\left(\frac{\eta_*}{\eta_{\rm{end}}}\right) \left(\frac{1-c_s^2}{c_s^2}\right)^2 \times \\
        & \times \int_{x_-}^{x_+} dx \,  \frac{(x-x_+)^2(x-x_-)^2}{x(x_+ + x_- - x)} \tilde{P}_v(x) \tilde{P}_v(x_++x_--x) ,
    \end{split}
\end{equation}
and 
\begin{equation}
    \begin{split}
        \tilde{P}^{\text{nlo, high}}_{\text{gw, }\Phi} (kR_*) = - \uptau_*^{-2} \, \frac{(1+\nu)(2+\nu)}{2\pi c_s^3 z} & \Upsilon_{3+2\nu}\left(\frac{\eta_*}{\eta_{\rm{end}}}\right) \left(\frac{1-c_s^2}{c_s^2}\right)^2 \times \\
        & \times \int_{x_-}^{x_+} dx \,  \frac{(x-x_+)^2(x-x_-)^2}{x^2(x_+ + x_- - x)^2} \tilde{P}_v(x) \tilde{P}_v(x_++x_--x) .
    \end{split}
\end{equation}

In $\mathcal{J}$ instead, we carry out the $y$-integration by parts. Remembering the fact that the geometric function $\rho(z, x, y)$, defined in equation~\eqref{eq:rho_1}, vanishes at the extrema of integration when $y = x+z$ or $y = \vert x-z\vert$, we can write
\begin{eqnarray}
    \mathcal{J} &=&  -\int_0^\infty dx \tilde{P}_v(x) \int_{\vert x - z\vert}^{x+z}dy \, \partial_{y} \left[ \rho(z, x, y) g(x,y) \tilde{P}_v(y)\right] \, \delta(z-c_s(x+y)) = \nonumber\\
    &=& -\frac{1}{c_s}\int_{x_-}^{x_+} dx \tilde{P}_v(x)  \partial_{y} \left[ \rho(z, x, y) g(x,y) \tilde{P}_v(y)\right]_{y = x_+ +x_- -x}.
\end{eqnarray}
Finally
\begin{subequations}
\begin{align} 
    \tilde{P}^{\text{nlo, gw}}_{\text{gw}} (kR_*)& = - \uptau_*^{-2}\, \frac{\nu(1+\nu)}{8 \pi c_s^2 z^2}\Upsilon_{3+2\nu}\left(\frac{\eta_*}{\eta_{\rm{end}}}\right)  \left(\frac{1-c_s^2}{c_s^2}\right)^2   \times \nonumber\\
    &  \times \int_{x_-}^{x_+} dx \frac{(x-x_+)^2(x-x_-)^2}{x(x_+ + x_- - x)^2} \tilde{P}_v(x) \tilde{P}_v(x_+ +x_- -x) \times \nonumber\\
    & \times \left\{ \frac{7y^4 - 6y^2(x^2+z^2) - (x^2-z^2)^2}{\left[y^2 -(x-z)^2\right] \left[y^2 - (x+z)^2\right]}  +2\left[1-\frac{3(y/2\pi)^6}{1+(y/2\pi)^6} \right]\right\}, \\
    &\nonumber\\
    \tilde{P}^{\text{nlo, high}}_{\text{gw, sw}} (kR_*)& = \uptau_*^{-2} \, \frac{4+\nu(3+\nu)}{8 \pi c_s^3 z} \Upsilon_{3+2\nu}\left(\frac{\eta_*}{\eta_{\rm{end}}}\right) \left(\frac{1-c_s^2}{c_s^2}\right)^2   \times \nonumber\\
    &  \times \int_{x_-}^{x_+} dx \frac{(x-x_+)^2(x-x_-)^2}{x^2(x_+ + x_- - x)^3} (x_+ +x_-)  \tilde{P}_v(x) \tilde{P}_v(x_+ +x_- -x) \times \nonumber \\
    & \times \left\{ \frac{7y^4 - 6y^2(x^2+z^2) - (x^2-z^2)^2}{\left[y^2 -(x-z)^2\right] \left[y^2 - (x+z)^2\right]} -\frac{x}{x+y} +2\left[1-\frac{3(y/2\pi)^6}{1+(y/2\pi)^6} \right]\right\}, 
\end{align}
\end{subequations}
with $y = x_+ +x_- -x$.

\subsection{Immediate return to pure radiation}\label{app_rad}

When the soft phase lasts for a sufficiently small amount of time, so that we can consider the equation of state to be that of pure radiation throughout the transition, the kernel~\eqref{eq:kern_app} can be integrated analytically across the entire gravitational wave frequency domain. Indeed, in pure radiation
\begin{equation}
    G^\prime_k(\eta, \eta_1)G^\prime_k(\eta, \eta_2)\bigg\vert_{\nu= 0} \simeq \frac{1}{2}  \cos\left[k(\eta_1-\eta_2) \right].
\end{equation}
Since this is exactly the leading-order contribution of equation~\eqref{eq:gpgp_result}, the calculation of the kernel functions proceeds very similarly to the case outlined in Section~\ref{app_large}. Taking the result in equation~\eqref{eqs:delta_appendix} and setting $\nu =0$, we write the kernel functions
\begin{subequations}\label{kernel_rad_app}
    \begin{eqnarray}
    \Delta^{\text{lo}}_{\text{sw}}\Big\vert_{\nu =0} &=& \frac{1}{8}\sum_{m, n = \pm 1} \int_{\uptau_*}^{\uptau_{\text{end}}} \frac{d\uptau_1 d\uptau_2}{\uptau_*^2}  \left(\frac{\uptau_*^2}{\uptau_1\uptau_2}\right) \times \nonumber \\
    && \qquad\qquad\qquad\quad \times \Big\{\cos(\omega_{m n}\uptau_1) \cos(\omega_{m n}\uptau_2) +\sin(\omega_{m n}\uptau_1) \sin(\omega_{m n}\uptau_2)\Big\},\\
    \Delta^{\text{nlo}}_{\text{gw}}\Big\vert_{\nu =0} &=& 0, \\
    \Delta^{\text{nlo}}_{\text{sw}}\Big\vert_{\nu =0} &=& -\frac{1}{4c_s}\sum_{m, n = \pm 1}  \left( \frac{n}{y} +\frac{m}{x}\right)\int_{\uptau_*}^{\uptau_{\text{end}}} \frac{d\uptau_1 d\uptau_2}{\uptau_*^2}   \left(\frac{\uptau_*^2}{\uptau_1\uptau_2}\right)    \left(\frac{1}{\uptau_1} - \frac{1}{\uptau_2}\right) \times \nonumber\\
    && \qquad\qquad\qquad\quad \times \Big\{\sin(\omega_{m n}\uptau_1) \cos(\omega_{m n}\uptau_2) - \cos(\omega_{m n}\uptau_1) \sin(\omega_{m n}\uptau_2) \Big\},\qquad\\
    \Delta^{\text{nlo}}_\Phi\Big\vert_{\nu =0} &=& -\frac{1}{4} \frac{1}{c_s^2} \sum_{m, n = \pm 1} \frac{mn}{xy} \int_{\uptau_*}^{\uptau_{\text{end}}} \frac{d\uptau_1 d\uptau_2}{\uptau_*^2}   \left(\frac{\uptau_*^2}{\uptau_1\uptau_2}\right)  \bigg(\frac{1}{\uptau_2^2} + \frac{1}{\uptau_1^2}\bigg)\times\nonumber\\
    && \qquad\qquad\qquad\quad\times \Big\{\cos(\omega_{m n}\uptau_1) \cos(\omega_{m n}\uptau_2) + \sin(\omega_{m n}\uptau_1) \sin(\omega_{m n}\uptau_2) \Big\}.
\end{eqnarray}
\end{subequations}
Finally, using equations~\eqref{eqs:sici}, we recover the expressions of the kernel contributions~\eqref{kernel_rad}. 

\section{Details on the numeric integration}\label{app:num}
We pursue the numeric integration of the kernel functions~\eqref{eq:gg_low_1} and~\eqref{eq:kernel_sub} using a \texttt{Cython} code and the \texttt{SciPy} library. However,
the generalized trigonometric functions $\operatorname{ci}_{-\alpha}(x)$ and $\operatorname{si}_{-\alpha}(x)$ are only defined in the library for the case $\alpha=0$. For the cases where $\alpha$ has a non-integer value, we need instead to relate the generalized trigonometric functions to other known geometric functions. This is done in the Appencix (C) of Ref.~\cite{Giombi:2024kju}, from which we take here the main result
\begin{subequations}\label{eqs:gen_trig}
    \begin{eqnarray}
        \operatorname{ci}_{-\alpha}(x) &=&  \operatorname{Re}\left\{ i^{\alpha}\left[ \Gamma(-\alpha) + \frac{(ix)^{-\alpha}}{\alpha}\, _1F_1(-\alpha,1-\alpha,-ix) \right]\right\}, \\
        \operatorname{si}_{-\alpha}(x) &=&  -  \operatorname{Im}\left\{ i^{\alpha}\left[ \Gamma(-\alpha) + \frac{(ix)^{-\alpha}}{\alpha}\, _1F_1(-\alpha,1-\alpha,-ix) \right]\right\},
    \end{eqnarray}
\end{subequations}
where $_1F_1(a,b,c)$ is the Krummer confluent hypergeometric function~\cite{abramowitz1948handbook}~\cite[\href{https://dlmf.nist.gov/8.5.E1}{(8.5.1)}]{NIST:DLMF}. The above expressions are ill-defined in the limit $x\rightarrow 0$. To avoid this problem we consider the limit $x\rightarrow 0$ of the original expressions~\eqref{eqs:trigo} as
\begin{equation}
    \operatorname{ci}_{-\alpha}(x\rightarrow 0) =  \frac{x^{-\alpha}}{\alpha}, \qquad \operatorname{si}_{-\alpha}(x\rightarrow 0) = 0.
\end{equation}
The apparent divergence in $\operatorname{ci}_{-\alpha}(x\rightarrow 0)$ is removed inside the kernels~\eqref{eq:gg_low_1} and~\eqref{eq:kernel_sub}, yielding always to finite results.

The expressions~\eqref{eqs:gen_trig} are well defined for non-integer orders $\alpha$ of the trigonometric integrals. In the kernel functions~\eqref{eq:gg_low_1} and~\eqref{eq:kernel_sub}, since $\nu\in [0, 1)$, trigonometric integrals of integer order only appear in the pure radiation case, where $\nu = 0$. The kernel functions in this case are given by equations~\eqref{kernel_rad_app}. For trigonometric integrals of integer orders we can derive a recurrence formula to reduce the order of the generalized trigonometric integrals. In particular
\begin{eqnarray}        
    \int_{\uptau_*}^{\uptau_{\text{end}}} \frac{\sin(\omega \uptau)}{\uptau^2} d\uptau^2  &=& -\operatorname{sign}(\omega) \vert \omega  \vert \left[ \frac{\sin(x)}{x} \bigg\vert_{x=\vert\omega\vert \uptau_*}^{x=\vert\omega\vert \uptau_{\text{end}}} -  \int_{\vert \omega\vert \uptau_*}^{\vert \omega\vert \uptau_{\text{end}}} \frac{\cos(y)}{y} dy \right]  =\nonumber\\
    &=& -\operatorname{sign}(\omega) \vert \omega \vert \left[ \frac{\sin(x)}{x} - \operatorname{Ci}(x)\right]_{x=\vert\omega\vert \uptau_*}^{x=\vert\omega\vert \uptau_{\text{end}}},
\end{eqnarray}
where in the first step we used integration by parts. We also recall that, in our notation, $\operatorname{ci}_0(x) = -\operatorname{Ci}(x)$.
In the same fashion we find
\begin{subequations}
    \begin{eqnarray}
        \int_{\uptau_*}^{\uptau_{\text{end}}} \frac{\cos(\omega \uptau)}{\uptau^2} d\uptau^2  &=& - \vert \omega \vert \left[ \frac{\cos(x)}{x} + \operatorname{si}(x)\right]_{x=\vert\omega\vert \uptau_*}^{x=\vert\omega\vert \uptau_{\text{end}}} ,  \\
        \int_{\uptau_*}^{\uptau_{\text{end}}} \frac{\sin(\omega \uptau)}{\uptau^3} d\uptau^2  &=& - \frac{\operatorname{sign}(\omega)}{2}\vert \omega \vert^2 \left[ \frac{\sin(x)}{x^2} + \frac{\cos(x)}{x} + \operatorname{si}(x)\right]_{x=\vert\omega\vert \uptau_*}^{x=\vert\omega\vert \uptau_{\text{end}}} ,\\
        \int_{\uptau_*}^{\uptau_{\text{end}}} \frac{\cos(\omega \uptau)}{\uptau^3} d\uptau^2  &=& - \frac{1}{2}\vert \omega \vert^2 \left[ \frac{\cos(x)}{x^2} - \frac{\sin(x)}{x} + \operatorname{ci}(x)\right]_{x=\vert\omega\vert \uptau_*}^{x=\vert\omega\vert \uptau_{\text{end}}}.
    \end{eqnarray}
\end{subequations}
Finally, implementing this results in the expressions~\eqref{kernel_rad_app}, we obtain
\begin{subequations}\label{eq:kernel_sub_rad}
    \begin{eqnarray}
        \Delta^{\text{lo}}_{\text{sw}} &=& \frac{1}{8} \sum_{m, n = \pm 1} \left[\left(\operatorname{Ci}(\omega_{m n} \uptau)\Big\vert^{\uptau_{\text{end}}}_{\uptau_*} \right)^2  + \left( \operatorname{si}(\omega_{m n} \uptau)\Big\vert^{\uptau_{\text{end}}}_{\uptau_*} \right)^2 \right],\\
        \Delta^{\text{nlo}}_{\text{gw}} &=& 0, \\
        \Delta^{\text{nlo}}_{\text{sw}} &=&  \frac{1}{2c_s} \sum_{m, n = \pm 1} \left(\frac{n}{y} + \frac{m}{x}\right) \left\vert\omega_{mn}\right\vert \operatorname{sign}(\omega_{mn})\times\qquad\qquad\qquad\qquad\qquad\qquad\qquad  \nonumber\\
        && \qquad\qquad\times\bigg\{ \operatorname{Ci}(\omega_{m n} \uptau)\Big\vert^{\uptau_{\text{end}}}_{\uptau_*}  \left[ \frac{\sin(\vert\omega_{mn}\vert \uptau)}{\vert\omega_{mn}\vert \uptau} - \operatorname{Ci}(\vert\omega_{mn}\vert \uptau)\right]_{\uptau_*}^{\uptau_{\text{end}}} - \nonumber\\
        && \qquad\qquad\qquad - \operatorname{si}(\omega_{m n} \uptau)\Big\vert^{\uptau_{\text{end}}}_{\uptau_*}\left[ \frac{\cos(\vert\omega_{mn}\vert \uptau)}{\vert\omega_{mn}\vert \uptau} + \operatorname{si}(\vert\omega_{mn}\vert \uptau)\right]_{\uptau_*}^{\uptau_{\text{end}}} \Bigg\},\\ 
        \Delta^{\text{nlo}}_{\Phi} &=& \frac{1}{4 c_s^2}  \sum_{m, n = \pm 1} mn \frac{\omega^2_{mn}}{xy} \times  \nonumber\\
        && \quad\times\bigg\{ \operatorname{Ci}(\omega_{m n} \uptau)\Big\vert^{\uptau_{\text{end}}}_{\uptau_*}  \left[ \frac{\cos(\vert\omega_{mn}\vert \uptau)}{(\vert\omega_{mn}\vert \uptau)^2} -  \frac{\sin(\vert\omega_{mn}\vert \uptau)}{\vert\omega_{mn}\vert \uptau} + \operatorname{Ci}(\vert\omega_{mn}\vert \uptau)\right]_{\uptau_*}^{\uptau_{\text{end}}} - \nonumber\\
        && \qquad\quad - \operatorname{si}(\omega_{m n} \uptau)\Big\vert^{\uptau_{\text{end}}}_{\uptau_*}\left[ \frac{\sin(\vert\omega_{mn}\vert \uptau)}{(\vert\omega_{mn}\vert \uptau)^2} + \frac{\cos(\vert\omega_{mn}\vert \uptau)}{\vert\omega_{mn}\vert \uptau} + \operatorname{si}(\vert\omega_{mn}\vert \uptau)\right]_{\uptau_*}^{\uptau_{\text{end}}} \Bigg\}.
    \end{eqnarray}
\end{subequations}

\bibliographystyle{JHEP}
\bibliography{biblio}
\end{document}